\documentclass[11pt,a4paper,longbibliography]{article}
\usepackage{jheppub}
\usepackage{graphicx}
\usepackage{latexsym}
\usepackage{amsfonts}
\usepackage{amssymb}
\usepackage{amsmath}
\usepackage{bm}
\usepackage{mathrsfs}
\usepackage{slashed}
\usepackage{ascmac}

\usepackage{color}

\newcommand{\rate}{D} % photon--to--dilepton conversion rate 
\renewcommand{\k}{\kappa} % label of the spin basis
\newcommand{\Gam}{ {\mathcal G}  } % Coefficient function
\renewcommand{\H}{ {\mathcal J}  } % Coefficient function
\newcommand{\prj}{ {\mathcal P} } %projection operator
\newcommand{\Ritus}{ {\mathcal R}} %Ritus basis
\newcommand{\qn}{  \chi  } %additional quantum number
\newcommand{\Q}{ {\mathcal Q} } %helicity-projection operators

\newcommand{\id}{\mbox{1}\hspace{-0.25em}\mbox{l}}
\newcommand{\sgn}{ {\rm sgn} }

\title{Di-lepton production from a single photon in strong magnetic fields: Vacuum dichroism}

\author[a]{Koichi Hattori}
\affiliation[a]{Yukawa Institute for Theoretical Physics, Kyoto University, Kyoto 606-8502, Japan}

\author[b]{Hidetoshi Taya}
\affiliation[b]{Research and Education Center for Natural Sciences, Keio University 4-1-1 Hiyoshi, Kohoku-ku, Yokohama, Kanagawa 223-8521, Japan}

\author[c,d]{Shinsuke Yoshida}
\affiliation[c]{Guangdong Provincial Key Laboratory of Nuclear Science, Institute of Quantum Matter, 
South China Normal University, Guangzhou 510006, China}
\affiliation[d]{Guangdong-Hong Kong Joint Laboratory of Quantum Matter, Southern Nuclear Science Computing Center, South China Normal University, Guangzhou 510006, China}

\preprint{YITP-20-136}
\abstract{ 
We study di-lepton production from a single photon in the presence of a strong constant magnetic field.  By the use of the Ritus-basis formalism, we analytically evaluate the photon--to--di-lepton conversion vertex with fully taking into account the non-perturbative interactions between the produced fermions and the strong magnetic field.  We show that the di-lepton spectrum becomes anisotropic with respect to the magnetic-field direction and depends on the photon polarization as a manifestation of the vacuum dichroism in a strong magnetic field.  According to the energy conservation in the presence of the Landau quantization, not only the transverse momentum of the produced fermions but also the longitudinal momentum is discretized, and the di-lepton spectrum exhibits spike structures as functions of the incident photon energy and the magnetic field strength.  We also show that the di-lepton production is strictly prohibited for massless fermions in the lowest Landau levels as an analogue of the so-called helicity suppression. 
}

\begin{document}

\maketitle
%\tableofcontents

\section{Introduction}

The vacuum structure of quantum electrodynamics (QED) is modified by the presence of electromagnetic fields, leading to non-trivial polarization of the vacuum.  In a seminal paper by Heisenberg and Euler \cite{Heisenberg:1935qt} as well as in several papers published in the same age \cite{sauter1931behavior, Euler:1935zz, Weisskopf:1996bu}, the authors addressed the vacuum polarization effects and showed their physical consequences such as the pair production from the vacuum by an electric field later called the Schwinger mechanism \cite{Schwinger:1951nm}.  The vacuum polarization by electromagnetic fields also affects propagation of probe particles, in particular photons.  Recall that, in the absence of electromagnetic fields, an on-shell real photon travels at the speed of light in vacuum without modification of the refractive index or conversion to di-lepton, even when the vacuum polarization effect is included.  On the contrary, Toll showed that the vacuum polarization effects in the presence of electromagnetic fields give rise to a non-trivial refractive index that can deviate from unity in a polarization-dependent manner \cite{Toll:1952rq}.  This phenomenon is called the {\it vacuum birefringence}, named after a similar optical property of materials.  In turn, the unitarity, or the optical theorem, implies the existence of an imaginary part in the refractive index, meaning that a single real photon can be converted to a di-lepton in electromagnetic fields.  According to the polarization dependence, this photon attenuation phenomenon may be called the {\it vacuum dichroism}.  There are a number of studies on the complex-valued refractive index (see, e.g., Refs.~\cite{Adler:1971wn, Tsai:1974fa, Tsai:1975iz, Shabad:1975ik, Melrose:1976dr, Urrutia:1977xb, Heyl:1997hr, Baier:2007dw, Shabad:2010hx, Hattori:2012je, Hattori:2012ny}).  

Experimental detection of the vacuum birefringence/dichroism, 
induced by the vacuum polarization with electromagnetic fields, has been quite challenging.  
Indeed, the vacuum birefringence/dichroism is strongly suppressed by the QED coupling constant 
higher than the fourth order as represented by box diagrams. 
Nevertheless, the QED coupling constant appears in a multiplicative form with the electromagnetic field strength, and thus this naive power counting is invalidated if the electromagnetic field strength compensates the suppression by the coupling constant.  Namely, strong electromagnetic fields are demanded for successful detection of the vacuum birefringence/dichroism.  

After more than a half century since publication of the classic papers, there is remarkable experimental progress to produce strong electromagnetic fields and/or to detect their signatures.  The intensity of high-power lasers has been increasing continuously and rapidly since the invention of the chirped pulse amplification \cite{DiPiazza:2011tq, Zhang:2020lxl, Ejlli:2020yhk}.  Also, as proposed in classic papers \cite{Fermi, Weizsacker, Williams, Breit:1934zz}, highly accelerated nuclei can be used as a source of strong electromagnetic fields.  Implementation of such an idea has been realized recently with Relativistic Heavy Ion Collider (RHIC) and the Large Hadron Collider (LHC).  In particular, much attention has been paid to ultra-peripheral collision events \cite{Adams:2004rz, Aaboud:2017bwk, Sirunyan:2018fhl, Aad:2019ock, Adam:2019mby} where the two colliding nuclei are distant enough from each other so that QED processes dominate over quantum chromodynamics (QCD).  Besides, strong magnetic fields are thought to be realized in astronomical systems such as neutron stars/magnetars \cite{Kouveliotou:2003tb, Harding:2006qn, Staubert:2018xgw, bignami2003magnetic} and the primordial Universe \cite{Grasso:2000wj, Giovannini:2003yn, Kandus:2010nw}.  There are several future programs planned around the world to detect strong-magnetic field effects in the astronomical systems through, e.g., observation of X-ray and gamma-ray photons.  

Motivated by those developments, it is timely to enrich theoretical foundation of the vacuum polarization effects on photon by tractable analytic methods.  In particular, we in this paper focus on the di-lepton production by a single photon in the presence of a strong magnetic field.  We emphasize that the differential cross section for the di-lepton production computed in this paper provides more information than the aforementioned imaginary part of the refractive index, which corresponds to the integrated cross section.  Our results will open a new avenue to study the energy and momentum distributions of di-leptons as signatures of the vacuum dichroism and implicitly of the vacuum birefringence, since the vacuum birefringence and dichroism are both sides of a coin.  Besides, the differential di-lepton measurement is more feasible than a photon-polarization measurement in the gamma regime.  We focus on effects of a magnetic field and do not consider those of an electric field.  In general, a magnetic field is more stable and can have larger spacetime extension than an electric field, since a (constant) magnetic field does not exert work on charged particles and is not Debye screened by charge distributions.  In addition, in actual physics systems such as ultra-peripheral heavy-ion collisions and magnetosphere of neutron stars/magnetars, it is plausible to assume that the strong electromagnetic fields are dominated by a magnetic-field component.  

The crucial point of the calculation in our problem is to include the interaction between fermions and a strong magnetic field to all orders in the QED coupling constant.  As mentioned above, this treatment is necessary when the magnetic-field strength is large enough to break down the naive perturbation theory.  Indeed, the di-lepton production from a single real photon is not allowed in the naive perturbation theory, since the energy-momentum conservation and the on-shell conditions for a photon and fermions are not compatible with each other.  One could obtain a finite di-lepton production rate from a single on-shell photon only after including a non-perturbative modification of the fermion dispersion relation by a strong magnetic field.  We accomplish such a computation for on-shell as well as off-shell photons by the use of the Ritus-basis formalism, which is constructed with the exact wave function of a fermion in a magnetic field.  Accordingly, the di-lepton spectrum is naturally specified by the Landau levels and a still continuous momentum along the magnetic field.  This may be regarded as an extension of previous works~\cite{Hattori:2012je, Hattori:2012ny} in which one of the present authors showed the Landau-level structure appearing in the complex refractive index.  Within a constant configuration of a magnetic field, we provide the most general analytic form of the photon--to--di-lepton conversion rate, or the lepton tensor in a magnetic field, with all-order Landau levels (see also Refs.~\cite{Sadooghi:2016jyf, Ghosh:2018xhh, Wang:2020dsr, Ghosh:2020xwp} for the Landau levels appearing in photon/di-lepton production from finite-temperature plasma).  

This paper is organized as follows.  In Sec.~\ref{sec-2}, we first briefly review the Ritus-basis formalism in a self-contained way.  Based on this formalism, we show an analytic form of the lepton tensor in a magnetic field and its square in Sec.~\ref{sec-3} and inspect basic properties of the di-lepton production rate in Sec.~\ref{sec--4} with some numerical plots.  Section~\ref{summary} is devoted to summary and outlook.  In appendices, we provide the wave function of a charged particle in a magnetic field as an ingredient of the Ritus-basis formalism in Appendix~\ref{app-a} and rigorous consistency checks of the computation such as the gauge invariance in Appendix~\ref{app-b}. 

Throughout the paper, we take the direction of a constant external magnetic field in the $z$-direction.  Accordingly, we decompose the metric $g^{\mu\nu} \equiv {\rm diag}(+1,-1,-1,-1)$ and four-vectors $v^\mu$ into the longitudinal and transverse parts, respectively, as $ g^{\mu\nu}_\parallel \equiv {\rm diag}(1,0,0,-1) $ and $ g^{\mu\nu}_\perp \equiv {\rm diag}(0,-1,-1,0) $, and $v^{\mu}_\parallel \equiv g^{\mu\nu}_\parallel v_\nu $ and $ v^{\mu}_\perp \equiv g^{\mu\nu}_\perp v_\nu $.

\section{Preliminaries: Ritus-basis formalism for a Dirac fermion} \label{sec-2}

To provide a self-contained construction, we first review the Ritus-basis formalism~\cite{Ritus:1972ky, Ritus:1978cj}\footnote{This part is based on a forthcoming review article \cite{HIO}.}.  In case of a constant external magnetic field, the energy spectrum of charged fermions is subjected to the Landau quantization and the Zeeman shift.  The resultant energy level has two-fold spin degeneracy except for the unique ground state.  The Ritus basis is, then, introduced as a superposition of (projection operators for) the two degenerate spin states.  An advantage of the Ritus basis is that it maps the Dirac equation in an external field into a free Dirac equation, which is easier to handle.  

We start with the Dirac equation in an external magnetic field $A_{\rm ext}^\mu$, 
\begin{align}
	\left( i \slashed D _{\rm ext} - m \right) \psi = 0 \, , \label{eq:Dirac}
\end{align}
where the covariant derivative $D^\mu_{\rm ext}$ is given by 
\begin{align}
	D^\mu_{\rm ext} \equiv \partial ^\mu + i q A_{\rm ext}^\mu \, . \label{eq:covariantD-QED}
\end{align}
The electric charge $q$ takes a positive (negative) value for positively (negatively) charged fermions.  Since we only have a constant magnetic field, the longitudinal components of the gauge potential $[A_{\rm ext}]_\parallel$ must be constant in spacetime.  Without loss of generality, one may set
\begin{align}
	0 = A_{\rm ext}^0 = A_{\rm ext}^3\, .  
\end{align}
Whereas there is still a residual gauge freedom in the transverse components $[A_{\rm ext}]_\perp$, we first discuss gauge-independent properties until we fix the residual gauge in Eq.~(\ref{eq:Landau-gauge}). 

To proceed, we discuss the energy level of a charged fermion in the presence of a constant magnetic field.  To this end, it is convenient to rewrite the Dirac equation (\ref{eq:Dirac}) into a Klein-Gordon type equation, by using an identity $\gamma^\mu \gamma^\nu = \frac{1}{2} [\gamma^\mu, \gamma^\nu] + \frac{1}{2} \{ \gamma^\mu, \gamma^\nu \}$, as
\begin{align}
	\left( D_{\rm ext}^2 + m^2 + \frac{q}{2} F_{\rm ext}^{\mu\nu} \sigma_{\mu\nu}\right) \psi = 0\, , \label{eq:K-GandZeeman}
\end{align}
where $\sigma_{\mu\nu} \equiv \frac{i}{2}[\gamma_\mu, \gamma_\nu]$ is a spin operator and $F_{\rm ext}^{\mu\nu} \equiv [ D^{\mu}_{\rm ext}, D^{\nu}_{\rm ext} ]/iq = \partial^\mu A_{\rm ext}^{\nu}-\partial^\nu A_{\rm ext}^{\mu}$ is field strength.  For a constant magnetic field pointing in the $z$-direction, only $(1,2)$- and $(2,1)$-components of the field strength $F_{\rm ext}^{\mu\nu}$ are nonvanishing, i.e., we have a nonvanishing commutation relation only for the transverse components of the covariant derivative 
\begin{align}
	[ i D_{\rm ext}^1 , iD_{\rm ext}^2] = - i q F_{\rm ext}^{12} \equiv iqB\, ,
\end{align}
with $B$ denoting the nonvanishing $ 3 $-component of the magnetic field.  Those transverse components of the covariant derivative may be regarded as a pair of the canonical variables in nonrelativistic quantum mechanics.  Motivated by this analogy, we introduce ``creation and annihilation operators,'' denoted by $\hat{a}$ and $\hat{a}^\dagger$, respectively, as 
\begin{align}\label{eq:aadagger}
	\hat a \equiv \frac{1}{ \sqrt{2 |qB|} } ( i D_{\rm ext}^1 - \sgn(qB)  D_{\rm ext}^2 )\ ,\quad 
	\hat a^\dagger \equiv \frac{1}{ \sqrt{2 |qB|} } ( i D_{\rm ext}^1 + \sgn(qB)  D_{\rm ext}^2 )\, ,
\end{align}
which satisfy the following ``canonical commutation relation'': 
\begin{align}
	1 = [\hat{a}, \hat{a}^\dagger]\ ,\quad 0 = [\hat{a}, \hat{a}] = [\hat{a}^\dagger, \hat{a}^\dagger]\, .  \label{eq:com}
\end{align}  
Using $\hat{a}$ and $\hat{a}^\dagger$, one may reexpress the Klein-Gordon operator as
\begin{align}
	D_{\rm ext}^2 + m^2  =  \partial_t^2 - \partial_z^2 + \left(2 \hat a^\dagger \hat a +1\right)  \vert qB \vert + m^2 \, .
\end{align}
This expression gives the relativistic energy spectrum of the Landau level for a charged scalar particle.  For a charged fermion, which obeys Eq.~(\ref{eq:K-GandZeeman}), we have another term $\frac{q}{2} F_{\rm ext}^{\mu\nu} \sigma_{\mu\nu}$ in addition to the Klein-Gordon operator.  This term is responsible for the Zeeman effect.  To confirm this point, we introduce spin projection operators along the magnetic field\footnote{These operators have useful properties: $  \prj_\pm^\dagger = \prj_\pm$, $ \prj_+ + \prj_- = 1 $, $ \prj_\pm \prj_\pm = \prj_\pm $, and $\prj_\pm \prj_\mp =0  $.  Also, $\prj_\pm \gamma^\mu \prj_\pm = \gamma_\parallel^\mu \prj_\pm  $ and $\prj_\pm \gamma^\mu \prj_\mp = \gamma_\perp^\mu \prj_\mp  $. We will use those properties below. }  
\begin{align}
	\prj _\pm 
		\equiv \frac{1}{2} \left(1 \pm   \sigma^{12} \sgn(qB)  \right) 
		= \frac{1}{2} \left(1 \pm  i \gamma^1 \gamma^2 \sgn(qB) \right)\, .
\end{align}
Using $ \frac{q}{2} F_{\rm ext}^{\mu\nu} \sigma_{\mu\nu}  =  \vert qB \vert ( - \prj_+ + \prj_-) $, one can reexpress the Klein-Gordon equation (\ref{eq:K-GandZeeman}) as 
\begin{align}
	\left[\, \partial_t^2 - \partial_z^2 + \left(2 \hat a^\dagger \hat a  + 1 - 2s \right)  \vert qB \vert  + m^2 \, \right] 
	\psi_{ s} = 0 \, ,\label{eq:rel_EoM}
\end{align}
with $s=\pm 1/2$ being the eigenvalue of the spin operator along the magnetic field, i.e., 
$2s \psi _{ s} = \sgn(qB) \sigma^{12} \psi_{ s}$ (the factor $2$ accounts for the Land\'{e} $g$-factor).  Namely, $s=+1/2$ and $-1/2$ ($s=-1/2$ and $+1/2$) if the spin direction is parallel and anti-parallel, respectively, with respect to the magnetic-field direction for positively (negatively) charged fermions.  Since the operators $\hat{a}$ and $\hat{a}^\dagger$ satisfy the commutation relation (\ref{eq:com}), one understands that the energy level is given by
\begin{align}
	\epsilon_n \equiv \sqrt{ m^2  +  2  n \vert qB \vert + p_z^2 } \, .\label{eq:fermion-rela}
\end{align}
The non-negative integer $n=0,1,2,\cdots \in {\mathbb N}$ is the resultant quantum number after the sum of the Landau level and the Zeeman shift.  Therefore, the energy level has two-fold spin degeneracy ($s=\pm 1/2$) except for the unique ground state ($s=+1/2$), which is often called the lowest Landau level (after the Zeeman shift is included).  $p_z$ is the longitudinal momentum, which is conserved because the longitudinal motion is not affected by a magnetic field.  Accordingly, one can factorize the eigenfunction $\psi_{s, n }$ (such that $i\partial_t\psi_{s,  n } = \pm \epsilon_n \psi_{s,  n}$) as
\begin{align}
	\psi_{s, n} \propto e^{-ip_\parallel \cdot x} \phi_{n} (x_\perp)\, , \label{eq:wave-functions}
\end{align}
where the transverse wave function $\phi_{n}$ depends on the transverse coordinates $x_\perp$ but not on the longitudinal coordinate $ x_\parallel $.  One can construct $\phi_{n}$ as eigenstates of the ``number operator'' as $\phi_n=\langle x | n\rangle$ with $\hat a^\dagger \hat a |n\rangle =n |n\rangle$ and $ \hat a |0\rangle=0 $.  For the moment, we do not need an explicit form of $\phi_{n}$ which can be obtained only after fixing the gauge, and just discuss gauge-invariant properties of $\phi_{n}$.  Precisely speaking, there exists another good quantum number $\qn$ [e.g., $\qn=p_y$ in the Landau gauge (\ref{eq:Landau-gauge}) which we adopt later].  The existence of this additional good quantum number is anticipated, since there is a constant of motion in classical picture of the cyclotron motion, that is, the center coordinate of the cyclotron motion.  In quantum theory, $  \qn$ corresponds to a label of the center coordinate, and each Landau level is degenerated with respect to $ \qn $ since shift of the center coordinate should not affect the energy level in a constant magnetic field.  For notational simplicity, we label the eigenfunction $\psi_{s,n}$ only with $n$ and suppress the additional label $\qn$ unless needed.  While $\qn$ is a gauge-dependent quantity, the Landau level constructed with the ``creation/annihilation operator'' is a gauge-invariant concept as clear in the construction of the operators (\ref{eq:aadagger}).  

Next, we examine the Dirac spinor structure of $\psi_{s,n}$ and introduce the Ritus basis.  In Eq.~(\ref{eq:rel_EoM}), we have seen that the two spin eigenstates such that $2s\psi_{s,n} = \sgn(qB)\sigma^{12}\psi_{s,n}$ ($s=\pm1/2$) provide appropriate bases for solutions of the Dirac equation.   It is, therefore, convenient to introduce a basis for a superposition of the two degenerated spin eigenstates, 
\begin{align}
\Ritus_{n} (x_\perp) \equiv \phi_{n} (x_\perp) \prj_+ +  \phi_{n-1} (x_\perp) \prj_- \, , \label{eq:Ritus}
\end{align}
where $  \phi_{-1} \equiv 0 $ is understood.  This is the so-called {\it Ritus basis}, which was introduced first for computation of the fermion self-energy in external fields \cite{Ritus:1972ky, Ritus:1978cj}.  Using the Ritus basis, one may write
\begin{align}
	\psi_{s,n} (x) = \left\{ \begin{array}{ll} \displaystyle e^{ - i p_\parallel \cdot x} \Ritus_{n} (x_\perp) \, u \vspace*{2mm} &\ {\rm for\ a\ positive\mathchar`-energy\ solution} \\  e^{ + i p_\parallel \cdot x} \Ritus_{n} (x_\perp)  \, v \vspace*{2mm} &\ {\rm for\ a\ negative\mathchar`-energy\ solution} \end{array} \right. \, , \label{eq:Dirac-Ritus}
\end{align}
where $u$ and $v$ are four-component spinors.  By noticing
\begin{align}
	\left[ i \slashed D _{\rm ext} -m  \right] \psi_{s,n}
		&= \left[ ( i \slashed \partial _\parallel -m) - \sqrt{2|qB|} \  \gamma^1\, \left(\hat a \prj_+ + \hat a^\dagger \prj_- \right) \right] \psi_{s,n} \nonumber\\
		&= \left\{ \begin{array}{l} \displaystyle e^{ - i p_\parallel \cdot x}  \Ritus_{n} (x_\perp)  \left( \slashed p _\parallel - \sqrt{2n |qB| }\ \gamma^1-m \right)  u \vspace*{2mm}\\ \displaystyle  e^{ + i p_\parallel \cdot x} \Ritus_{n} (x_\perp)  \left( -\slashed p _\parallel - \sqrt{2n |qB| }\ \gamma^1-m \right) v \end{array} \right. \, , \label{Dirac_Ritus}
\end{align}
one finds that the ansatz (\ref{eq:Dirac-Ritus}) satisfies the Dirac equation (\ref{eq:Dirac}) if the spinors $u$ and $v$ satisfy the ``free" Dirac equations 
\begin{subequations}  \label{eq:free}
\begin{align}
	0 &=  ( {\slashed p}_n -m ) u(p_n)\ , \label{eq:free_u} \\
	0 &=  ( \bar{\slashed p}_n +m ) v(\bar{p}_n) \, , \label{eq:free_v}
\end{align}
\end{subequations}
where
\begin{align}
	p^\mu_n \equiv ( \epsilon_n, \sqrt{2n |qB| },0,p_z) \, ,\quad 
	\bar p^\mu_{ n} \equiv ( \epsilon_n , - \sqrt{2n |qB| },0,p_z).
\end{align} 
Each ``free'' Dirac equation has two solutions $u=u_\k, v=v_\k$, 
corresponding to two spin degrees of freedom that we label with $\k = 1,2$.  
We normalize the solutions $u_\k$ and $v_\k$ in such a way that they satisfy the following completeness relation
\begin{align}\label{eq:spin-sum}
	\sum_{\k=1,2}  u_\k (p_n)  \bar u_\k (p_n) = ( \slashed p_n + m)
	\, , \quad
	\sum_{\k=1,2}    v_\k (\bar p_n)  \bar{v}_\k (\bar p_n)  = ( \bar{\slashed p}_n - m)
	\, .  
\end{align}
The choice of $\k$ is arbitrary, and one could choose $\k$ different from 
the spin label $s=\pm 1/2$ defined with respect to the magnetic field direction [see Eq.~(\ref{eq:rel_EoM})].  
In general, $\k$ is a superposition of $s=\pm1/2$.  
This is understood by explicitly writing down the normalized solutions 
(although we do not use the explicit forms in our study). 
In the chiral representation, they read \cite{Peskin:1995ev}
\begin{eqnarray}
\label{eq:freesol}
u_\k (p_n ) = ( \sqrt{ p_n \cdot  \sigma} \, \xi_\k, \sqrt{ p_n \cdot \bar \sigma} \,  \xi_\k) 
\, , \quad 
v_\k (\bar p_n ) = ( \sqrt{ \bar p_n \cdot  \sigma} \, \eta_\k, - \sqrt{ \bar p_n \cdot  \bar \sigma} \,  \eta_\k) 
\, ,
\end{eqnarray}
where $ \sigma^\mu = (1, \sigma^i) $ and $ \bar \sigma^\mu = (1, -\sigma^i) $. 
The free Dirac equation (\ref {eq:free}) is satisfied with any two-component spinors $\xi_\k$ and $\eta_\k$, 
resulting in the aforementioned arbitrariness in the choice of $  \k$.   
One of the most convenient choices is to take the eigenvectors of $\sigma^3$ as $ \xi_1 = \eta_1 = (1,0) $ and $ \xi_2 = \eta_2 = (0,1) $.  Note for this case that $u_\k$ and $v_\k$ are in general not eigenstates of $\sigma^{12} = {\rm diag}(\sigma^3, \sigma^3)$ because of the presence of nonvanishing $p_n^1, \, \bar p_n^1$ for the higher Landau levels $ n\geq 1 $. 
%That is, $\k$ is not necessarily the same as the spin label $s=\pm 1/2$ defined with respect to the magnetic field direction [see Eq.~(\ref{eq:rel_EoM})].  In general, $\k$ is a superposition of $s=\pm1/2$.  

We perform canonical quantization of the Dirac field operator $\psi$ to define creation and annihilation operators.  We expand $\psi$ in terms of the solution of the Dirac equation (\ref{eq:Dirac-Ritus}) as
\begin{subequations}
\label{eq:Ritus-mode}
\begin{align}
	\psi(x) &= \sum_{\k=1,2} \sum_{n=0}^{\infty} \int \frac{d p_z dp_y}{(2\pi)^2 \sqrt{2 \epsilon_n}}
\Ritus_{n,p_y}  (x_\perp)  \left[ \, 
 a_{p_n,p_y,\k}  e^{ - i p_\parallel \cdot x}    u_\k (p_n)
 +  b^{\dagger}_{\bar{p}_n,p_y,\k} e^{ i p_\parallel \cdot x}  v_\k(\bar p_n) \,\right]
 ,
 \\
	\bar{\psi}(x) &=\sum_{\k=1,2}  \sum_{n=0}^{\infty} \int \frac{d p_z dp_y}{(2\pi)^2 \sqrt{2 \epsilon_n}}
\left[ \, 
  b_{\bar{p}_n,p_y,\k} e^{-i p_\parallel \cdot x}  \bar v_\k (\bar{p}_n)  
  + a^{\dagger}_{p_n,p_y,\k}  e^{ i p_\parallel \cdot x} \bar u_\k (p_n)    \,\right] 
\Ritus_{n,p_y}^\dagger(x_\perp)  
 .
\end{align}
\end{subequations}
Here and hereafter, we assume the Landau gauge, 
\begin{align}
	A^\mu_{\rm ext} (x) = ( 0, 0, Bx, 0)\, . \label{eq:Landau-gauge}
\end{align}
It is clear that one of the canonical transverse momenta $p_y$ is conserved and specifies the Landau degeneracy.  $p_y$ is a gauge-dependent quantity and hence is not an observable, while the energy $\epsilon_n$ (or the Landau level $n$) and the longitudinal momentum $p_z$ are gauge-independent and observable quantities\footnote{While one could define a (gauge-invariant) {\it kinetic} transverse momentum, its expectation value taken for the eigenstates of the Landau levels should be vanishing due to the closed cyclotron orbit. }.  We chose the Landau gauge just for simplicity, and the mode expansion (\ref{eq:Ritus-mode}) as well as the calculations below can be done in a similar way in other gauges.  Importantly, we explicitly prove that our final physical results are gauge-independent; see Appendix~\ref{app-b2}.  Next, we impose the canonical commutation relations on the Dirac field operator $\psi$ (see Appendix~\ref{sec:commutation} for details) and normalize the transverse wave function $\phi_{n,p_y}$, which fixes the normalization of the Ritus basis $\Ritus_{n,p_y}$, as
\begin{align}
	\int d^2x_\perp \phi^*_{n,p_y}(x_\perp) \phi_{n',p_y'}(x_\perp) = 2\pi \delta(p_y-p'_y)  \delta_{n,n'} \, .
\end{align}
Then, $a_{p_n,p_y,\k}$ and $b_{p_n,p_y,\k}$ are quantized as
\begin{align}
	\{ a_{p_n,p_y,\k},  a^{\dagger}_{p_{n'},p_y',\k'} \} 
		= \{ b_{\bar{p}_n,p_y,\k},  b^{\dagger}_{\bar{p}'_{n'},p_y',\k'}  \} 
		= (2\pi)^2 \delta(p_y-p'_y) \delta( p_z - p'_z)  \delta_{n,n'} \delta_{\k,\k'} \, , \label{eq:commutation} 
\end{align}
while the other commutations are vanishing.  The operators $a_{p_n,p_y,\k}$ and $b_{p_n,p_y,\k}$ can now be interpreted as annihilation operators of a fermion and an anti-fermion state, respectively, which are specified with the energy $\epsilon_n$, longitudinal momentum $p_z$, and spin label $\k$.  Note again that $\k$ is a label for the spin basis for the free Dirac spinors $u_\k$ and $v_\k$, which can be chosen arbitrarily [see discussions above Eq.~(\ref{eq:spin-sum})] and is projected to the physical spin states $s=\pm1/2$ by $\prj_\pm$ in the mode expansion (\ref{eq:Ritus-mode}).  After the quantization, one can identify the vacuum state $| 0 \rangle$ as 
\begin{align}
	0 = a_{p_n,p_y,\k} |0\rangle = b_{\bar{p}_n,p_y,\k} |0\rangle \quad {\rm for\ any\ } p_n,\bar{p}_n, p_y, \k \, , 
\end{align}
and construct multi-particle states as
\begin{align}\label{eq:multi}
	&| f_{p_{n,1}, p_{y,1} \k_1} f_{p_{n,2}, p_{y,2} \k_2} 
	\cdots \bar{f}_{\bar{p}'_{n',1}, p'_{y,1} \k'_1} \bar{f}_{\bar{p}'_{n',2}, p'_{y,2} \k'_2} \cdots \;\rangle  \\
	&\equiv (\sqrt{2 \epsilon_{n,1}} \hat a^{\dagger}_{p_{n,1}, p_{y,1} \k_1})(\sqrt{2 \epsilon_{n,2}} \hat a^{\dagger}_{p_{n,2}, p_{y,2} \k_2}) \cdots
	 (\sqrt{2 \epsilon'_{n',1}} b^{\dagger}_{\bar{p}'_{n',1}, p'_{y,1} \k'_1})(\sqrt{2 \epsilon'_{n',2}} \hat b^{\dagger}_{\bar{p}'_{n',2}, p'_{y,2} \k'_2})  | 0 \rangle 
	 \nonumber\, ,
\end{align}
where $ f $  $(\bar f) $ is a fermion (anti-fermion) state carrying the quantum numbers $p_n,p_y,\k$ ($\bar{p}'_{n'},p'_y,\k'$).  We normalized the multi-particle states as 
\begin{align}
	&\| \, | f_{p_{n,1}, p_{y,1} \k_1} f_{p_{n,2}, p_{y,2} \k_2} \cdots \bar{f}_{\bar{p}'_{n',1}, p'_{y,1} \k'_1} \bar{f}_{\bar{p}'_{n',2}, p'_{y,2} \k'_2} \cdots \;\rangle\, \|^2 \nonumber\\
	&= \left[ \prod_{f \; \rm states} (2\epsilon_n)(2\pi)^2 \delta^{(2)}(0)\right]  \left[ \prod_{ \bar f \; \rm states} (2\epsilon^{\prime}_{n'})(2\pi)^2 \delta^{(2)}(0)\right] \, ,
\end{align}  
where the products on the right-hand side are taken over the multi-fermion and anti-fermion states 
%and $\epsilon'_{n} = \sqrt{ m^2 + 2n|qB| + (p'_z)^2} $ denotes the energy level of an anti-fermion.  
and we use an abbreviation $\epsilon'_{n} \equiv \epsilon_{n} (p_z')= \sqrt{ m^2 + 2n|qB| + (p'_z)^2} $.

\section{Lepton tensor with all-order Landau levels} \label{sec-3}

\begin{figure}
     \begin{center}
              \includegraphics[width=0.55\hsize]{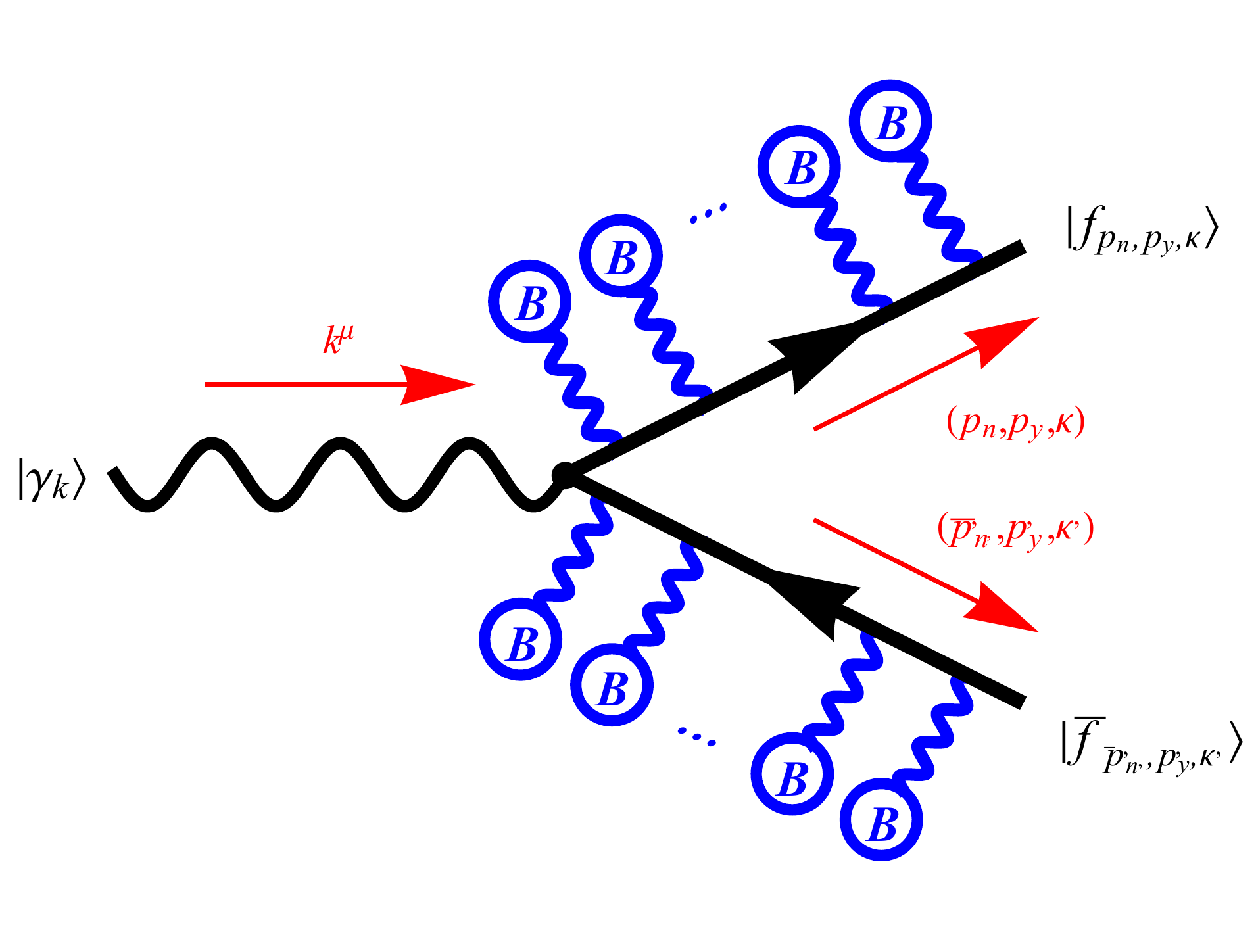}
     \end{center}
\vspace*{-10mm}
\caption{The photon--to--di-lepton vertex in a magnetic field (\ref{eq:matrixelement}).  The wave functions of the produced fermion and anti-fermion are non-perturbatively dressed by the magnetic field.  
}
\label{fig:diagram}
\end{figure}

We analytically evaluate the photon--to--di-lepton vertex in a strong constant magnetic field.  Since the magnetic field is assumed to be strong, we treat the interaction with the magnetic field non-perturbatively, which can be conveniently achieved by the Ritus-basis formalism reviewed in the preceding section.  On the other hand, we treat the interaction with a dynamical photon $A$ perturbatively.  At the leading order in $A$, one can write down the amplitude explicitly as (see Fig.~\ref{fig:diagram}) 
\begin{align}
	\varepsilon_\mu q {\mathcal M}^\mu 
		&\equiv \langle f_{p_n,p_y,\k}, \bar{f}_{\bar p_{n^\prime}^\prime,p_y',\k'}  | \, q \int d^4 x \, \bar \psi (x) \slashed A(x) \psi(x)  \, | \gamma_k \rangle \nonumber \\
		&= \varepsilon_\mu  q \int d^4 x \, e^{ - ik\cdot x + i (p_\parallel + p'_\parallel)\cdot x } \bar u_\k(p_n) 
		 \Ritus_{n,p_y}^\dagger(x_\perp)    \gamma^\mu   \Ritus_{n',p_y'}  (x_\perp) v_{\k'}(\bar p'_{n'}) \, , \label{eq:matrixelement}
\end{align}
where the one-photon state $| \gamma_k \rangle$ is normalized as 
\begin{align}	
	\| \, |\gamma_k \rangle \, \|^2 = 2k^0 (2\pi)^3  \delta^{(3)}(0) \label{eq:gammanorm}
\end{align}
and $ \varepsilon_\mu $ is a polarization vector of the initial dynamical photon (which is not necessarily on-shell).  
The fermion field is expanded with the Ritus basis (\ref{eq:Ritus-mode}) 
and hence is dressed by the magnetic field non-perturbatively. 
On the other hand, the mode expansion for the dynamical photon field $A$ is the usual Fourier decomposition because $A$ is charge neutral and still in a momentum eigenstate.  As a consequence, the three-point vertex between the fermion current and the dynamical photon field is given by a convolution of two Ritus bases and a plane wave.  Inserting the explicit form of the Ritus basis (\ref{eq:Ritus}) into the amplitude (\ref{eq:matrixelement}), we find 
\begin{align}
\label{eq:Ritus-form}
	{\mathcal M}^\mu 
		&= (2\pi)^2 \delta^{(2)}( k_\parallel - p_\parallel - p_\parallel') \\
		&\quad \times \bar u_\k(p_n) \big[ \gamma_\parallel^\mu \left( \prj_+ \Gamma_{n,n'} + \prj_-  \Gamma_{n-1,n'-1} \right) + \gamma_\perp^\mu \left( \prj_+  \Gamma_{n-1,n'} + \prj_-  \Gamma_{n,n'-1} \right)  \big] v_{\k'}(\bar p_{n'}^\prime) \, , \nonumber
\end{align}
where the scalar form factor $\Gamma_{n, n'}$ is defined 
as an overlap between the two fermion wave functions in the transverse plane: 
\begin{align}
	\Gamma_{n, n'}(p,p';k) \equiv \int \!\! d^2 x_\perp \, e^{  i (k_x x + k_y y)} \phi^\ast _{n,p_y} (x_\perp)  \phi _{n',p_y'} (x_\perp)   \label{eq:Gam0}\, .
\end{align}
We will discuss properties of $\Gamma_{n, n'}$ in detail in Sec.~\ref{sec:3.3}.  In Eq.~(\ref{eq:Ritus-form}), the first term coupled to $\gamma_\parallel$ gives the amplitude for spin-zero lepton pair production.  This observation is based on the facts that the same spin projection operators are acting on the spinors $u$ and $v$ (cf. $ \gamma_\parallel^\mu  \prj_\pm  =\prj_\pm \gamma_\parallel^\mu \prj_\pm   $) and that the anti-fermion spinor $ \prj_\pm v $ corresponds to the opposite spin direction to that of the fermion spinor $ \prj_\pm u $ \cite{Peskin:1995ev}.  In the same manner, one understands that the second term coupled to $\gamma_\perp$ is responsible for spin-one lepton pair production according to a property $   \gamma_\perp^\mu \prj_\pm =\prj_\mp \gamma_\perp^\mu \prj_\pm   $.

\subsection{Scalar form factor $\Gam_{n,n'}$ in the Landau gauge}

We explicitly evaluate the scalar form factor $\Gamma_{n,n'}$ (\ref{eq:Gam0}) in the Landau gauge (\ref{eq:Landau-gauge}).  
While we choose a particular gauge, we keep track of the gauge dependence carefully 
and confirm that the gauge invariance is finally restored in the squared amplitude (see the next subsection). 
Moreover, we show that our amplitude (\ref{eq:FF-Landau}), and thus the squared amplitude, 
satisfies the Ward identity $0=k_\mu {\mathcal M}^\mu$ for each pair of Landau levels, 
which is the manifestation of the gauge invariance for the dynamical photon field; see Appendix~\ref{app-b1}.

In the Landau gauge (\ref{eq:Landau-gauge}), one can explicitly write down the wave function $\phi_n$ obtained as eigenstates of the number operator (see below Eq.~(\ref{eq:wave-functions}) and Appendix~\ref{sec:wf_Landau} for derivation):
\begin{align}
	\phi_{n,p_y} (x_\perp) 		= e^{i p_y y }\,  i^n \sqrt{ \frac{1}{ 2^n n! \pi^{\frac{1}{2}} \ell} } e^{- \frac{\xi^2}{2} }  H_n (\xi) \, , \label{eq:WF_Landau}
\end{align}
where $H_n$ is the Hermite function $H_n(z) \equiv ( -1)^n e^{ +z^2 } \partial_z^n e^{ -z^2 }$ 
and $\xi \equiv (x-x_{\rm c})/\ell$ with $x_{\rm c}\equiv p_y/qB$ and $\ell \equiv 1/\sqrt{|qB|}$ 
representing the center and the typical radius of the cyclotron motion (called the magnetic length), respectively.  
By substituting this expression (\ref{eq:WF_Landau}) into Eq.~(\ref{eq:Gam0}), one obtains\footnote{Note that one may extract the positive-power dependence on $ |\bar {\bm k}_\perp| $ by using $L_m^{-\ell} (\rho) = \frac{(m-\ell)!}{m!} (-\rho)^\ell L_{m-\ell}^{\ell} (\rho)$ as
\begin{align}
 {\Gam}_{n, n'} 
		&= e^{  i \frac{k_x (p_y + p^\prime_y)}{2 qB} } \, e^{-  \frac{1}{2} |\bar {\bm k}_\perp|^2} (-1)^{ n' - \min( n,n') } \sqrt{ n! / n^\prime!} ^{ \, \sgn(\Delta n)} e^{  - i \;\sgn(qB) \Delta n \, \theta_\perp} |\bar {\bm k}_\perp|^{|\Delta n|} L_{\min( n,n')}^{|\Delta n|} ( | \bar {\bm k}_\perp| ^2 )\, .  \label{eq:FF-Landau-2}
\end{align} 
}
\footnote{The square of the scalar form factor $|{{\Gam}}_{n,n'}|^2$ has precisely the same form as $C_\ell^m(\eta) $ defined in Eq.~(38) of Ref.~\cite{Hattori:2012je} if one identifies the variables as $ \eta = |\bar {\bm k}_\perp|^2 $, $ m = - \Delta n $ and $ \ell=n' $; namely, $| {\Gam}_{n , n'} |^2 = C_n'^{- \Delta n} (\eta) $.  } 
\begin{align}
	&\Gamma_{n ,n^\prime}(p,p';k) \nonumber\\
		&= 2 \pi \delta(k_y - p_y + p'_y) \times e^{  i \frac{k_x(p_y + p^\prime _y)}{2 qB} } \, e^{-  \frac{1}{2} |\bar {\bm k}_\perp|^2} (-1)^{ \Delta n} \sqrt{ \frac{n!}{ n^\prime!}}  \, e^{  - i \;\sgn(qB) \Delta n \, \theta_\perp} |\bar {\bm k}_\perp|^{\Delta n} L_{ n }^{ \Delta n } ( | \bar {\bm k}_\perp| ^2 ) \nonumber\\
		&\equiv 2 \pi \delta(k_y - p_y + p'_y)  \times  {\Gam}_{n, n^\prime}(p,p';k) \, ,
\label{eq:FF-Landau}
\end{align}
where the scalar form factor $  {\Gam}_{n, n^\prime}(p,p';k) $ is defined for the Landau gauge 
after factorizing the delta function, and $ \Gam_{n, n^\prime} = 0$ for $n$ or $n'<0$ is understood. 
We also defined 
\begin{align}
	\Delta n \equiv n^\prime - n, \quad
	|\bar {\bm k}_\perp|^2 \equiv \frac{ |  {\bm k}_\perp|^2}{ 2| qB|} ,\quad 
	\theta_\perp \equiv \arg(k_x + i k_y) ,
\end{align}
where $\theta_\perp$ is the azimuthal angle of the photon momentum $ {\bm k} $.  
$L_n^k$ is the associated Laguerre polynomial such that $L_n^k(z) \equiv \partial_z^k(e^{+z} \partial_z^n (z^n e^{-z}))$.  While systems exposed to a constant magnetic field should maintain the gauge symmetry and the translational and rotational symmetries in the transverse plane, the gauge configuration $ A^\mu_{\rm ext} $ partially or completely hides those symmetries.  Indeed, the scalar form factor ${\Gam}_{n, n'}$ itself does not possess the gauge symmetry due to the exponential phase factor $\exp\left[  i k_x(p_y + p^\prime _y)/2 qB \right]$ (the so-called Schwinger phase) because  $p_y$ is a gauge-dependent label in the Landau gauge.  The rotational symmetry also seems to be apparently broken by the dependence on the azimuthal angle $ \theta_\perp $.  Nevertheless, the Schwinger factor does not depend on $n$ and thus is canceled out when the amplitude is squared.  Similar to this, one can also show that the rotational symmetry is restored after squaring the amplitude; see Appendix~\ref{app-b2}.

\subsection{Analytic form of the lepton tensor} 
\label{sec:general}

We turn to evaluate the lepton tensor in a magnetic field $L^{\mu\nu}_{n,n'}$ which is defined via the squared amplitude as 
\begin{align}
	\sum_{\k,\k'} \frac{ \left| \varepsilon_\mu q {\mathcal M}^\mu  \right|^2}{2k^0 (2\pi)^3 \delta^{(3)}(0)}
		\equiv  \frac{T}{L_x} \frac{q^2}{2k^0}(2\pi)^3 \delta^{(2)}( k_\parallel - p_\parallel - p'_\parallel) \delta(k_y - p_y + p'_y) [\, \varepsilon_\mu  \varepsilon^\ast_\nu L^{\mu\nu} _{n,n'} \, ] \, , \label{eq:sq}
\end{align}
where $T \equiv 2\pi \delta(p^0=0)$ and $ L_x $ are the whole time-interval and the system length in the $x$ direction, respectively.  The factor of $1/[2k^0(2\pi)^3 \delta^{(3)}(0)]$ is inserted so as to cancel the uninterested normalization factor coming from the one-photon state (\ref{eq:gammanorm}).  For notational brevity, we rewrite the amplitude ${\mathcal M}^\mu$ as 
\begin{align}
	{\mathcal M}^\mu 
		&= (2\pi)^3  \delta^{(2)}( k_\parallel - p_\parallel - p_\parallel') \delta( k_y - p_y + p'_y) \times \bar u_\k(p_n) [ \gamma_\parallel^\mu \H_0 + \gamma_\perp^\mu \H_1  ] v_{\k'}(\bar p_{n'}^\prime) \, ,
\end{align}
with 
\begin{align}
	\H_0 \equiv   \prj_+ {\Gam}_{n, n'} + \prj_- {\Gam}_{n-1,n'-1} \  , \quad
	\H_1 \equiv  \prj_+ {\Gam}_{n-1,n'} + \prj_-  {\Gam}_{n,n'-1} \, .
\end{align}
Note that $\H_0$ and $\H_1$ control the amplitudes for spin-zero and spin-one lepton pair productions, respectively, as we remarked below Eq.~(\ref{eq:Gam0}).  Then, after the spin summation $\sum_{\k,\k'}$, one can express $L^{\mu\nu}_{n,n'}$ as 
\begin{align}
	L^{\mu\nu}_{n,n'}  
		&= {\rm tr} \left[ \,  (\slashed p_n - m)  ( \gamma_\parallel^\mu \H_0 + \gamma_\perp^\mu  \H_1 ) (\bar {\slashed p}'_{n'} + m) ( \H^\dagger_0 \gamma_\parallel^\nu  + \H^\dagger_1\gamma_\perp^\nu  ) \, \right] \nonumber\\
		&= T_1 - 2 |qB|   \sqrt{ n n'} T_2 - \sqrt{ 2 n |qB|}  T_3 + \sqrt{ 2 n' |q B|}  T_4\, , \label{eq:leptontensor1}
\end{align}
where  
\begin{subequations}
\label{eq:traces}
\begin{align}
	T_1 &\equiv {\rm tr} \left[ \, (\slashed p_\parallel - m)  ( \gamma_\parallel^\mu \H_0 + \gamma_\perp^\mu  \H_1 ) ({\slashed p}'_\parallel+ m) ( \H^\dagger_0 \gamma_\parallel^\nu  + \H^\dagger_1 \gamma_\perp^\nu  ) \, \right]\, , \\
	T_2 &\equiv {\rm tr} \left[ \, \gamma^1 ( \gamma_\parallel^\mu \H_0 + \gamma_\perp^\mu  \H_1 )\gamma^1  ( \H^\dagger_0 \gamma_\parallel^\nu  + \H^\dagger_1 \gamma_\perp^\nu  ) \, \right]\, , \\
	T_3 &\equiv {\rm tr} \left[ \, \gamma^1( \gamma_\parallel^\mu \H_0 + \gamma_\perp^\mu  \H_1 ) ( {\slashed p}'_\parallel+ m) ( \H^\dagger_0 \gamma_\parallel^\nu  + \H^\dagger_1 \gamma_\perp^\nu  ) \, \right]\, , \\
	T_4 &\equiv {\rm tr} \left[ \,  (\slashed p_\parallel - m)  ( \gamma_\parallel^\mu \H_0 +\gamma_\perp^\mu  \H_1 ) \gamma^1 ( \H^\dagger_0 \gamma_\parallel^\nu  + \H^\dagger_1 \gamma_\perp^\nu  ) \, \right]\, .
\end{align}
\end{subequations}
The gauge-dependent Schwinger phase goes away by the squaring operation, and thus the gauge and translational invariances have been restored explicitly in Eq.~(\ref{eq:leptontensor1}).  
The rotational symmetry has also been restored here, although it may not be obvious at a glance;  
see Appendix~\ref{app-b2} for an explicit demonstration.

Before proceeding, we introduce several notations in order to simplify the traces (\ref{eq:traces}) in a physically transparent way.  We first introduce photon's circular polarization vectors with respect to the direction of the magnetic field, 
\begin{align}
	\varepsilon^\mu_\pm \equiv -( g^{\mu1} \pm i\, \sgn(qB)  g^{\mu2} )/\sqrt{2} = ( 0, 1 , \pm i\, \sgn(qB)  , 0)/\sqrt{2} \, , \label{eq:pol}
\end{align}
which are ortho-normalized as $ \varepsilon^\mu_\pm \varepsilon^*_{\mp\mu} = 0 $ and $ \varepsilon^\mu_\pm \varepsilon^*_{\pm\mu}  = -1 $ and satisfy $\varepsilon^{\mu*}_\pm=\varepsilon^\mu_\mp$.  We inserted a sign function in the above definition (\ref{eq:pol}) because the direction of fermion's spin changes depending on $\sgn(qB)$.  Those polarizations couple to di-leptons carrying total spin $s+s'= \pm 1 $ states as we see below.  Next, we introduce helicity-projection operators for circularly polarized photons 
\begin{align}
  \Q_\pm^{\mu\nu} 
		\equiv - \varepsilon_\pm^\mu  \varepsilon_\pm^{\nu\ast} 
		= ( g_\perp^{\mu\nu} \pm i\,\sgn(qB) \varepsilon_\perp^{\mu\nu} )/2  \, ,
\end{align}
where $\varepsilon_\perp^{\mu\nu} \equiv g^{\mu1} g^{\nu2}  -g^{\mu2} g^{\nu1}$.  Those operators have eigenvectors $ \varepsilon^\mu_\pm $ which satisfy $\Q_\pm^{\mu\nu} \varepsilon_{\pm \nu} = \varepsilon_\pm^\mu $ and $ \Q_\pm^{\mu\nu} \varepsilon_{\mp \nu}  =0$.  Finally, we introduce scalar and longitudinal photon polarization vectors 
\begin{align}
	\varepsilon_0^\mu \equiv (1,0,0,0) \ ,\quad
	\varepsilon_\parallel^\mu \equiv (0,0,0,1)\, , 
\end{align}
respectively, which are ortho-normalized as $\varepsilon_0^\mu\varepsilon^*_{\parallel\mu}=0$, $\varepsilon_0^\mu\varepsilon^*_{0\mu}=+1$, and $\varepsilon_\parallel^\mu\varepsilon^*_{\parallel\mu}=-1$.  Clearly, those vectors are orthogonal to the circular polarization vectors as $0=\varepsilon_\pm^\mu \varepsilon_{0\mu}=\varepsilon_\pm^\mu \varepsilon_{\parallel\mu}$ and do not couple to the helicity-projection operators, i.e., $0=\Q_\pm^{\mu\nu} \varepsilon_{0\mu}=\Q_\pm^{\mu\nu} \varepsilon_{\parallel\mu}$.  The scalar and longitudinal photons couple to $s+s'=0$ channel of di-leptons, as we see below.  Note that the four polarization vectors $\varepsilon_{0,\pm, \parallel}$ satisfy the following completeness relation 
\begin{align}
	g^{\mu\nu} = \varepsilon^\mu_{0}\varepsilon^{\nu*}_{0} - \varepsilon^\mu_{+}\varepsilon^{\nu*}_{+} -\varepsilon^\mu_{-}\varepsilon^{\nu*}_{-} -\varepsilon^\mu_{\parallel}\varepsilon^{\nu*}_{\parallel}\, , \label{eq:completeness}
\end{align}
and form a complete basis for the photon polarization vector $\varepsilon^\mu$.

Now, we are ready to simplify the traces in Eq.~(\ref{eq:traces}).  For $T_1$, a straightforward calculation yields
\begin{align}
	T_1 
		&= {\rm tr} \left[ \,  (\slashed p_\parallel - m)  \gamma_\parallel^\mu  ({\slashed p}'_\parallel+ m)  \gamma_\parallel^\nu  | \H_0 |^2\, \right] + {\rm tr} \left[ \, (\slashed p_\parallel - m)   \gamma_\perp^\mu  ({\slashed p}'_\parallel+ m) |  \H_1|^2 \gamma_\perp^\nu  \, \right] \\
		&= ( |  {\Gam}_{n , n'} |^2   + |{\Gam}_{n-1 , n'-1 } |^2 ) L_\parallel^{\mu\nu} - 4 (p_\parallel \cdot p'_\parallel + m^2)  \left( \,   |  {\Gam}_{n-1 , n'} |^2 \Q_+^{\mu\nu} + |{\Gam}_{n , n'-1 } |^2  \Q_-^{\mu\nu} \, \right) \nonumber \, .
\end{align}
Here, we introduced the lepton tensor in the $(1+1)$-dimensional form 
\begin{align}
	L_\parallel^{\mu\nu} = 2 \left[ \, p_\parallel^\mu p_\parallel^{\prime \nu} +  p_\parallel^\nu p_\parallel^{\prime \mu} - (p_\parallel \cdot p'_\parallel + m^2)  g_\parallel^{\mu\nu} \, \right] \, ,
\end{align} 
which couples only to the scalar and longitudinal photon polarizations, i.e., $0 \neq L^{\mu\nu}_{\parallel} \varepsilon_{0\mu}, L^{\mu\nu}_{\parallel} \varepsilon_{\parallel\mu}$ and $0 = L^{\mu\nu}_{\parallel} \varepsilon_{\pm\mu}$.  Remember that the terms proportional to $\H_0 $ and $\H_1$ are originated from spin-zero and spin-one di-lepton configurations.  Therefore, the first term is responsible for spin-zero di-lepton production and is coupled to the photon mode longitudinal to the direction of the magnetic field.  On the other hand, the last two terms describe spin-one di-lepton production and are coupled to circularly polarized photons.  Note that, among all the terms, only the term proportional to $ | {\Gam}_{0, 0}|^2 $ survives in the lowest Landau level approximation.  Next, we turn to evaluate the remaining terms $T_2, T_3$, and $T_4$: 
\begin{align}
	T_2 
		&= {\rm tr} \left[ \, \gamma^1  \gamma_\parallel^\mu (\H_0 \gamma^1  \H^\dagger_0 ) \gamma_\parallel^\nu  \, \right] + {\rm tr} \left[ \, \gamma^1 \gamma_\perp^\mu ( \H_1 \gamma^1  \H^\dagger_1)  \gamma_\perp^\nu  \, \right] \nonumber\\
		&= 4 \left[ \ {\rm Re}[    {\Gam}_{n , n'}     {\Gam}_{n-1 , n'-1} ^\ast  ] g_\parallel^{\mu\nu} + {\Gam}_{n-1, n'}  {\Gam}_{n, n'-1}^\ast    \varepsilon_+^\mu  \varepsilon_-^{\nu \ast} + [  {\Gam}_{n-1, n'} {\Gam}_{n, n'-1}^\ast  ]^\ast  \varepsilon_-^\mu  \varepsilon_+^{\nu \ast} \ \right]\, .
\end{align}
The $ \gamma^1 $'s in between $\H_0, \H_0\dagger$ and $\H_1, \H_1\dagger$ induce a spin flip, unlike the case in $T_1$.  Therefore, the first term in the last line and the others are responsible for the interferences between the amplitudes for two spin-zero and spin-one di-lepton configurations, respectively, which consist of fermion pairs with the same energy level but distinct spin combinations due to the spin flipping (recall the spin degeneracy in each energy level).  In the remaining two traces $T_3$ and $T_4$, the mass terms do not survive, since one cannot hold an even number of $ \gamma_\parallel^\mu $ and $ \gamma_\perp^\mu $ simultaneously.  Therefore, 
\begin{subequations}
\begin{align}
	T_3 
		&= {\rm tr} \left[ \, \gamma^1 \gamma_\parallel^\mu  {\slashed p}'_\parallel \H_0 \H^\dagger_1 \gamma_\perp^\nu  \, \right] + {\rm tr} \left[ \, \gamma^1 \gamma_\perp^\mu  {\slashed p}'_\parallel \H_1 \H^\dagger_0 \gamma_\parallel^\nu 
\, \right] \nonumber \\
		&= - 2 \sqrt{2} \big[ \  {\Gam}_{n,n'}  {\Gam}_{n-1,n'} ^\ast  p_\parallel^{\prime \mu} \varepsilon_-^\nu + {\Gam}_{n,n'} ^\ast 
		{\Gam}_{n-1,n'}   \varepsilon_+^\mu p_\parallel^{\prime \nu} \nonumber \\
 		&\quad + {\Gam}_{n-1,n'-1} {\Gam}_{n,n'-1} ^\ast  p_\parallel^{\prime \mu} \varepsilon_+^\nu +{\Gam}_{n-1,n'-1} ^\ast  {\Gam}_{n,n'-1}  \varepsilon_-^\mu p_\parallel^{\prime \nu} \ \big]
\, , \\
	T_4 
		&= {\rm tr} \left[ \, \slashed p_\parallel   \gamma_\parallel^\mu \H_0  \gamma^1  \H^\dagger_1 \gamma_\perp^\nu  \, \right] + {\rm tr} \left[ \, \slashed p_\parallel  \gamma_\perp^\mu  \H_1 \gamma^1 \H^\dagger_0 \gamma_\parallel^\nu   \, \right] \nonumber\\
		&= - 2 \sqrt{2} \big[ \  {\Gam}_{n,n'} {\Gam}_{n,n'-1} ^\ast p_\parallel^\mu \varepsilon_+^\nu  +  {\Gam}_{n,n'}^\ast {\Gam}_{n,n'-1} \varepsilon_-^\mu p_\parallel^\nu \nonumber \\
		&\quad +  {\Gam}_{n-1,n'-1} {\Gam}_{n-1,n'}^\ast p_\parallel^\mu \varepsilon_-^\nu +{\Gam}_{n-1,n'-1}^\ast  {\Gam}_{n-1,n'} \varepsilon_+^\mu   p_\parallel^\nu\ \big] \, .
\end{align}
\end{subequations}
As clear from the appearance of $ \H_0 $ coupled to $ \H_1 $ and vice versa, all those terms are for the interferences between the amplitudes for two spin-zero and spin-one di-lepton configurations having distinct fermion spin contents.

Getting all the above contributions together, we arrive at the lepton tensor in a magnetic field: 
\begin{align}
	L^{\mu\nu}_{n,n'} 
	&= ( | {\Gam}_{n , n'} |^2   + |  {\Gam}_{n-1 , n'-1 } |^2 ) L_\parallel^{\mu\nu} 	- 4 (p_\parallel \cdot p'_\parallel + m^2)  \left( \,   |   {\Gam}_{n-1 , n'} |^2 \Q_+^{\mu\nu} + |  {\Gam}_{n , n'-1 } |^2  \Q_-^{\mu\nu} \, \right)	\nonumber \\
		&\quad - 4 |qB| \sqrt{nn'}  \;{\rm Re}\left[ \ 2 {\Gam}_{n , n'} {\Gam}_{n-1 , n'-1} ^\ast  g_\parallel^{\mu\nu} + {\Gam}_{n-1, n'} {\Gam}_{n, n'-1}^\ast    \varepsilon_+^\mu  \varepsilon_+^{\nu }  \ \right] \nonumber \\
		&\quad + 8 \sqrt{ n|qB|} \;{\rm Re} \left[ \ \left( {\Gam}_{n,n'}^\ast  {\Gam}_{n-1,n'}  + {\Gam}_{n-1,n'-1} {\Gam}_{n,n'-1} ^\ast \right) p_\parallel^{\prime \mu} \varepsilon_+^\nu \ \right] \nonumber \\
		&\quad -  8 \sqrt{ n' |qB|} \;{\rm Re} \left[ \ \left( {\Gam}_{n,n'} {\Gam}_{n,n'-1} ^\ast  + {\Gam}_{n-1,n'-1}^\ast {\Gam}_{n-1,n'} \right) p_\parallel^\mu \varepsilon_+^\nu \ \right]  \label{eq:L}\, .
\end{align}
The general form of the lepton tensor~(\ref{eq:L}), together with the analytic expression for the scalar form factor (\ref{eq:FF-Landau}), is one of the main results of the present paper.  We will further inspect its basic behaviors and apply it to compute the di-lepton yields.  To verify the correctness of the expression (\ref{eq:L}), we provide some consistency checks in Appendix~\ref{app-b2}.  We have confirmed the following three points: (i) $ |\epsilon_\mu {\mathcal M}^\mu_{n,n'}|^2 \propto \varepsilon_\mu \varepsilon_\nu^\ast L^{\mu\nu}_{n,n'} $ is a real-valued quantity; (ii) $ L^{\mu\nu}_{n,n'}  $ satisfies the Ward identity, i.e., $k_\mu  L^{\mu\nu}_{n,n'} = k_\nu  L^{\mu\nu}_{n,n'}=0  $; (iii) The rotational symmetry in the transverse plane is restored in $ \varepsilon_\mu \varepsilon_\nu^\ast L^{\mu\nu}_{n,n'}  $ for an arbitrary photon polarization $ \varepsilon_\mu $ in spite of the use of the Landau gauge, which apparently breaks the rotational symmetry.

\subsection{Polarization-projected lepton tensors} \label{sec:3.3}

In the matrix element squared (\ref{eq:leptontensor1}), 
the lepton tensor (\ref{eq:L}) appears in contraction with photon polarization vectors $\varepsilon^\mu, \varepsilon^{\ast}_{\nu}$, and here we discuss the basic behaviors of the contracted lepton tensors.  The magnitudes of the contracted lepton tensors are controlled by the square of the scalar form factor $ | {\Gam}_{n, n'} |^2 $.  
For each photon polarization $\varepsilon^\mu = \varepsilon^\mu_{0,\pm,\parallel}$, we find 
\begin{align}
	\varepsilon_{\mu}^0  \varepsilon^{0\ast}_{\nu} L_{n,n'}^{\mu\nu} 
			&=	2  ( \epsilon_n \epsilon'_{n'}  + p_z p'_z -m^2 )    ( | {\Gam}_{n , n'} |^2   + |  {\Gam}_{n-1 , n'-1 } |^2 ) \nonumber\\
				&\quad -4 |q B| \left[ - |{\bm k}_\perp| ^2 |{\Gam}_{n-1,n'}|^2 + n |  {\Gam}_{n, n'} |^2 +  n' | {\Gam}_{n-1,n'-1}|^2 \right] \, , \nonumber\\
	\varepsilon_{\mu}^+  \varepsilon^{+\ast}_{\nu} L_{n,n'}^{\mu\nu} 
			&= 4  ( \epsilon_n \epsilon'_{n'}  - p_z p'_z + m^2 )    | {\Gam}_{n , n'-1 } |^2  \, ,\nonumber\\
	\varepsilon_{\mu}^-  \varepsilon^{-\ast}_{\nu} L_{n,n'}^{\mu\nu}  
			&= 4   ( \epsilon_n \epsilon'_{n'}  - p_z p'_z + m^2 )   | {\Gam}_{n-1 , n'} |^2   \, , \nonumber\\
	\varepsilon_{\mu}^\parallel  \varepsilon^{\parallel\ast}_{\nu} L_{n,n'}^{\mu\nu} 
		&= 2   ( \epsilon_n \epsilon'_{n'}  + p_z p'_z + m^2 )  ( | {\Gam}_{n , n'} |^2   + |{\Gam}_{n-1 , n'-1 } |^2 ) \nonumber\\
			&\quad+4 |q B| \left[ - | {\bm k}_\perp| ^2 |{\Gam}_{n-1,n'}|^2 + n |  {\Gam}_{n, n'} |^2 +  n' |  {\Gam}_{n-1,n'-1}|^2 \right]  \, , \label{eq:projected-L}
\end{align}
where we used an identity [which follows from Eq.~(\ref{eq:identities})]: 
\begin{align}
\label{eq:GG}
2 \sqrt{n n'} \,  {\Gam}_{n, n'}  {\Gam}^\ast_{n-1,n'-1} 
	= - | {\bm k}_\perp| ^2 |{\Gam}_{n-1,n'}|^2 + n | {\Gam}_{n, n'} |^2 +  n' |  {\Gam}_{n-1,n'-1}|^2\, .
\end{align}
Note that the two circularly ($\pm$) polarized photons give distinct contracted lepton tensors (and thus distinct production number of di-leptons) because $ | {\Gam}_{n , n'-1 } |^2  \neq  | {\Gam}_{n-1 , n'} |^2   $ unless $ n=n' $ [cf. identities (\ref{eq:identities})].  Note also that the contracted lepton tensors depend only on the photon transverse momentum $ |{\bm k}_\perp| $ and do not depend explicitly on the longitudinal variables $k_0$ and $k_z$.  One may understand, therefore, that basic behaviors of the di-lepton production are determined by the transverse variables in the system, rather than the longitudinal ones.  This might look reasonable in the sense that magnetic fields never affect the longitudinal motion.  Nevertheless, the longitudinal variables do affect the production through the kinematics, i.e., the delta function in front of the contracted lepton tensor in the matrix element squared (\ref{eq:leptontensor1}).  We postpone to discuss effects of the longitudinal variables and the kinematics until Sec.~\ref{sec4.2}.

To get a qualitative understanding of the contracted lepton tensors, we discuss basic behaviors of the square of the scalar form factor $| {\Gam}_{n, n'} |^2 $, whose explicit expression is given by
\begin{align}
\label{eq:GG2}
	| {\Gam}_{n, n'} |^2 
		= |\bar {\bm k}_\perp|^{2|\Delta n|} e^{- |\bar {\bm k}_\perp|^2} \left( \frac{n!}{n'!} \right)^{\sgn(\Delta n)} \left| L^{|\Delta n|}_{\min(n,n')} ( |\bar {\bm k}_\perp|^2)  \right|^2 \ .
\end{align}
This form factor $ |{\Gam}_{n, n'}|^2 $ behaves differently depending on the strength of the magnetic field compared to the typical resolution scale set by the transverse photon momentum, i.e., $ |\bar {\bm k}_\perp|^2 = |{\bm k}_\perp|^2/|2qB| $.  $ |{\Gam}_{n, n'}|^2 $ has peaks in the $|\bar {\bm k}_\perp|$ dependence originating from an oscillation of the associated Laguerre polynomial, which is a reminiscent of the transverse momentum conservation (recall the definition of $ {\Gam}_{n, n'}  $ with the overlap among the wave functions (\ref{eq:Gam0}) that yields a delta function in the absence of a magnetic field).  The structure of the peaks deviates from a delta function due to the fermion's dressed wave functions in a magnetic field.  Further basic properties are summarized as follows.

When $|\bar {\bm k}_\perp| \lesssim 1$ (i.e., $|{\bm k}_\perp|$ is small and/or $|qB|$ is large relative to each other), $|{\Gam}_{n, n'} |^2$ is suppressed by the power factor $|\bar {\bm k}_\perp|^{2|\Delta n|}$.  The suppression becomes larger for a larger $ |\Delta n| $, and $ |  {\Gam}_{n, n'} |^2 $ eventually vanishes in the limit $|\bar {\bm k}_\perp| \to 0$, unless $\Delta n = 0$.  Indeed, one can show that 
\begin{align}
	\lim_{|\bar {\bm k}_\perp|\to0 } | {\Gam}_{n, n'} |^2 
		= \delta_{n,n'} \Big[ \,   L_{\min(n,n')} ( 0 )  \, \Big]^2 
		= \delta_{n,n'} \, , \label{eq:G(k->0)}
\end{align}
where $L^{0}_{n} ( 0 ) = 1$.  This behavior is anticipated from the definition (\ref{eq:Gam0}), which just reduces to the orthonormal relation for the transverse wave function $\phi_{n,p_y}$ in the limit $ |\bar {\bm k}_\perp| \to 0 $.  Intuitively, this property can be understood from the reminiscent of the transverse-momentum conservation mentioned above.  A dynamical photon with $|{\bm k}_\perp| \neq 0$ ($|{\bm k}_\perp| = 0$) produces a di-lepton carrying a nonzero transverse momentum in total, and hence the produced fermion and anti-fermion would have distinct (the same) magnitudes of transverse momentum.  This implies that a larger transverse-momentum difference in the produced fermions requires a larger photon transverse momentum.  Therefore, production of di-lepton with a large $|\Delta n|$ is suppressed in a small $|\bar{\bm k}_\perp|$.

In the opposite regime $|\bar {\bm k}_\perp| \gtrsim 1$, $|{\Gam}_{n, n'} |^2$ is suppressed exponentially by the factor of $e^{- |\bar {\bm k}_\perp|^2}$.  This suppression originates from the exponentially small overlap between the fermion and anti-fermion wave functions in the transverse plane, which are squeezed around the center of the cyclotron motion with length scale $\sim 1/\sqrt{|qB|}$  (i.e., the Landau quantization).  To be specific, let us take the Landau gauge as an example.  In this gauge, the fermion momentum, and thus the photon momentum, is related to the center coordinate of the cyclotron motion as $ k_y = p_y - p_y' = qB ( x_c - x_c')$ [see Eq.~(\ref{eq:WF_Landau})].  Thus, peaks of the fermions' transverse wave functions recede from each other when the photon momentum $ k_y $ becomes large.  The other component $k_x $ appears in Eq.~(\ref{eq:Gam0}) as the Fourier mode of the overlap between the fermions' transverse wave functions.  The Fourier power spectrum is exponentially suppressed when the resolution scale $k_x $ is much larger than the structure of the fermions' transverse wave functions, whose characteristic scale is given by the inverse of the cyclotron radius $ \sim 1/ \sqrt{|qB|} $.  Thus, we get the factor of $e^{- |\bar {\bm k}_\perp|^2}$.

\section{Di-lepton production by a single photon} \label{sec--4}

We provide more detailed discussions about the di-lepton production by a single photon with fixed momentum and polarization in a magnetic field.  The di-lepton spectrum is given by the squared amplitude (\ref{eq:sq}) with the lepton tensor obtained in Eq.~(\ref{eq:L}).  We demonstrate that the di-lepton spectrum becomes anisotropic with respect to the magnetic-field direction and exhibits discrete and spike structures due to the kinematics in the Landau quantization and that the production with a lowest-Landau level fermion or anti-fermion is strictly prohibited, depending on the photon polarization and/or the fermion mass due to the conservation of spin or chirality.  Note that in realistic situations such as ultra-peripheral heavy-ion collisions, one should consider a photon source or distribution and convolute it with the di-lepton production rate to make some predictions, which will be discussed in a forthcoming publication.  

Before proceeding, we recall that the transverse momentum $ p^{(\prime)}_y $ is {\it canonical} momentum and is a gauge-dependent quantity in the Landau gauge (\ref{eq:Landau-gauge}).  Also, the transverse components of the {\it kinetic} momentum are not conserved, as one can imagine from the classical cyclotron motion.  These facts mean that each component of the transverse fermion momenta $p^{(\prime)}_x$ and $p^{(\prime)}_y$ is not a good quantum number nor measurable.  Within the current set-up of problem with a constant magnetic field, one can only measure the norm of the kinetic transverse momentum, assuming that the magnetic field is adiabatically damped out in the asymptotic future.

\subsection{Spikes in the longitudinal-momentum distribution} \label{sec4.2}

We first discuss how the Landau quantization manifests itself in kinematics (or more specifically, the energy-momentum conservation) in the photon--to--di-lepton conversion process.  We show that the fermion and anti-fermions' longitudinal momenta can only take discrete values because of the kinematics, resulting in spike structures in the distribution.  

The kinematical constraints are incorporated in the delta function in the squared amplitude  (\ref{eq:sq}).  Because of the delta function, the di-lepton production occurs only when 
\begin{align}
k_0 = \epsilon_n + \epsilon'_{n'}
	\label{eq:energy}
\end{align}
is satisfied, which is nothing but the energy conservation.  Noting the longitudinal momentum conservation $p'_z=k_z-p_z$, one can explicitly solve the condition (\ref{eq:energy}) and finds that the kinematically allowed $p_z$ for given $ k_0,k_z $ reads 
\begin{align}
	p_z 
		= \frac{ k_z ( k_\parallel^2 +  m_n^2- m_{n'}^2  ) \pm k^0 \sqrt{ \big( \, k_\parallel^2 - (m_n^2 + m_{n'}^2) \, \big)^2 -  4 m_n^2  m_{n'}^2 }  }{2k_\parallel^2} 
		\equiv p_{n,n'}^\pm \ ,  \label{eq:discpz} 
\end{align}
where 
\begin{align}	
	m_n \equiv \epsilon_n(p_z=0) =\sqrt{m^2 + 2 n|qB| } .   
\end{align}
It is evident that only two discrete values are allowed for $p_z=p_{n,n'}^+$ and $p_{n,n'}^-$ (accordingly, $p'_z = k_z-p_{n,n'}^\pm \equiv p_{n,n'}^{\prime \pm}$), once photon momenta $ k_0,k_z $ and the Landau levels $ n,n' $ are specified.  In other words, while magnetic fields do not directly quantize fermions' longitudinal momenta, the energy-momentum conservation and the Landau quantization force the longitudinal momenta to take discrete values.  For $p_{n,n'}^\pm$ to be real-valued, the inside of the square root must be non-negative.  This condition sets a threshold energy of the incident photon as\footnote{This semi-positivity condition itself admits another region $ k_\parallel^2  \leq  ( m_n - m_{n'})^2 $, which is, however, not compatible with the energy conservation (\ref{eq:energy}).  The energy conservation tells us that $  k_\parallel^2  = ( \sqrt{p_z^2 + 2  n \vert qB \vert + m^2} + \sqrt{(k_z-p_z)^2 + 2 n' \vert qB \vert + m^2} )^2 \geq ( m_n + m_{n'})^2 $, where we evaluated the boost-invariant quantity $  k_\parallel^2$ in the Lorentz frame such that $ k_z=0 $. } 
\begin{align}
	k_\parallel^2  \geq ( m_n + m_{n'})^2 \, . \label{eq:thres}
\end{align}
The right-hand side is the smallest possible invariant mass of a di-lepton.  As the system is boost invariant along the magnetic field, $k_\parallel^2$ on the left-hand side is boost invariant and gives the minimum photon energy for the di-lepton production in the Lorentz frame such that $k_z=0 $.  Note that both of the two on-shell momenta $p_{n,n'}^\pm$ take the same limiting value at the threshold energy (\ref{eq:thres}) as
\begin{align}
	p_{n,n'}^\pm \to k_z \frac{m_n}{ m_n + m_{n'} } \, . \label{eq:limpz}
\end{align}

\begin{figure}
\begin{center}
	\mbox{\small For $  (k_z/m, |qB|/m^2) = (0, 1)$}\\
	\includegraphics[width=0.49\hsize]{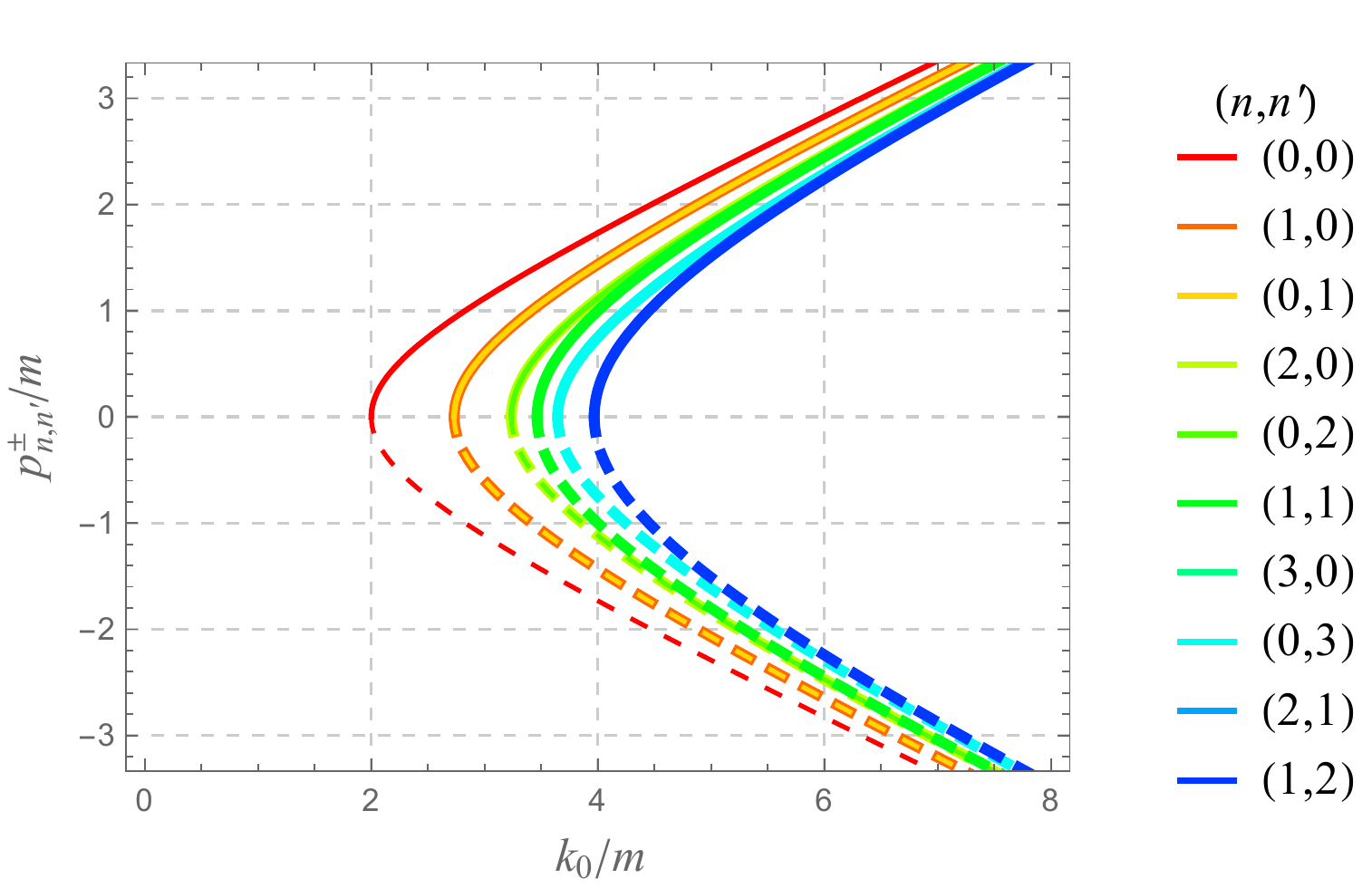}
	\includegraphics[width=0.49\hsize]{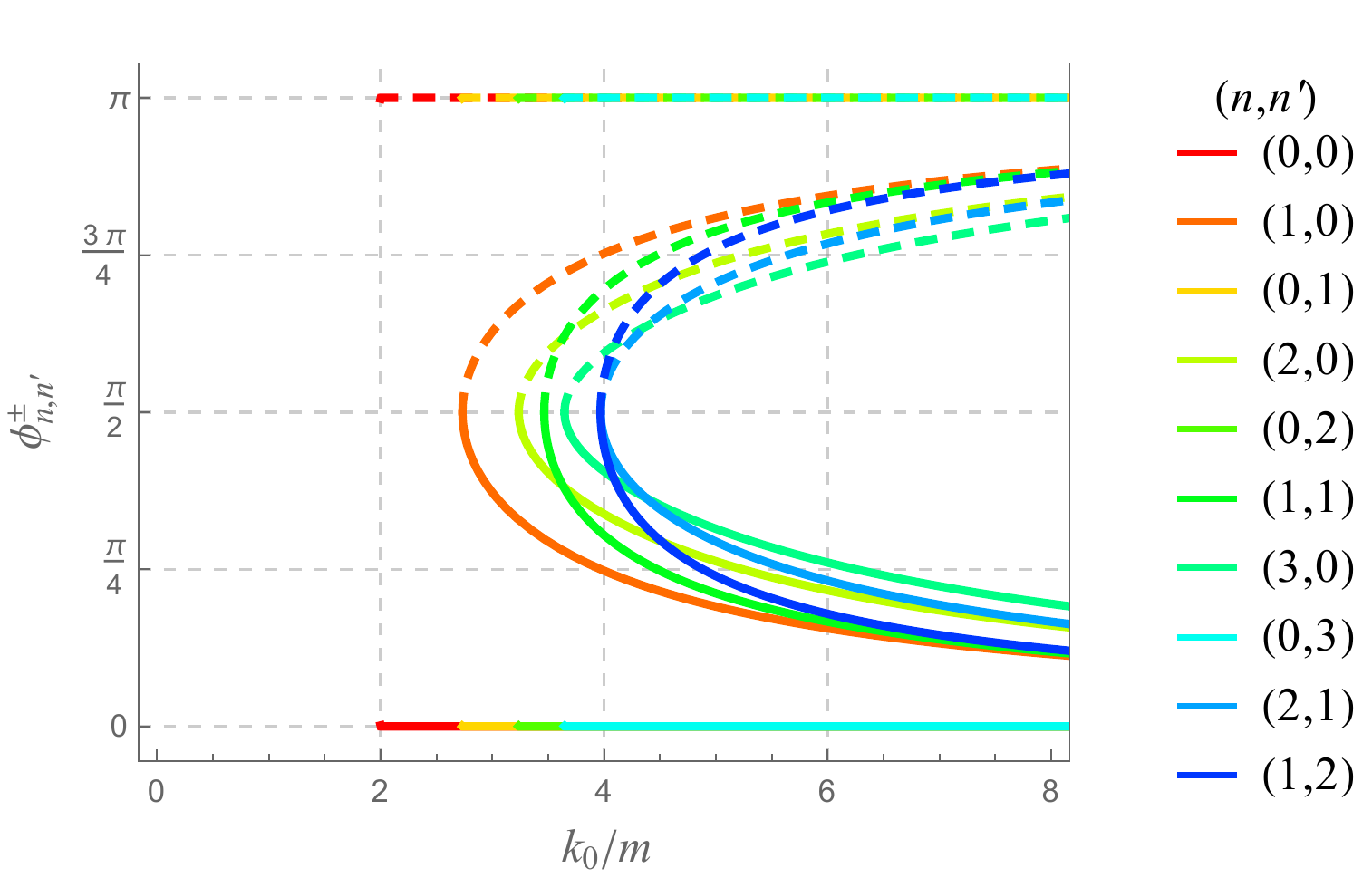} 
\\
	\mbox{\small For $  (k_z/m, |qB|/m^2) = (0, 3)$}\\
	\includegraphics[width=0.49\hsize]{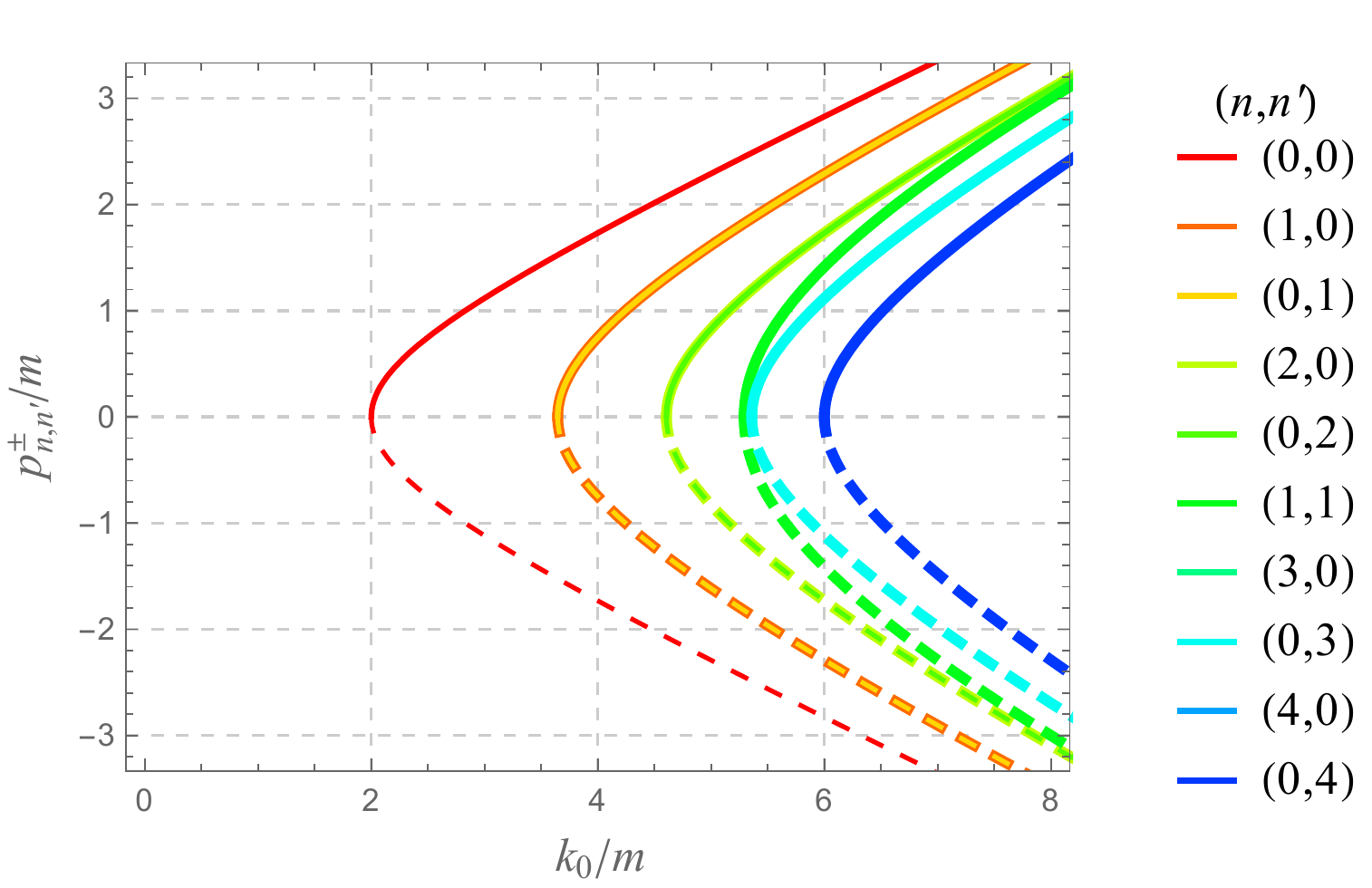}
	\includegraphics[width=0.49\hsize]{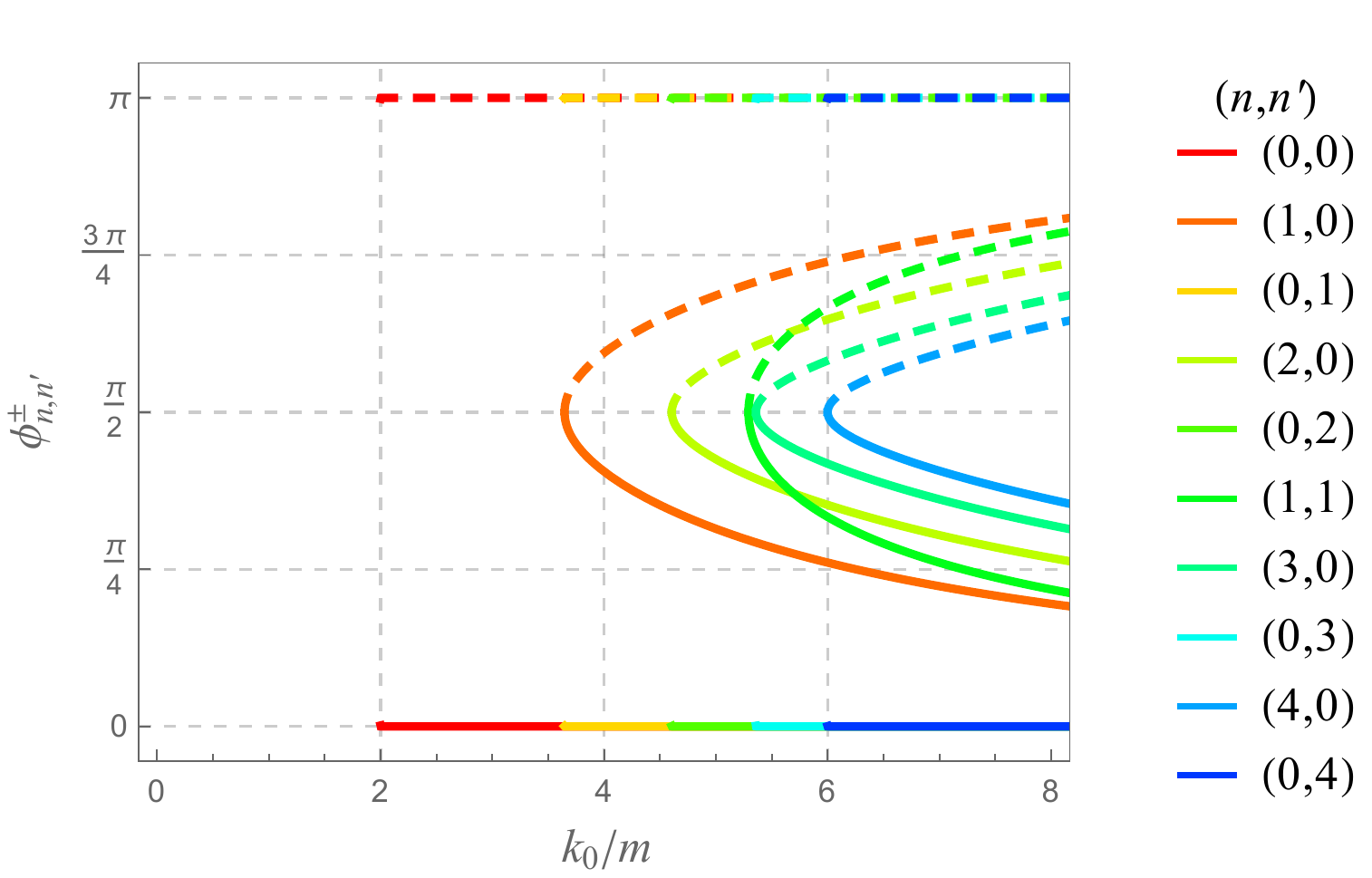}
\\
	\mbox{\small For $  (k_z/m, |qB|/m^2) = (3, 3)$}\\
	\includegraphics[width=0.49\hsize]{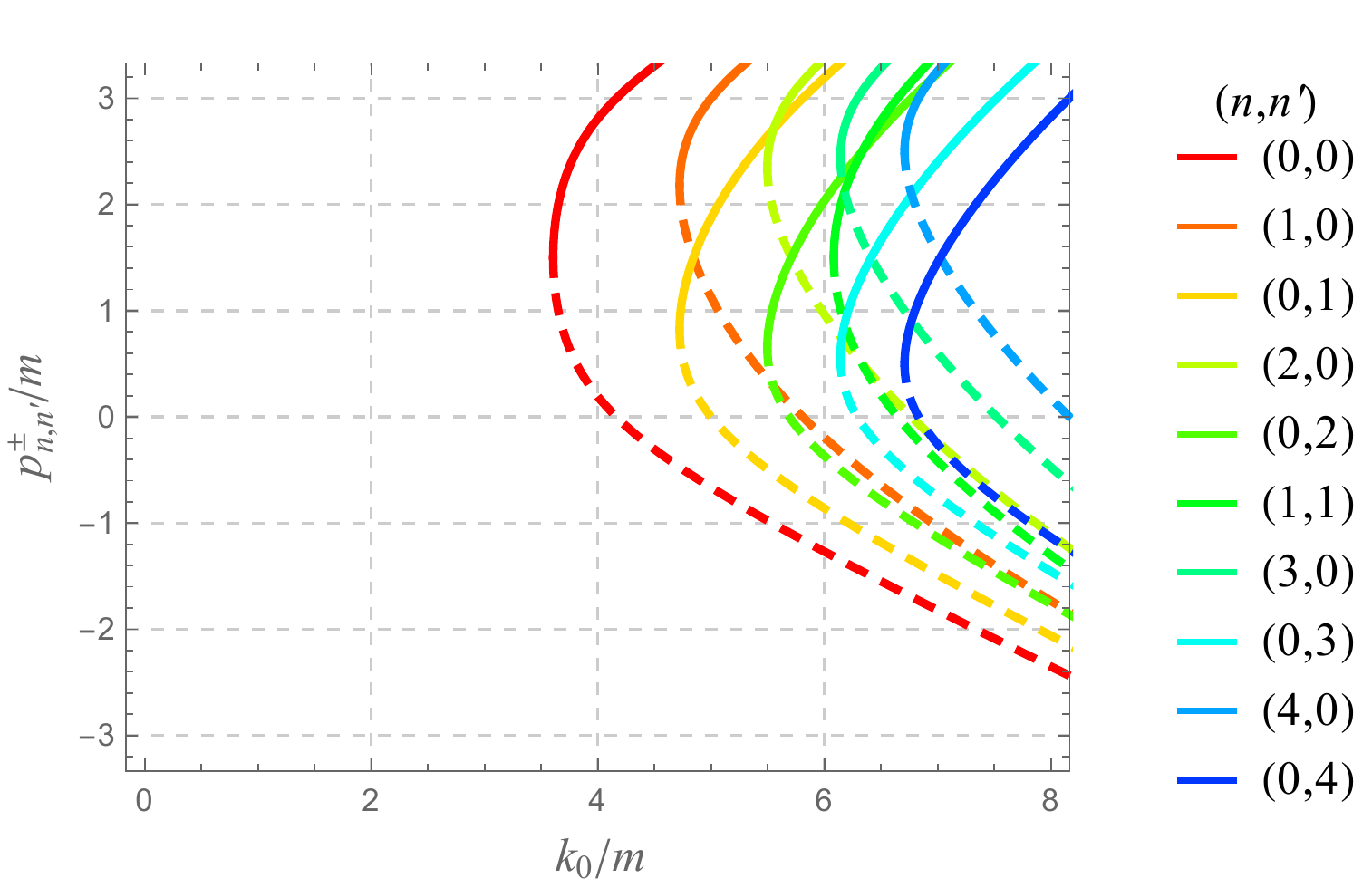}
	\includegraphics[width=0.49\hsize]{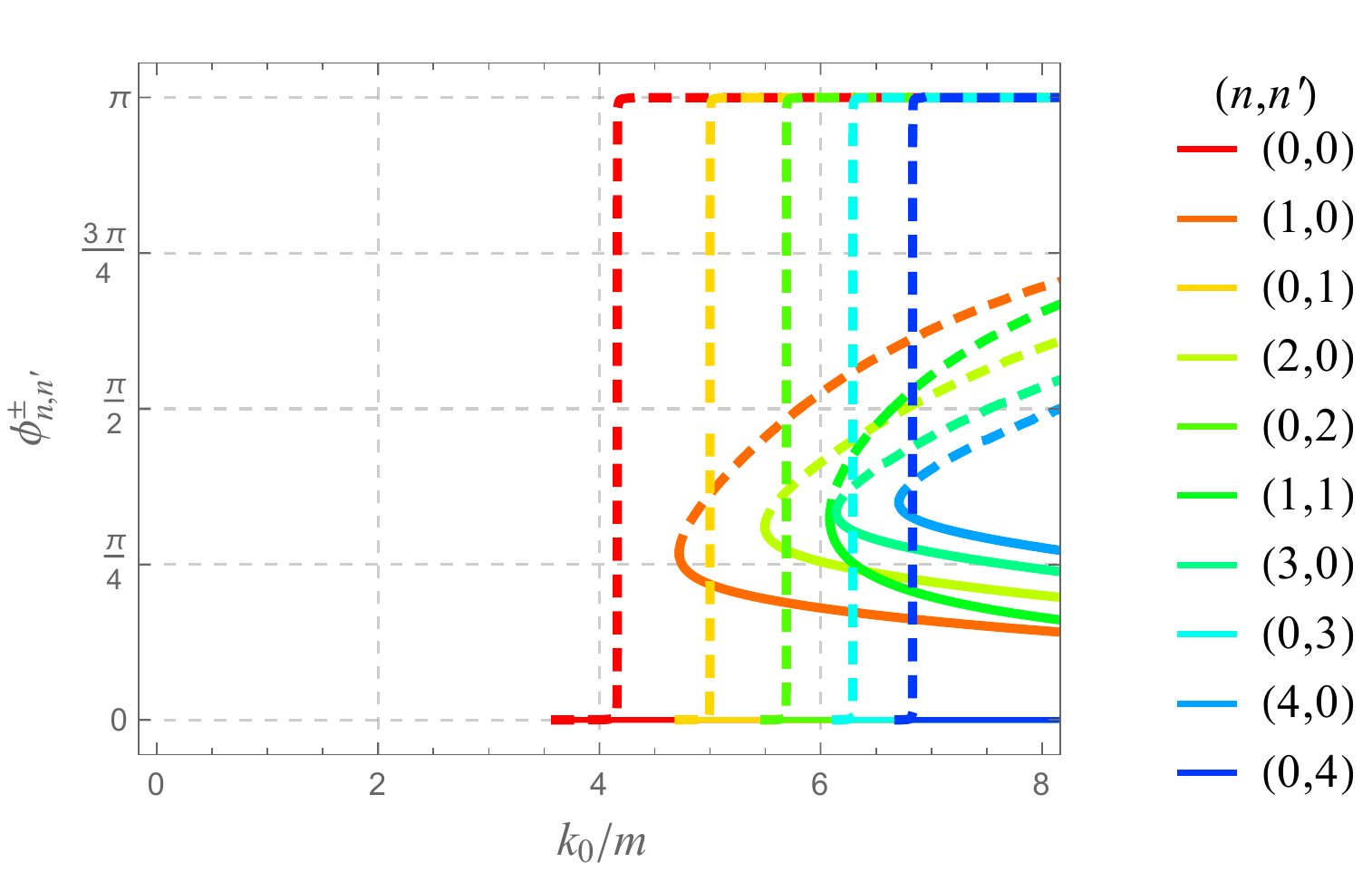}
\end{center}
\caption{
The longitudinal momentum $p_{n,n'}^\pm$ (left) and zenith angle $\phi_{n,n'}^\pm$ (right) allowed for the di-lepton to take, when converted from a photon carrying energy $ k_0 $ and momentum $ k_z $ with parameter sets: $  (k_z/m, |qB|/m^2) = (0, 1)$ [first row], $ (0, 3)$ [second row], and $ (3, 3)$ [third row].  Only the first ten pairs of Landau levels are shown in ascending order with respect to the threshold energy (\ref{eq:thres}) from red (which is always the lowest Landau level pair $n=n'=0$) to blue.  }
\label{fig:pz-phi}
\end{figure}

\begin{figure}
\begin{center}
		\includegraphics[width=0.45\hsize]{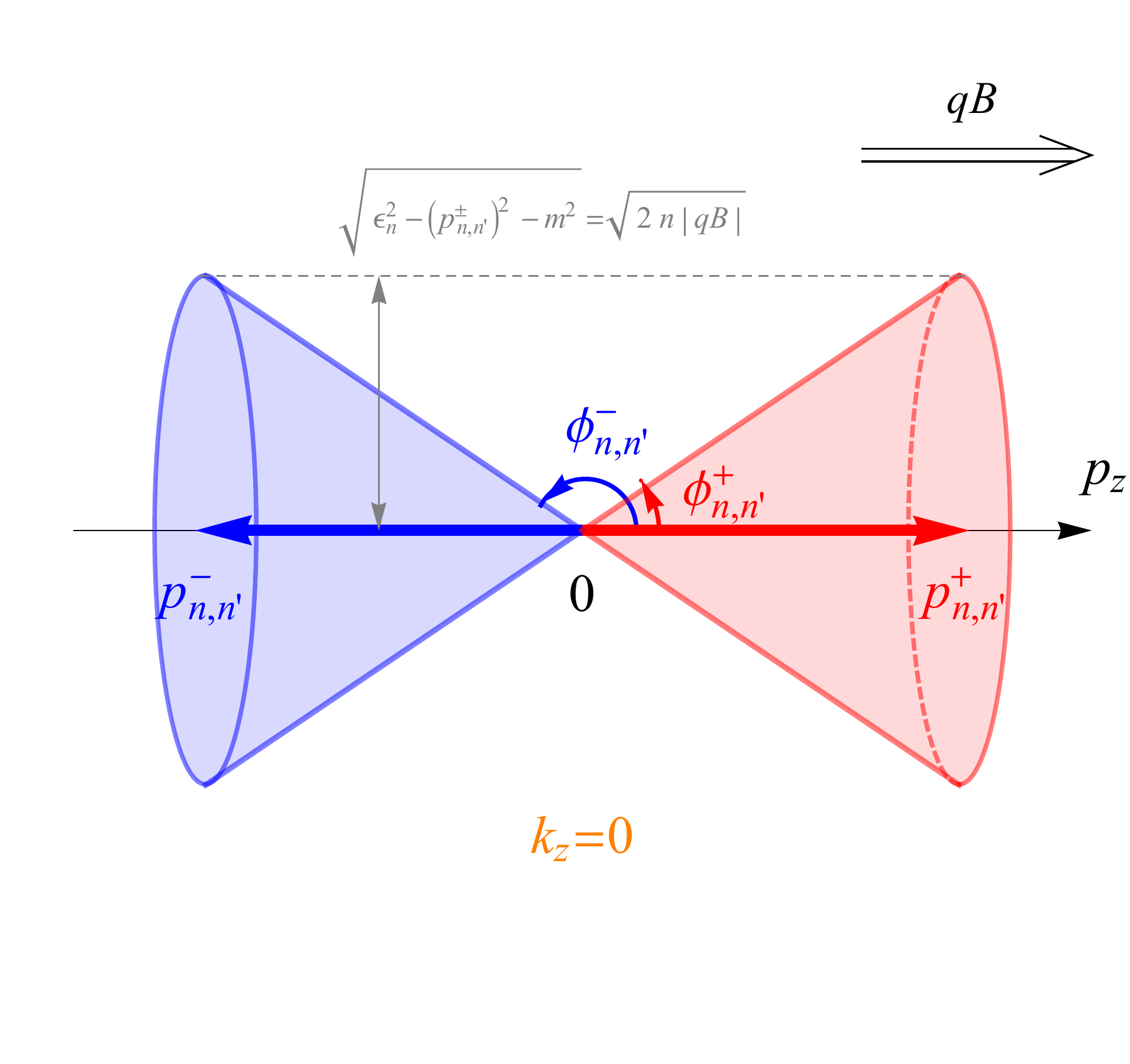} \hspace*{5mm}
		\includegraphics[width=0.45\hsize]{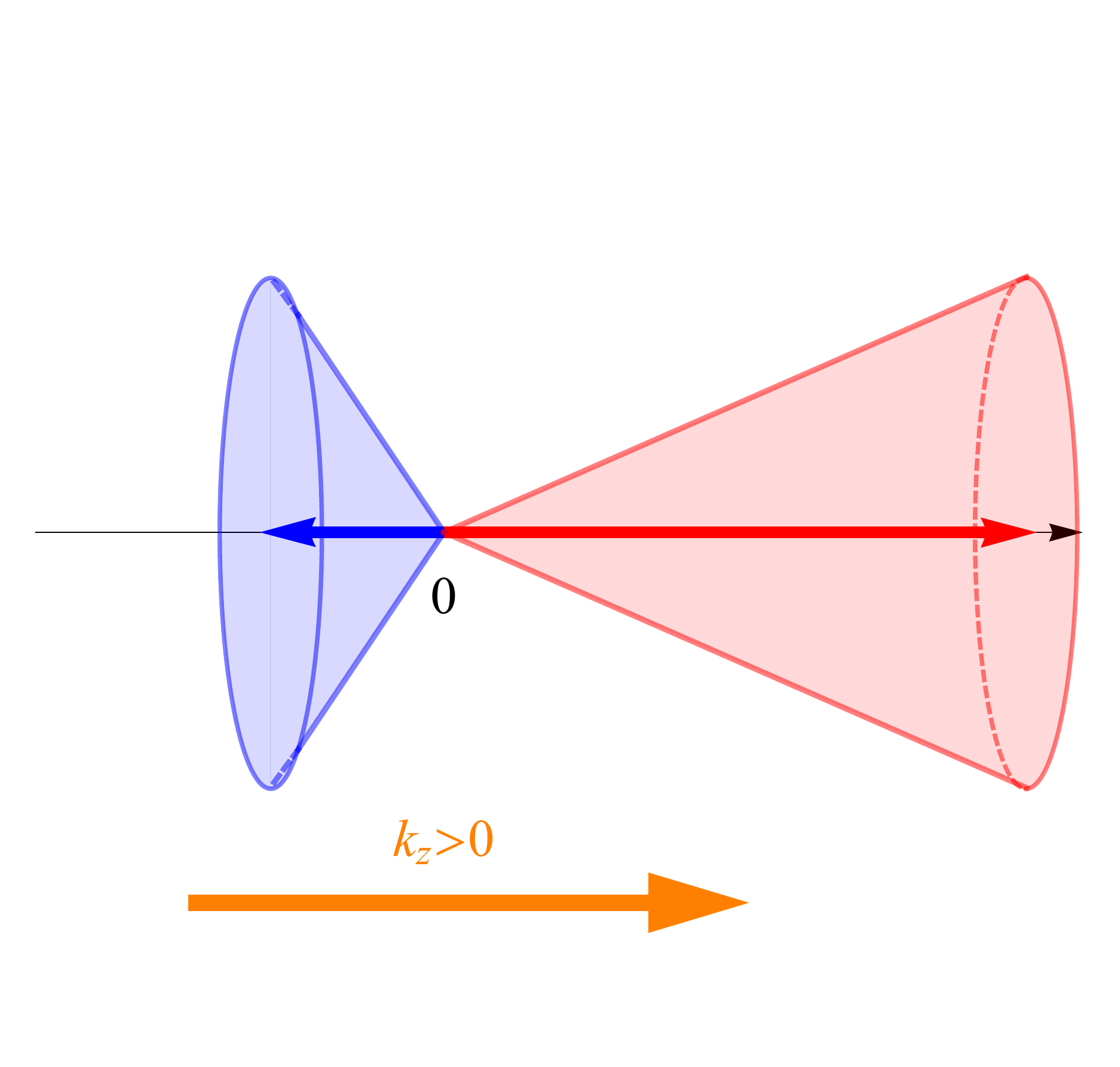}
\end{center}
\vspace{-7mm}
\caption{A schematic picture of the on-shell fermion momentum $p_{n,n'}^\pm$ and the corresponding zenith angle $\phi^\pm_{n,n'}$ for a vanishing photon longitudinal momentum $k_z=0$ (left) and the boost effect by a nonzero $ k_z > 0 $ (right).  Shown is the case for $ n=n' $ where the two cones appear in the same size at $  k_z=0$.}
\label{fig:boost}
\end{figure}

We discuss more details about the discretized fermion's on-shell longitudinal momentum $p_{n,n'}^\pm$.  The left column in Fig.~\ref{fig:pz-phi} shows $p_{n,n'}^\pm$ as a function of the photon energy normalized by the fermion mass $ k_0 /m$.  The photon momentum $ k_z $ and the magnetic field strength $ |qB| $ are also normalized in the same manner and are fixed in each plot as $  (k_z/m, |qB|/m^2) = (0, 1)$ [first row], $  (0, 3)$ [second row], and $ (3, 3)$ [third row].  Each colored curve corresponds to each pair of Landau levels $ (n,n') $.  The solid and dotted curves show $ p_{n,n'}^+ $ and $ p_{n,n'}^- $, respectively.  The threshold energy (\ref{eq:thres}) is the point where $p_{n,n'}^+=p_{n,n'}^-$ holds [cf. Eq.~(\ref{eq:limpz})], i.e., where the solid and dotted curves merge.  We find the following: 
\begin{itemize}
\item The di-lepton spectrum converted from a photon carrying a fixed energy $ k_0/m $ is given as a superposition of $p_{n,n'}^\pm $ allowed for each pair of $ (n,n') $.  The spectrum is given as a set of intersections among the curves and a vertical line at $ k_0/m = {\rm const.}$, and looks like a bunch of spikes with a vanishing width. 
Namely, the produced di-leptons exhibit a discrete spike spectrum in the longitudinal direction.  As we vary other continuous parameters and/or consider convolution with some photon source, the spikes may acquire a finite width.  An incident photon carrying a larger energy can be converted to di-leptons in higher Landau levels.  Those are the common features in the three panels.

\item When $ k_z =0$ (first and second rows), the plots are symmetric in the reflection with respect to the horizontal axis because $ p_{n,n'}^+ + p_{n,n'}^- = k_z = 0$ due to the momentum conservation.  When $ k_z > 0 $ (third row), the spectrum is shifted to the positive $ p_z $ direction.  This is a boost effect by the finite $ k_z $ with respect to the center-of-momentum frame of a di-lepton where $ k_z = p_z + p_z' =0$ (see Fig.~\ref{fig:boost}).  Also, there are two-fold degeneracies $p_{n,n'}^\pm = p_{n',n}^\pm$ for $k_z=0$ because of the reflection symmetry, while they are resolved by nonzero $ k_z \neq 0$.

\item Since the spacings of the energy levels in the Landau quantization increase with $ |qB| $ (which we call Landau-level spacings), we find larger spacings between adjacent $p_{n,n'}^{\pm}$'s in the second and third rows ($|qB|/m^2=3$) as compared to the first row ($|qB|/m^2=1$).  In particular, in a very strong magnetic field such that $|qB|>[(\sqrt{|k_\parallel^2|}-m)^2-m^2]/2$, only the lowest Landau level $n=n'=0$ can contribute to the di-lepton production, and there are only two discrete spikes in the longitudinal $p_z$-distribution at $p_z = p_{0,0}^{\pm}=[k_z \pm \sgn(k^2_\parallel) k^0\sqrt{1-4m^2/k_\parallel^2}]/2$.  Note that the lowest-Landau energy level $\epsilon_0$ is independent of $qB$, and $p_{0,0}^{\pm}$ stays in the low-energy regime even in the (infinitely) strong-field limit.  For weak magnetic fields, the spacings between $p_{n,n'}^{\pm}$'s get smaller, i.e., many Landau levels can contribute to the di-lepton production, which smears out the spike structures.  

\end{itemize}

\subsection{Zenith-angle distribution} \label{sec:4.2}

One can predict the zenith-angle distribution measured from the direction of the magnetic field (see Fig.~\ref{fig:boost}):  
\begin{align}	
	\phi_{n,n'}^\pm (k_\parallel) \equiv 
		\tan^{-1} \frac{   \sqrt{ \epsilon_n^2 - (p_{n,n'}^\pm)^2 -m^2 } }{p_{n,n'}^\pm} \ , \quad 
	\phi_{n,n'}^{\prime\pm} (k_\parallel) 		\equiv 
		\tan^{-1} \frac{   \sqrt{ \epsilon_{n'}^{\prime2} - (p_{n,n'}^{\prime\pm})^2 -m^2 }}{p_{n,n'}^{\prime\pm}} \ ,\label{eq:phil}
\end{align}
for a fermion and anti-fermion, respectively.  The numerator $\sqrt{ \epsilon^{(\prime) \, 2}_{n^{(\prime)}}  - (p_{n,n'}^{(\prime)\pm})^2 -m^2} = \sqrt{2n^{(\prime)}|qB|}$ corresponds to the magnitude of the transverse momentum under the Landau quantization.  The range of the angle is defined as $ 0 \leq \phi_{n,n'}^{(\prime)\pm} \leq \pi $.  Both of $\phi_{n,n'} ^\pm$ and $\phi_{n,n'}^{\prime\pm} $ are discrete quantities for a given photon momentum because of the discretization of $\epsilon^{(\prime)}$ and $p_z^{(\prime)}$, and each of them takes two values corresponding to $ p_{n,n'}^{(\prime)\pm}$. 
Note that $ \phi_{n,n'}^{\prime \pm}$ is obtained from $ \phi_{n,n'}^{\pm}$ 
just by exchanging the labels as $n \leftrightarrow n'$ and $+ \leftrightarrow -$, 
because $ p^{\prime\pm}_{n,n'} = p_{n',n}^\mp $ under those exchanges as a manifestation of the CP symmetry. 
Thus, it is sufficient to focus on $ \phi_{n,n'}^{\pm}$ in the following.  Remark again that one cannot predict the azimuthal-angle distribution, which is equivalent to predicting ${\bm k}_\perp$-distribution (not $|{\bm k}_\perp|$-distribution), within the current set-up of the problem with a constant magnetic field.  To get this information, one needs to know when the produced fermions are released from the cyclotron motion, implying that one has to go beyond the constant magnetic field and solve the dynamics with a time-dependent magnetic field damped out in time.  This is a dynamical and process-dependent issue beyond the scope of the present work.

The right column in Fig.~\ref{fig:pz-phi} shows the zenith-angle distribution $\phi_{n,n'}^\pm$.  The plotting style is the same as the left column for $p_{n,n'}^\pm$, and one may enjoy correspondences between $p_{n,n'}^\pm$ and $\phi_{n,n'}^\pm$ in the left and right columns.  We find the following: 
\begin{itemize}

\item In the lowest Landau level $n=0$, the produced fermion moves precisely along the magnetic-field direction, and thus $\phi _{0,n'}^{\pm}= 0$ and $\pi$ for ${\rm sgn}\,\phi _{0,n'}^{\pm}>0$ and $<0$, respectively.  On the other hand, fermions with higher Landau levels are emitted with a finite transverse momentum, and hence $ 0 < \phi_{n,n'}^\pm (k_\parallel) < \pi $.  Note that there are discontinuous jumps in the lowest Landau level $n=0$ for $k_z \neq 0$ in the third row.  This behavior just originates from the change of the sign of $ p_{n,n'}^- $ ($ p_{n,n'}^+$) for $k_z>0$ ($k_z<0$) due to the boost effect; see the left column.  Those jumps, however, do not occur for massless fermions in the lowest Landau level, as the Lorentz boost cannot change the sign of their momenta.  

\item The zenith-angle distribution is limited to few discrete directions when the photon energy is small, where fermions can take only a few number of low-lying Landau levels due to the threshold condition (\ref{eq:thres}).  As we increase the photon energy, more number of higher Landau levels start contributing to the production.  This results in the smearing of the spike structures in the zenith-angle distribution with narrower spacings. 

\item Comparing panels in the first ($|qB|/m^2=1$) and second ($|qB|/m^2=3$) rows, we understand that major effects of strong magnetic fields are two-fold: (i) shift of $ \phi_{n,n'}^{\pm}$ to higher photon energies, and (ii) squeezing of $ \phi_{n,n'}^{\pm} \to \pi/2 $.  Namely, (i) the photon threshold energy (\ref{eq:thres}) is lifted up except for the lowest Landau level when $ |qB|$ is increased.  Therefore, the contributions from higher Landau levels are shifted to higher photon energies (rightward), and we find only the lowest-Landau level contribution in the strong field limit $\sqrt{|qB|} \gg k_\parallel^2$.  (ii) Under a stronger magnetic field, a larger portion of the photon energy need to be converted to the fermion transverse energy $ \sqrt{2 n|qB| } $.  Only the remaining portion can be converted to the longitudinal momentum $ p_z $, and thus the magnitude of $p_z$ is reduced.   Therefore, the angle $\phi_{n,n'}^\pm$ for each pair of Landau levels approaches $ \pi/2 $ as we increase $ |qB| $.  This tendency is seen as squeezing of a bunch of curves toward the center at $ \pi/2 $.

\item When $ k_z=0 $ (the first and second rows) and thus $p_{n,n'}^+ = -p_{n,n'}^-$, 
we find that $ \phi_{n,n'}^+ + \phi_{n,n'}^- = \pi $ holds.  This is a consequence of the reflection symmetry with respect to the transverse plane or a flip of the magnetic-field direction (cf. Fig.~\ref{fig:boost}).  When $ k_z > 0$ (the third row), the reflection symmetry is broken and the curves in the plot become asymmetric with respect to the horizontal axis.  This originates from the positive increase of $ p_{n,n'}^\pm $ in the left column and is a consequence of the Lorentz boost of the cones in the longitudinal direction (see Fig.~\ref{fig:boost}), i.e., one of the cones shrinks while the other expands.  

\end{itemize}
We emphasize that the kinematics in the Landau quantization is essential in the above observations.  Thus, the results obtained in the last two subsections \ref{sec4.2} and \ref{sec:4.2} are insensitive to the size of the transverse photon momentum $ |{\bm k}_\perp| $, as the kinematics is determined by the longitudinal variables $k_0$ and $k_z$ only.  The di-lepton production acquires $ |{\bm k}_\perp| $-dependence only via the scalar form factor ${\Gam}_{n,n'}$, as we have discussed in Sec.~\ref{sec:3.3} and will further demonstrate below.

\subsection{Inclusive photon--to--di-lepton conversion rate} \label{sec:conversion}

Having explained the basic behaviors of the contracted lepton tensors in Sec.~\ref{sec:3.3} and the kinematics of the di-lepton production in Secs.~\ref{sec4.2} and \ref{sec:4.2}, we discuss more details about the di-lepton production by investigating the inclusive photon--to--di-lepton conversion rate $\rate$.  It is obtained by integrating the squared amplitude (\ref{eq:sq}) as
\begin{align}
\rate_{\lambda} (k) 
	&\equiv \sum_{n,n'} \sum_{s,s'}  \int \frac{dp_z dp_y}{(2\pi)^2(2\epsilon_n)}  \int \frac{dp'_z dp'_y}{(2\pi)^2(2\epsilon_{n'})} \frac{ |\varepsilon_\mu q {\mathcal M}^\mu|^2}{2k^0(2\pi)^3\delta^{(3)}(0)}   \nonumber\\
	&= \sum_{n,n'}\left[ q^2 T  \frac{|qB|}{2\pi} \sum_{ p_z = p_{n,n'}^\pm} \frac{ \Theta( \, k_\parallel^2 - ( m_n + m_{n'})^2 \,) }{  8 k^0 |p_z\epsilon'_{n'} - (k_z - p_z)  \epsilon_n | }  \left.   \ [ \varepsilon_\mu^\lambda  \varepsilon^{\lambda\ast}_\nu L^{\mu\nu}_{n,n'} ] 		\right|_{p'_z=k_z-p_z}  \right] \nonumber\\
	&\equiv	 \sum_{n,n'} \rate_{\lambda}^{n,n'} (k) \, ,  \label{eq:num}
\end{align}
where we used $\int dp_y = |qB| L_x$ and introduced $ \lambda =0,\pm,\parallel$ and $\Theta$, representing the four photon polarization mode and a step function, respectively.  We also used
\begin{align} 
	\delta ( \epsilon_n + \epsilon'_{n'} - k^0 ) 
	&= \sum_{\alpha = \pm} \delta(p_z-p_{n,n'}^\alpha) 
	\left|  \frac{\epsilon_n \epsilon'_{n'}}{p_z\epsilon'_{n'} - (k_z - p_z)  \epsilon_n }  \right| 
	\Theta( \, k_\parallel^2 - ( m_n + m_{n'})^2 \, )
	\label{eq:delta-function}
	\ ,
\end{align}
which accounts for the kinematics discussed in the previous subsections.  As explained in Sec.~\ref{sec-3}, the coupling between the lepton tensor (\ref{eq:L}) and an incident photon depends on the photon polarization mode via the tensor structures such as $L_\parallel^{\mu\nu}$ and $\Q_\pm^{\mu\nu}$.  Plugging Eq.~(\ref{eq:projected-L}) into Eq.~(\ref{eq:num}), we obtain the polarization-projected conversion rates 
\begin{subequations}\label{eq:N-lambda} 
\begin{align}
	\rate_{0}^{n,n'} 
		&\!=   q^2 T \frac{|qB|}{2\pi} \!\!\!\sum_{ p_z = p_{n,n'}^\pm}\!\!\! \frac{  \Theta( \, k_\parallel^2 - ( m_n + m_{n'})^2 \,) }{  4 k^0|p_z\epsilon'_{n' }- (k_z - p_z)  \epsilon_n | } \Big[ ( \epsilon_n \epsilon'_{n' } + p_z p'_z - m^2 ) ( | {\Gam}_{n , n'} |^2   + |{\Gam}_{n-1 , n'-1 } |^2 )   \nonumber \\
			&\quad\quad\quad\quad\quad\quad\quad\quad 	-  2 |q B| \left( - |\bar {\bm k}_\perp| ^2 |{\Gam}_{n-1,n'}|^2 + n |   {\Gam}_{n, n'} |^2 +  n' |  {\Gam}_{n-1,n'-1}|^2 \right) \Big]_{p'_z=k_z-p_z} \, ,	\label{eq:N0}  \\
	\rate_{+}^{n,n'}    
		&\!= q^2 T  \frac{|qB|}{2\pi}  \!\!\!\sum_{ p_z = p_{n,n'}^\pm}\!\!\! \frac{\Theta( \, k_\parallel^2 - ( m_n + m_{n'})^2 \,) }{4 k^0|p_z\epsilon'_{n' }- (k_z - p_z)  \epsilon_n | } \Big[ 2 ( \epsilon_n \epsilon'_{n' } - p_z p'_z + m^2 )   | {\Gam}_{n , n'-1 } |^2\Big]_{p'_z=k_z-p_z} \, , \label{eq:Np} \\
	\rate_{-}^{n,n'}   		
		&\!= q^2 T  \frac{|qB|}{2\pi} \!\!\!\sum_{ p_z = p_{n,n'}^\pm} \!\!\! \frac{\Theta( \, k_\parallel^2 -( m_n + m_{n'})^2 \,) }{4 k^0|p_z\epsilon'_{n' }- (k_z - p_z)  \epsilon_n |} \Big[ 2 ( \epsilon_n \epsilon'_{n' } - p_z p'_z + m^2 )   | {\Gam}_{n-1 , n' } |^2\Big]_{p'_z=k_z-p_z} \, , \label{eq:Nm} \\
	\rate_{\parallel}^{n,n'}  
		&\!= q^2 T  \frac{|qB|}{2\pi} \!\!\!\sum_{ p_z = p_{n,n'}^\pm} \!\!\! \frac{ \Theta( \, k_\parallel^2 - ( m_n + m_{n'})^2 \,) }{4 k^0 |p_z\epsilon'_{n' }- (k_z - p_z)  \epsilon_n |} \Big[  ( \epsilon_n \epsilon'_{n' } + p_z p'_z + m^2 )   ( |  {\Gam}_{n , n'} |^2   + |{\Gam}_{n-1 , n'-1 } |^2 )   \nonumber \\
			&\quad\quad\quad\quad\quad\quad\quad\quad +	 2 |q B| \left( - |\bar {\bm k}_\perp| ^2 |{\Gam}_{n-1,n'}|^2 + n |  {\Gam}_{n, n'} |^2 +  n' |   {\Gam}_{n-1,n'-1}|^2 \right) \Big]_{p'_z=k_z-p_z} \, .   \label{eq:Npara}
\end{align}
\end{subequations}
Note that $ \rate_{+}^{n,n'} = \rate_{-}^{n',n} $ (but $ \rate_{+}^{n,n'} \neq  \rate_{-}^{n,n'} $ in general) and that $ \rate_{+} = \rate_{-} $ after the summation over $ n, \, n' $.  This is a natural manifestation of the fact that the lepton tensor does not contain any parity-breaking effect.  At the algebraic level, one can show those identities by using the facts that $ |{\Gam}_{n , n'-1 } |^2  =  |  {\Gam}_{n'-1 , n} |^2   $ [cf. Eq.~(\ref{eq:FF-Landau-2})] and that the denominator $ |p_z\epsilon'_{n' }- (k_z - p_z)  \epsilon_n | $ as well as the other parts is invariant with respect to simultaneous interchanges between $ n , n'  $ and between $ p^\pm _{n,n'} , p^\mp _{n,n'} $ (cf. $ p^\pm _{n',n}  = k_z - p^\mp_{n,n'}$).

In the following, we examine dependences of the conversion rates $ \rate_\lambda $ on the physical parameters such as the photon momentum and the magnetic field strength.  We normalize all the dimensionful parameters by the fermion mass assuming that $m\neq 0$, except in Sec.~\ref{sec434} where we discuss the massless limit ($m\to 0$).

\subsubsection{Photon-energy dependence}\label{sec4.3.1}

\begin{figure}
\begin{center}
	\mbox{\small For $  (k_z/m, |qB|/m^2) = (0, 1)$}\\
	\includegraphics[width=0.34\hsize]{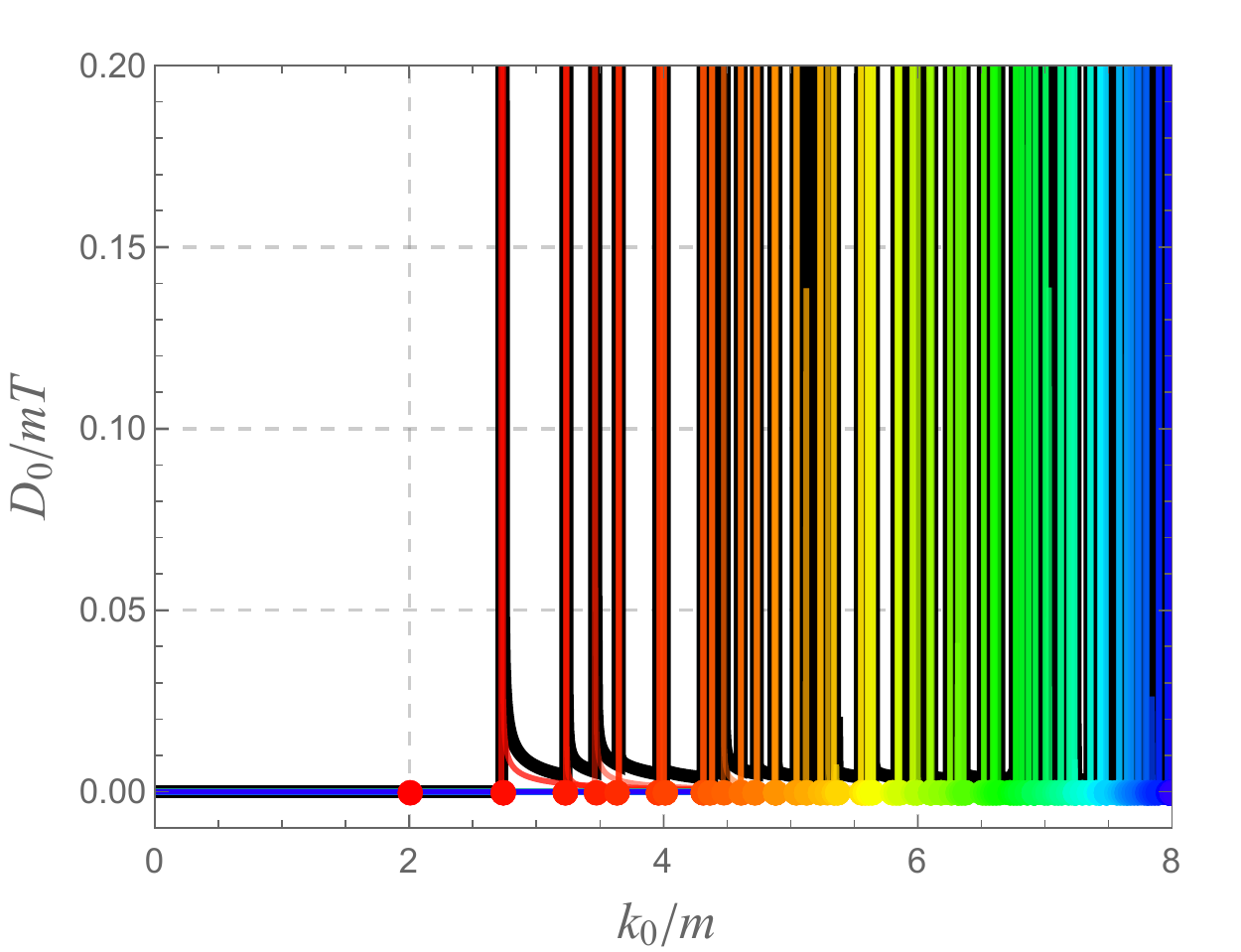} \hspace*{-4mm}
	\includegraphics[width=0.34\hsize]{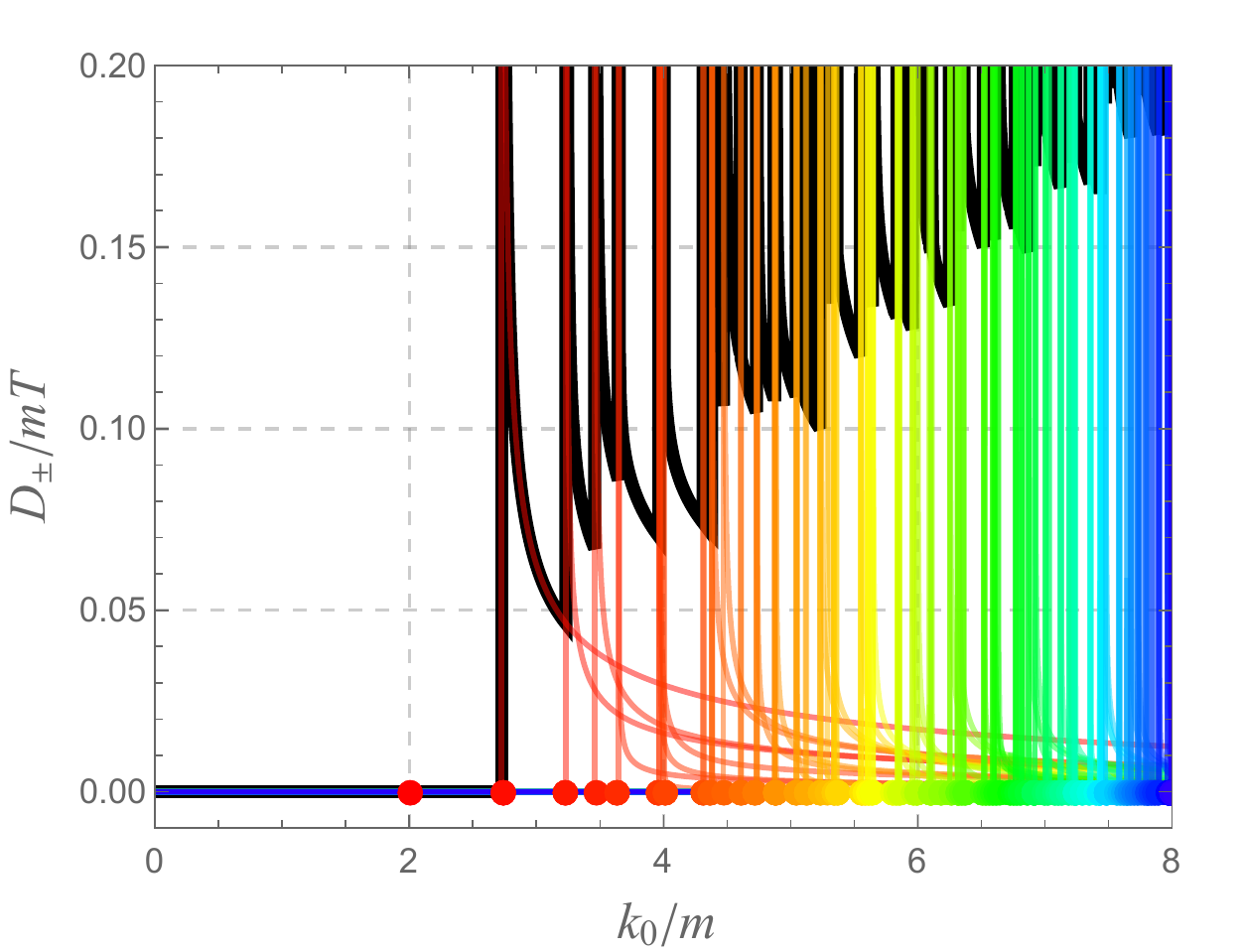} \hspace*{-4mm}
	\includegraphics[width=0.34\hsize]{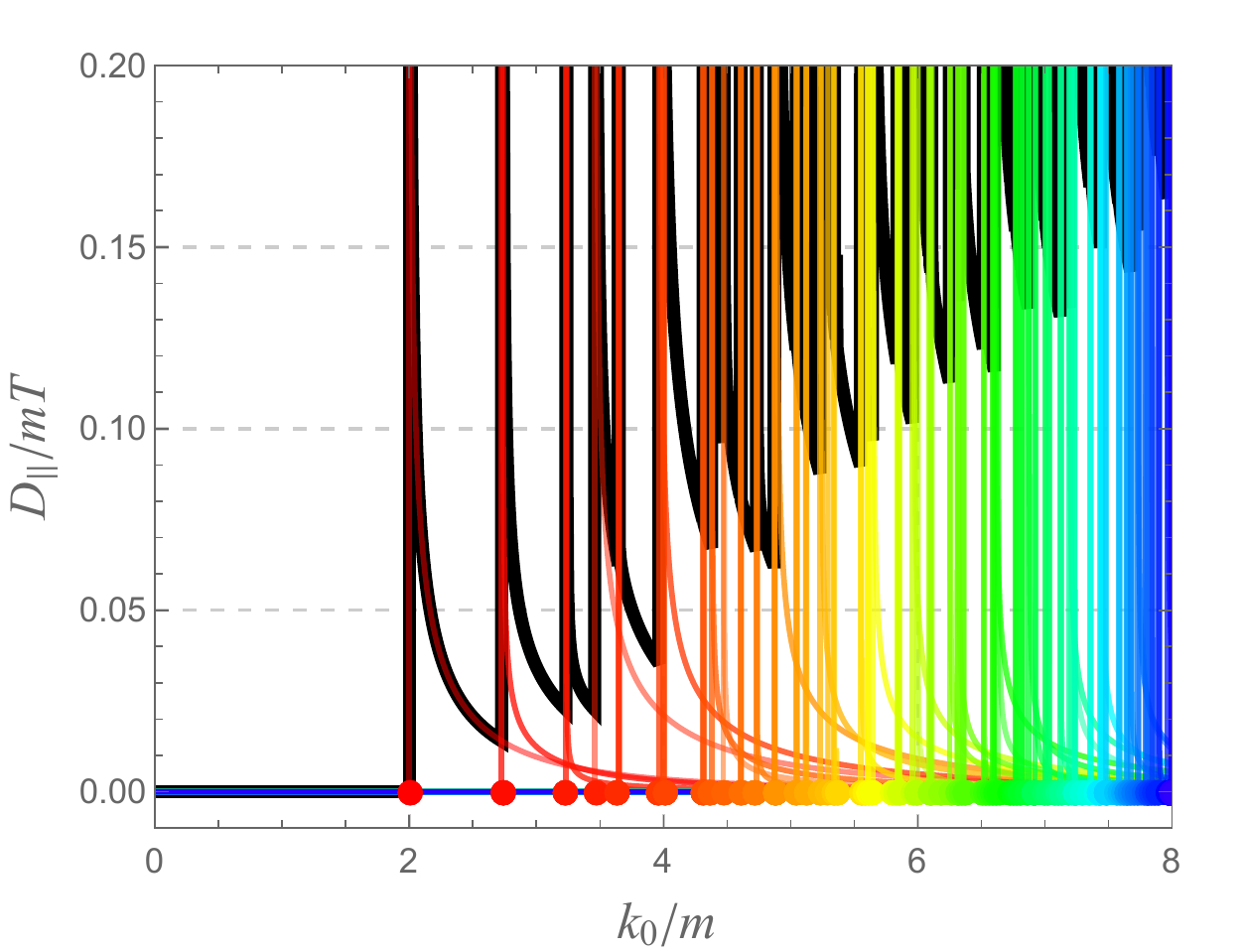} 
\\
	\mbox{\small For $  (k_z/m, |qB|/m^2) = (0, 3)$}\\
	\includegraphics[width=0.34\hsize]{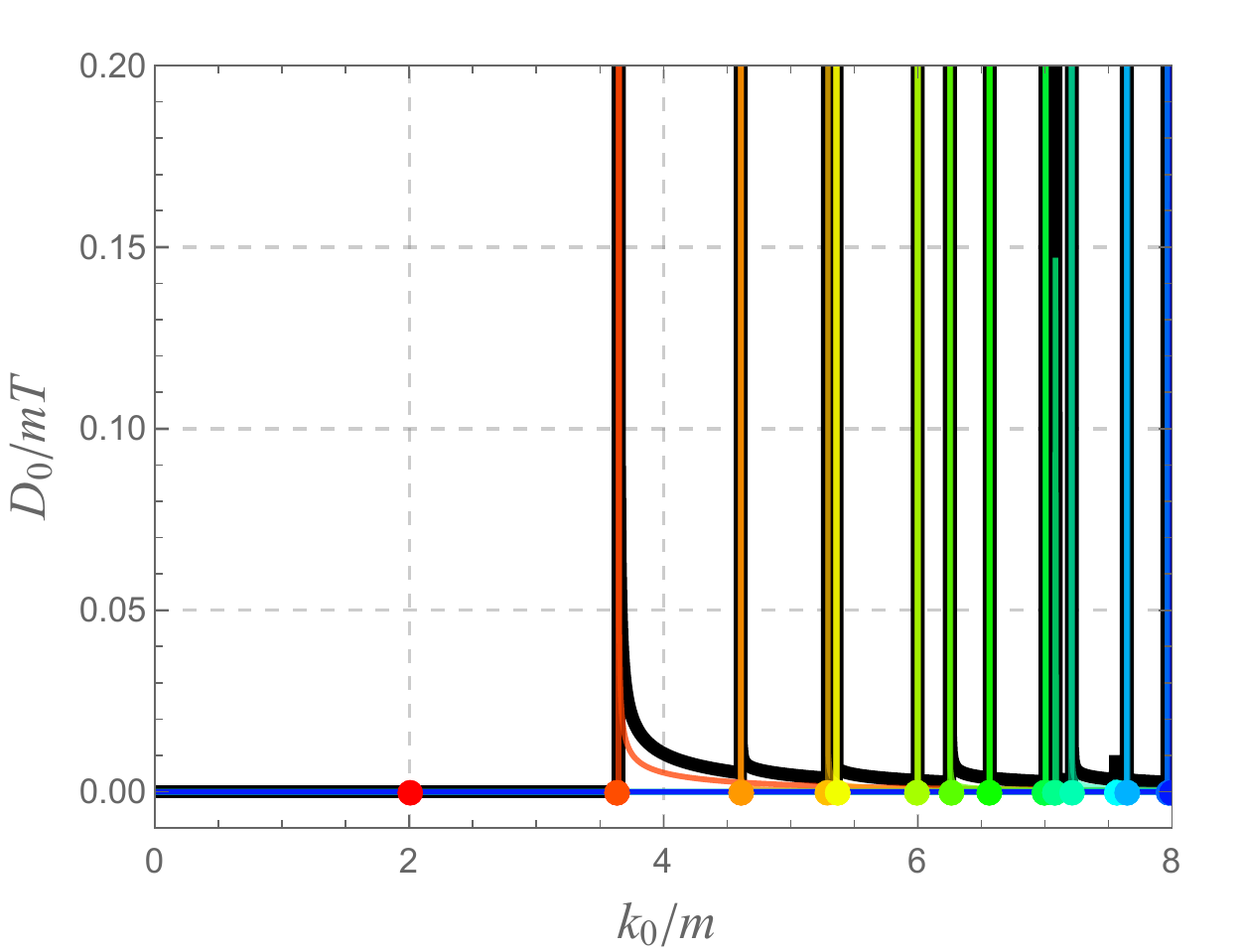} \hspace*{-4mm}
	\includegraphics[width=0.34\hsize]{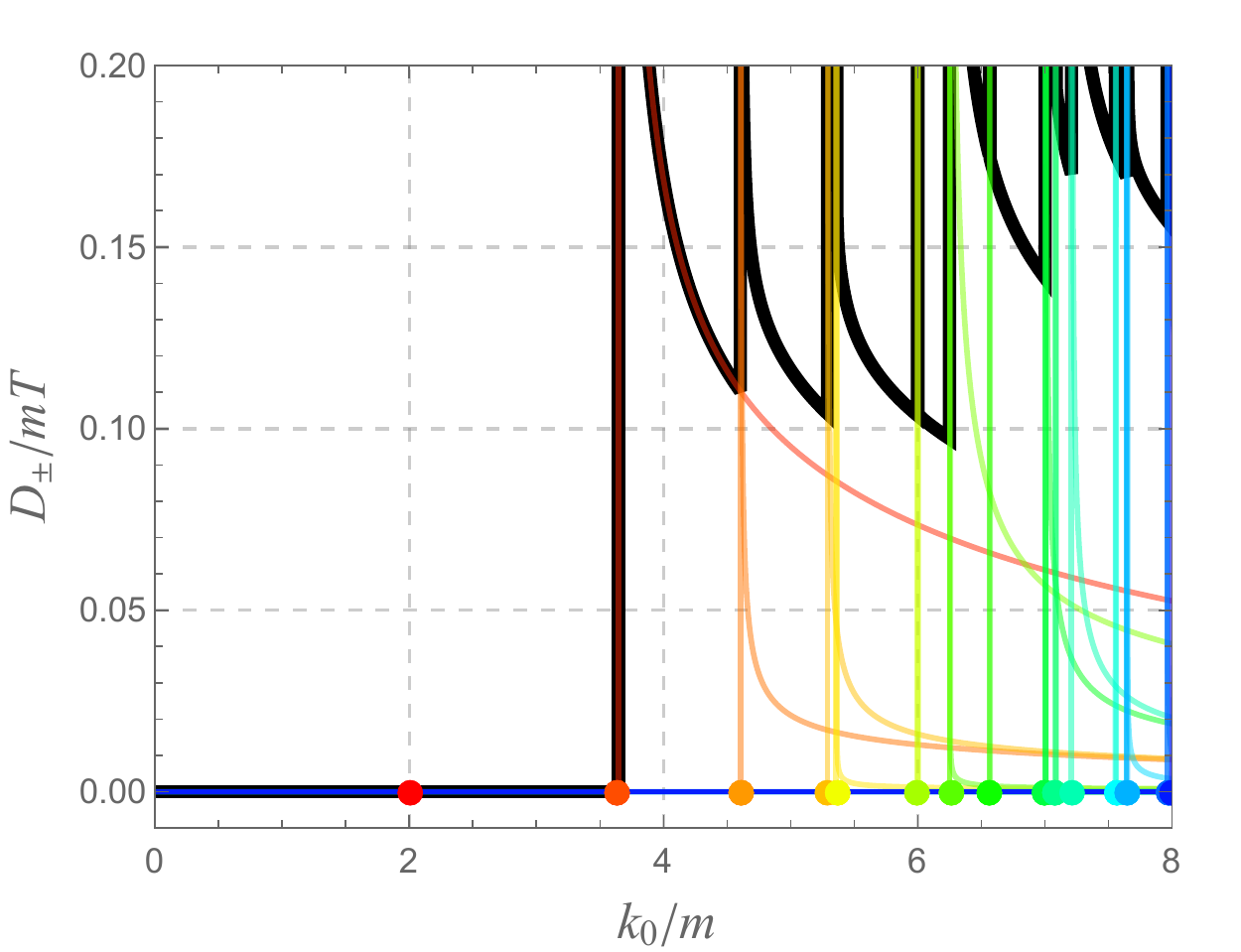} \hspace*{-4mm}
	\includegraphics[width=0.34\hsize]{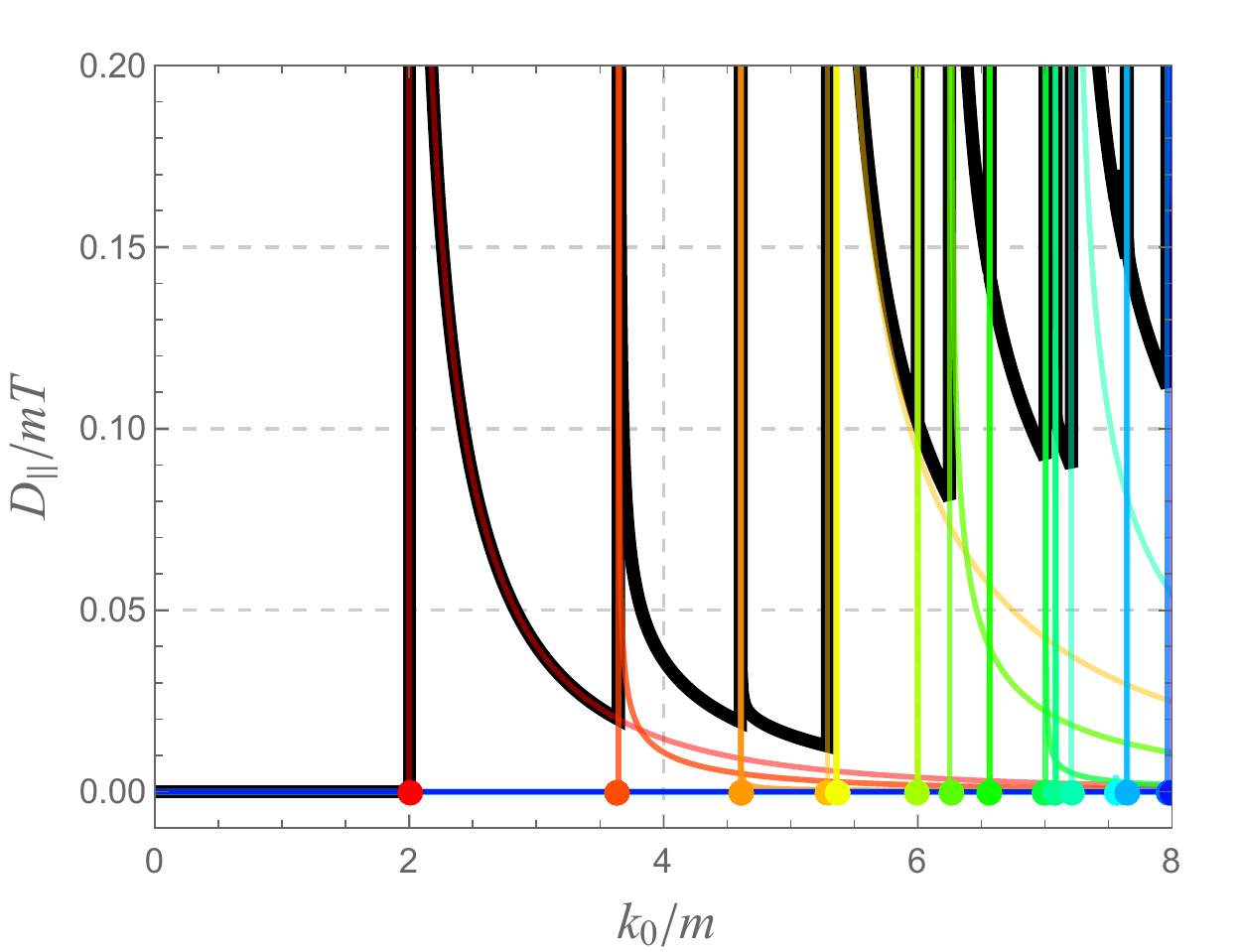}
\\
	\mbox{\small For $  (k_z/m, |qB|/m^2) = (3, 3)$}\\
	\includegraphics[width=0.34\hsize]{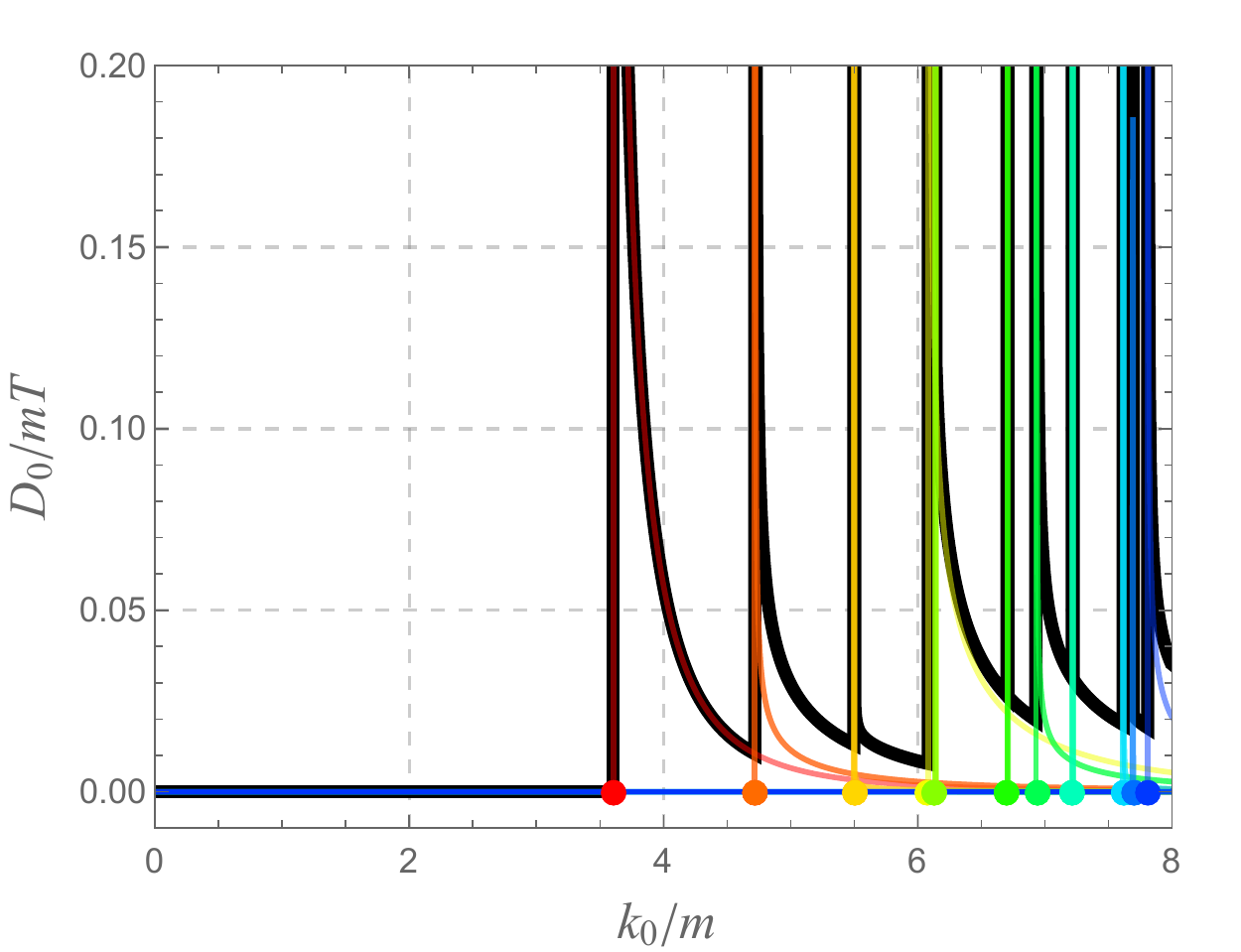} \hspace*{-4mm}
	\includegraphics[width=0.34\hsize]{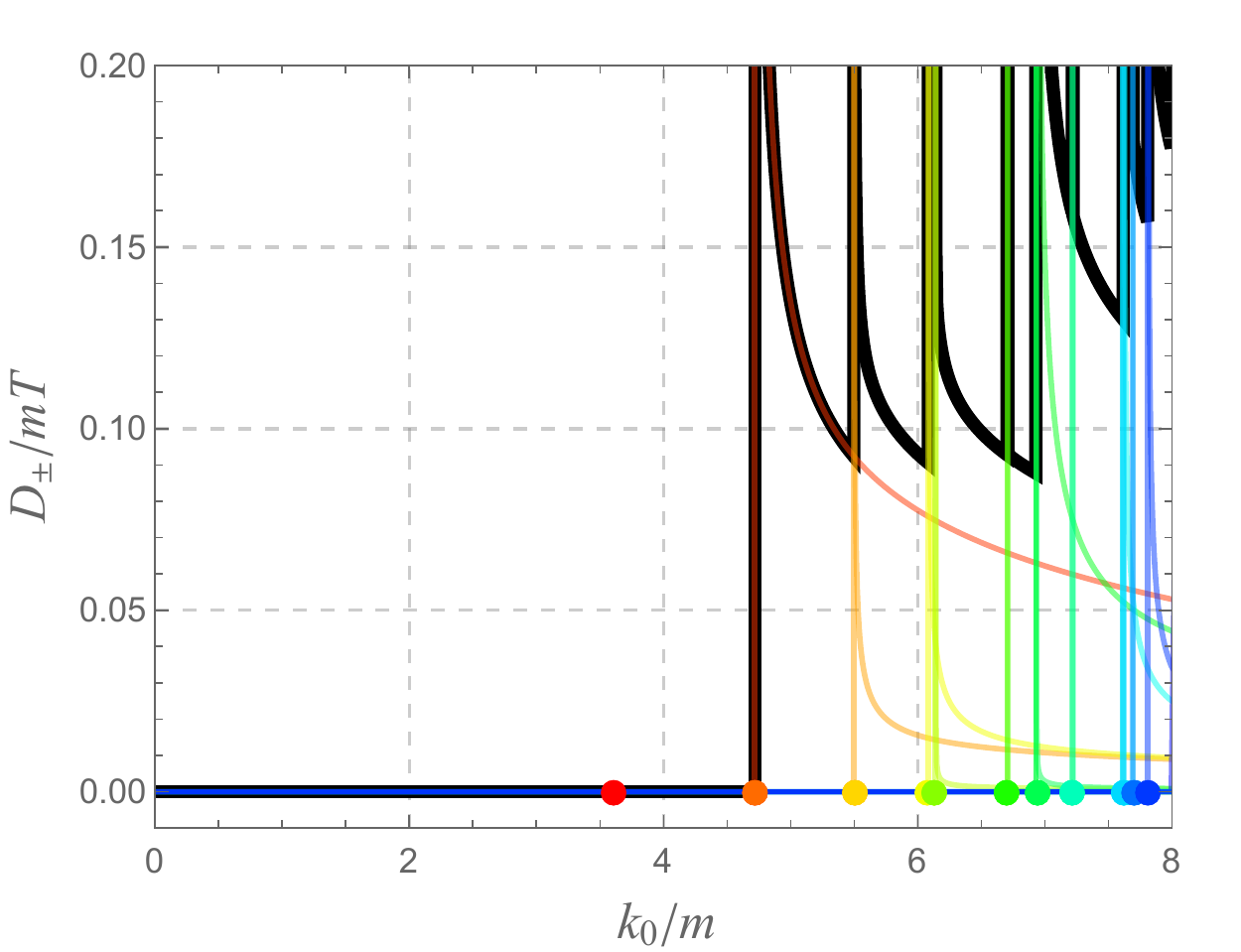} \hspace*{-4mm}
	\includegraphics[width=0.34\hsize]{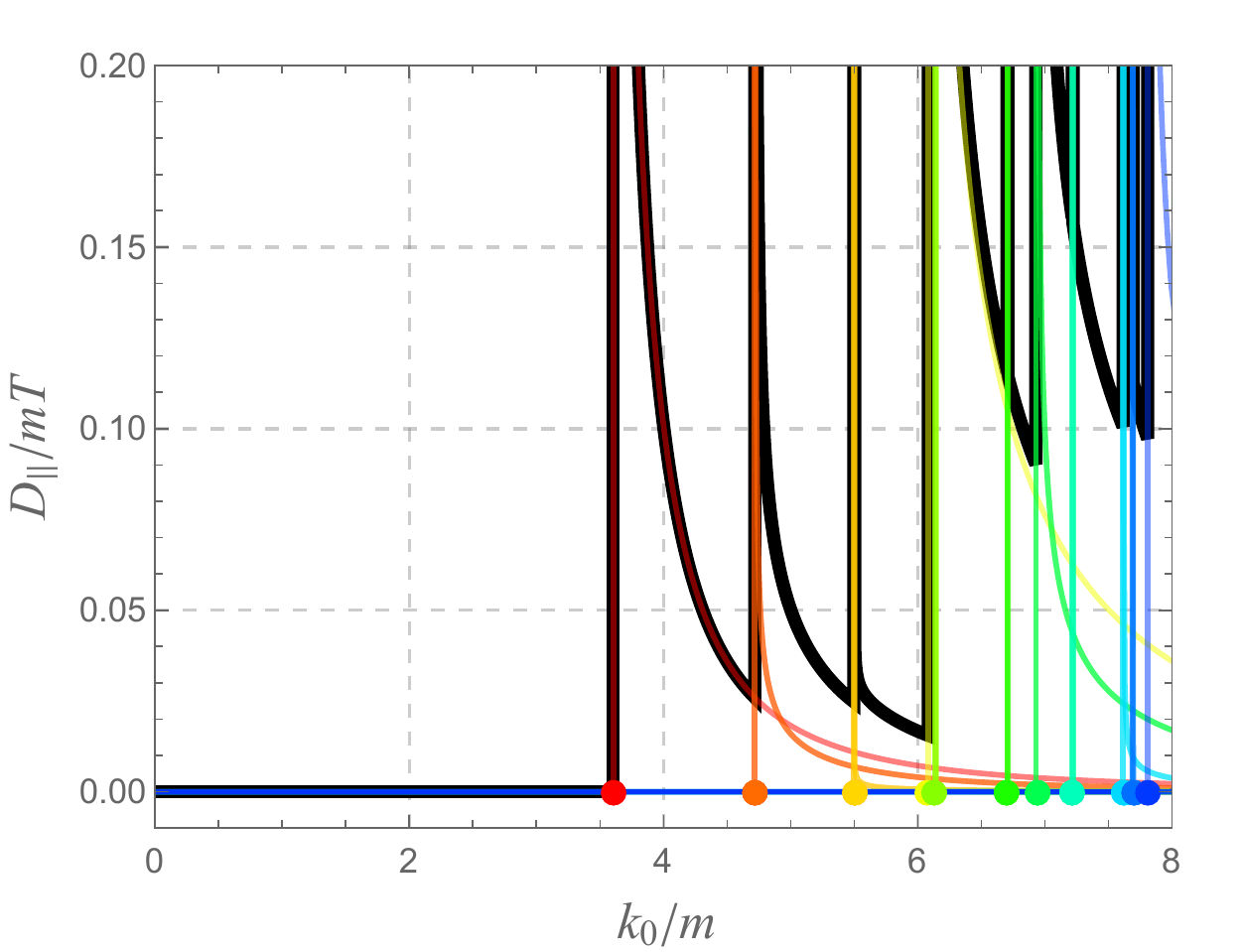}
\end{center}
\caption{Photon-energy dependences of the conversion rates: $ \rate_0$ (left), $ \rate_\pm$ (middle), and $ \rate_\parallel$ (right).  The photon transverse momentum is fixed at $|{\bm k}_\perp|/m=1 $, and the other parameters are at $ (k_z/m, |qB|/m^2) = (0, 1)$ [first row], $(0, 3)$ [second row], and $ (3, 3)$ [third row].  The colored dots on the horizontal axes indicate the threshold energies (\ref{eq:thres}) in ascending order from red, representing the lowest threshold for the lowest Landau level pair $n=n'=0$, to blue for higher Landau level pairs.  The colored lines originating from each dot show the contributions from each Landau level pair $ \rate_\lambda^{n,n'} $.  The black lines show the total contribution summed over the Landau levels $ \rate_\lambda$.  
}
\label{fig:Npm(k0)}
\end{figure}

We show the photon-energy dependence of the conversion rates $ \rate_\lambda $ in Fig.~\ref{fig:Npm(k0)}, with different sets of photon longitudinal momentum $ k_z /m$ and magnetic field strength $|qB|/m^2$.  The photon transverse momentum is fixed at $  |{\bm k}_\perp|/m =1 $, and dependences on $  |{\bm k}_\perp| $ will be discussed in Sec.~\ref{sec4.3.2}.  In each plot, the colored curves show contributions from each pair of Landau levels $ D_\lambda^{n,n'}$ (the lowest Landau level pair $n=n'=0$ is in red, and the color changes to blue as we go to higher Landau level pairs), while the black curve shows the total contribution summed over the Landau levels $ D_\lambda $.  Note that we plotted $D_\pm$ in a single plot because $\rate_{+}^{n,n'} = \rate_{-}^{n', n}$ and $D_+ = D_-$, as we remarked below Eq.~(\ref{eq:N-lambda}).

The most important message of Fig.~\ref{fig:Npm(k0)} is that there are an infinite number of thresholds for the di-lepton production, at which points the conversion rates exhibit resonant behaviors, i.e., the spike structures as a function of $k_0$.  This is essentially the same as the cyclotron resonances in quantum mechanics under a weak magnetic field, and the presence of such thresholds is a direct manifestation of the Landau quantization.  The locations of the thresholds are given by Eq.~(\ref{eq:thres}) and are specified by the colored dots on the horizontal axis in the plots.  One observes that the interval between two adjacent thresholds, that is nothing but the Landau-level spacing, increases as one increases the magnetic field strength, as is evident from the comparison between the first row ($|qB|/m^2=1$) and the second and third rows ($|qB|/m^2=3$) in Fig.~\ref{fig:Npm(k0)}.  Also, comparing the second ($k_z/m=0$) and third ($k_z/m=3$) rows in Fig.~\ref{fig:Npm(k0)}, one notices that the locations of the thresholds shift to higher photon energies when the photon longitudinal momentum is nonvanishing $ k_z \neq 0$.  This is because a nonzero $ k_z $ requires a nonzero di-lepton longitudinal momentum for the momentum conservation, which costs an additional energy for the di-lepton production.

Next, we look more closely into the resonant behaviors at the thresholds and analytically show that the heights of the spikes are divergent if $m\neq 0$.  As the boost-invariant photon energy $ k_\parallel^2 $ approaches each threshold (\ref{eq:thres}) from the above $k_\parallel^2 \to ( m_n + m_{n'})^2 + 0^+$, the on-shell fermion longitudinal momentum (\ref{eq:discpz}) behaves as 
\begin{align}
\label{eq:pz-expansion}
	p_{n,n'}^\pm 
		=   k_z    \frac{ m_n }{ m_n + m_{n'}  } \pm  k_0  \frac{  \sqrt{ m_n m_{n'}}}{ ( m_n + m_{n'})^2 }  \delta k_\parallel + {\mathcal O}(  \delta k_\parallel^2) \, ,
\end{align}
where $\delta k_\parallel^2 \to 0^+$ is the deviation from the threshold such that 
\begin{align}
	\delta k_\parallel \equiv \sqrt{  k_\parallel^2 -  ( m_n + m_{n'})^2 } \, .
\end{align}
Then, the common factor in the denominators of the conversion rates (\ref{eq:N-lambda}) goes to zero as
\begin{align}
\label{eq:denom-expand}
	|p_z\epsilon'_{n'} - (k_z - p_z)  \epsilon_n | 
		= \sqrt{ m_n m_{n'}} \delta k_\parallel + {\mathcal O}( \delta k_\parallel ^3) \to 0^+ \, .
\end{align}
Therefore, the conversion rates diverge $ \sim ({\rm numerator})/\delta k_\parallel \to \infty $ as the photon energy approaches every threshold, unless the numerator is ${\mathcal O}(\delta k_\parallel)$.  The numerator is always ${\mathcal O}(1)$ for $m\neq 0$ but can be ${\mathcal O}(\delta k_\parallel)$ for $m=0$ because of the linear dispersion of the lowest Landau level, as we will discuss in Sec.~\ref{sec434}.  This divergent behavior is seen as the spike structures in Fig.~\ref{fig:Npm(k0)}, and its inverse square-root dependence $ (\sim 1/\delta k_\parallel) $ is a typical threshold behavior in the (1+1) dimensions.

We remark that the di-lepton production with a lowest Landau level fermion and/or anti-fermion (i.e., $n$ or $n'=0$) is prohibited for particular photon polarization modes, and the corresponding conversion rates are vanishing even above the threshold\footnote{Note that we here concentrate on the massive case ($m \neq 0$). There are further prohibitions in the massless limit $m\to 0$ due to a chirality reason, which we will show in Sec.~\ref{sec434}. }.  Namely, circularly polarized photons with $ \lambda=+ $ ($ \lambda=- $) do not couple to $n'=0$ anti-fermion ($n=0$ fermion), and $D_+^{n,0}=D_-^{0,n'}=0$ for any $n,n'$.  Physically, this is because those photon polarization modes carry nonzero spin components $ \pm1 $ along the magnetic field, whereas the di-lepton states with $n$ or $n'=0$ carry either spin zero or $\mp1$ [recall discussions below Eq.~(\ref{eq:pol})].  On the other hand, photons with $ \lambda=0,\parallel $ can couple to di-leptons with $n$ or $n'=0$, since those photons can have the same spin state as that of di-leptons.  Yet the case of $D_0^{0,0}$ is somewhat exceptional in that it is vanishing at $k_z=0$.  Indeed, when $ n=n'=0 $, a factor in the numerator of $ D_0^{0,0} $ can be evaluated as $ \epsilon_n \epsilon'_{n' } + p_z p'_z - m^2 = k_z^2/2 +{\mathcal O} (\delta k_\parallel^2)$ and the other numerator factors are ${\mathcal O}(1)$.  Taking into account the denominator factor (\ref{eq:denom-expand}), we find $D_0^{0,0} \propto k_z^2/( \delta k_\parallel\sqrt{m_n m_{n'}} )$, which is vanishing at $k_z= 0$.

\subsubsection{Transverse-momentum dependence} \label{sec4.3.2}

\begin{figure}
	\begin{center} 
		\includegraphics[width=0.49\hsize]{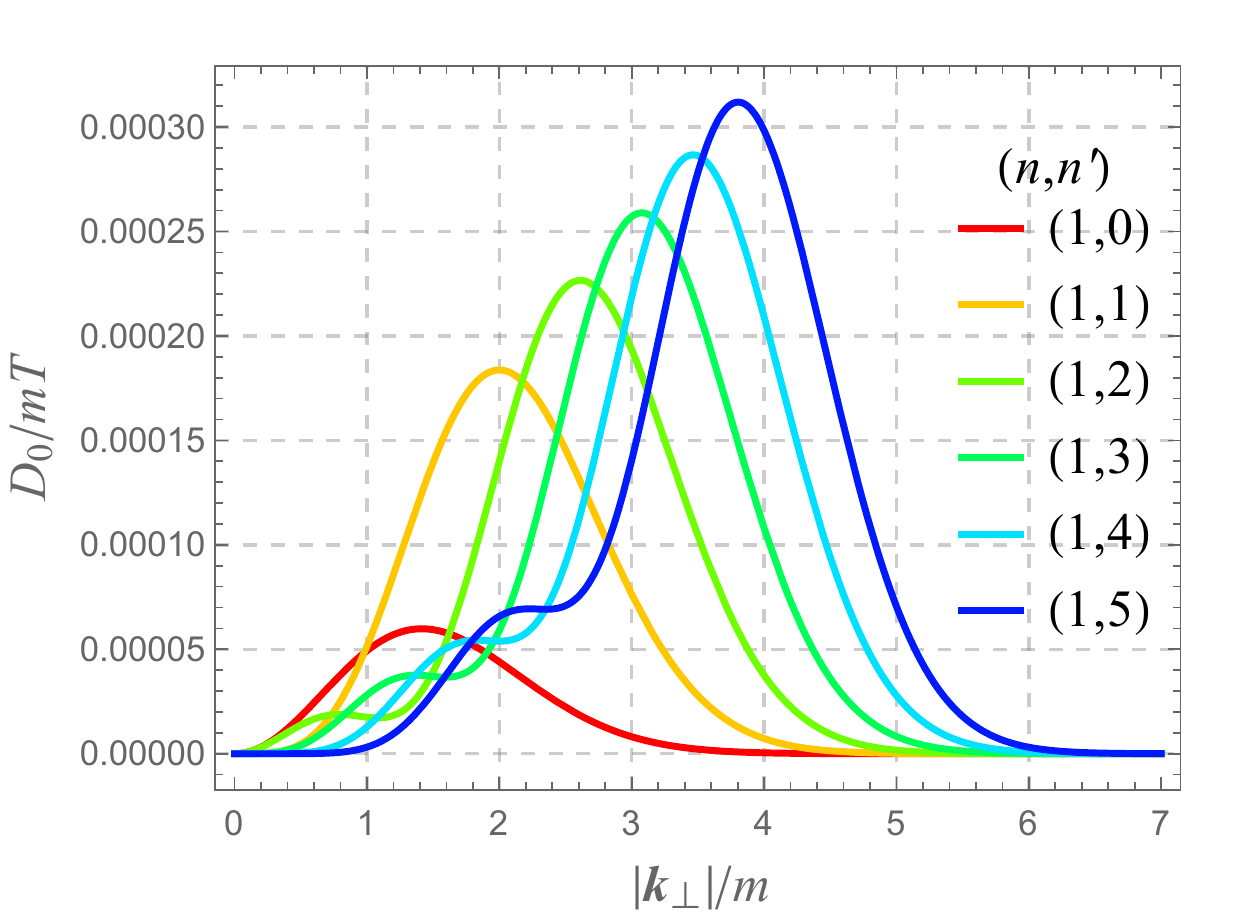}
		\includegraphics[width=0.49\hsize]{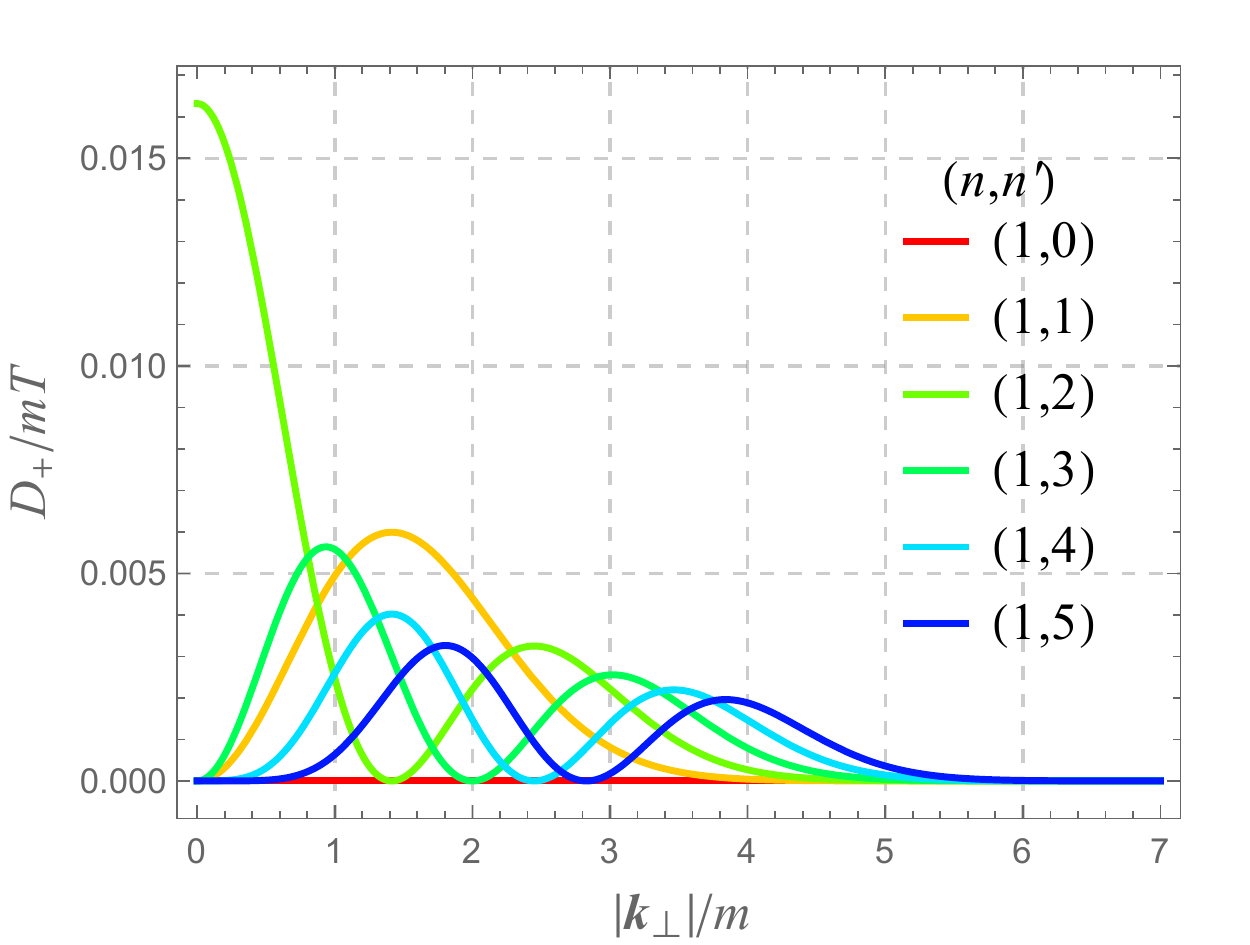}
		\includegraphics[width=0.49\hsize]{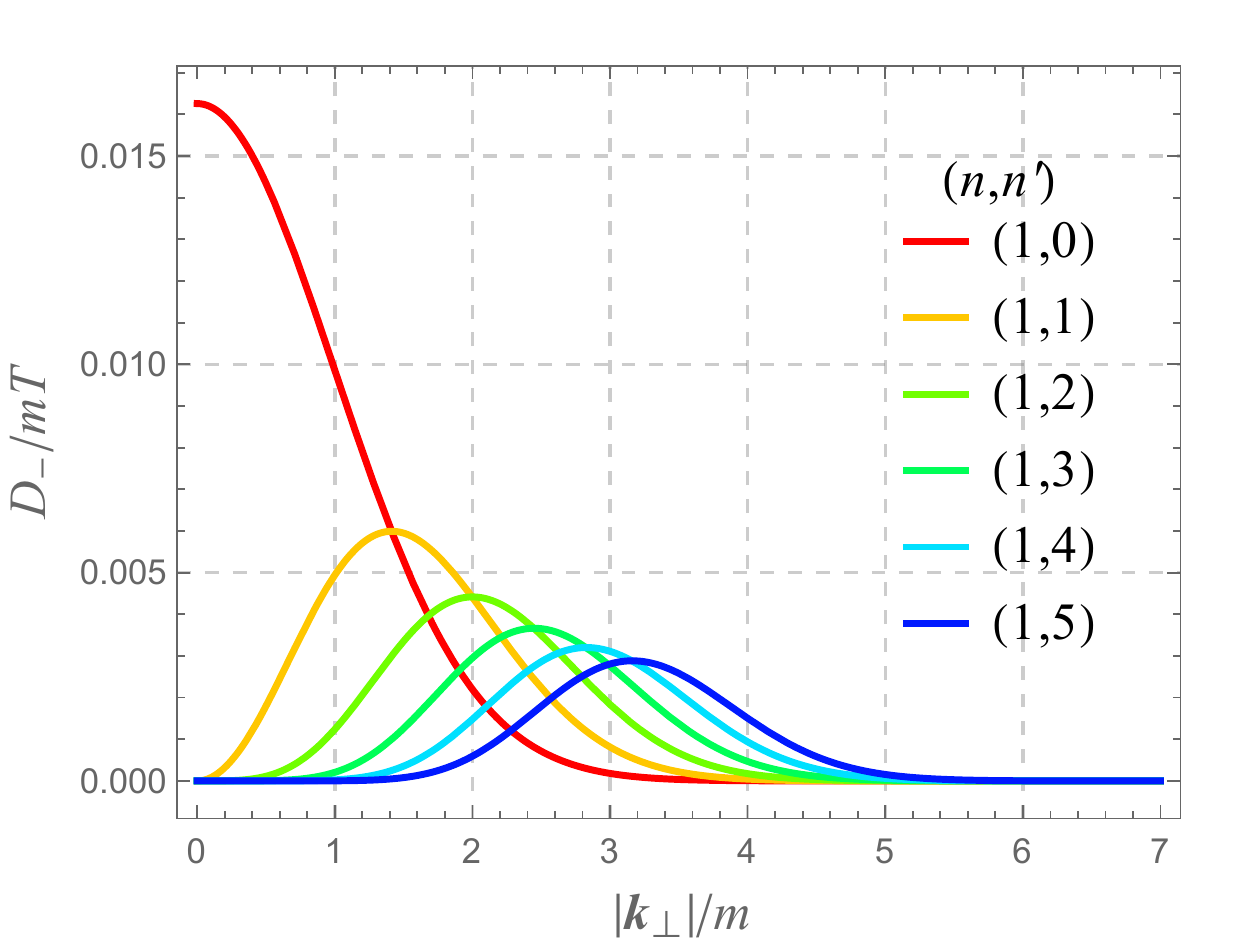}
		\includegraphics[width=0.49\hsize]{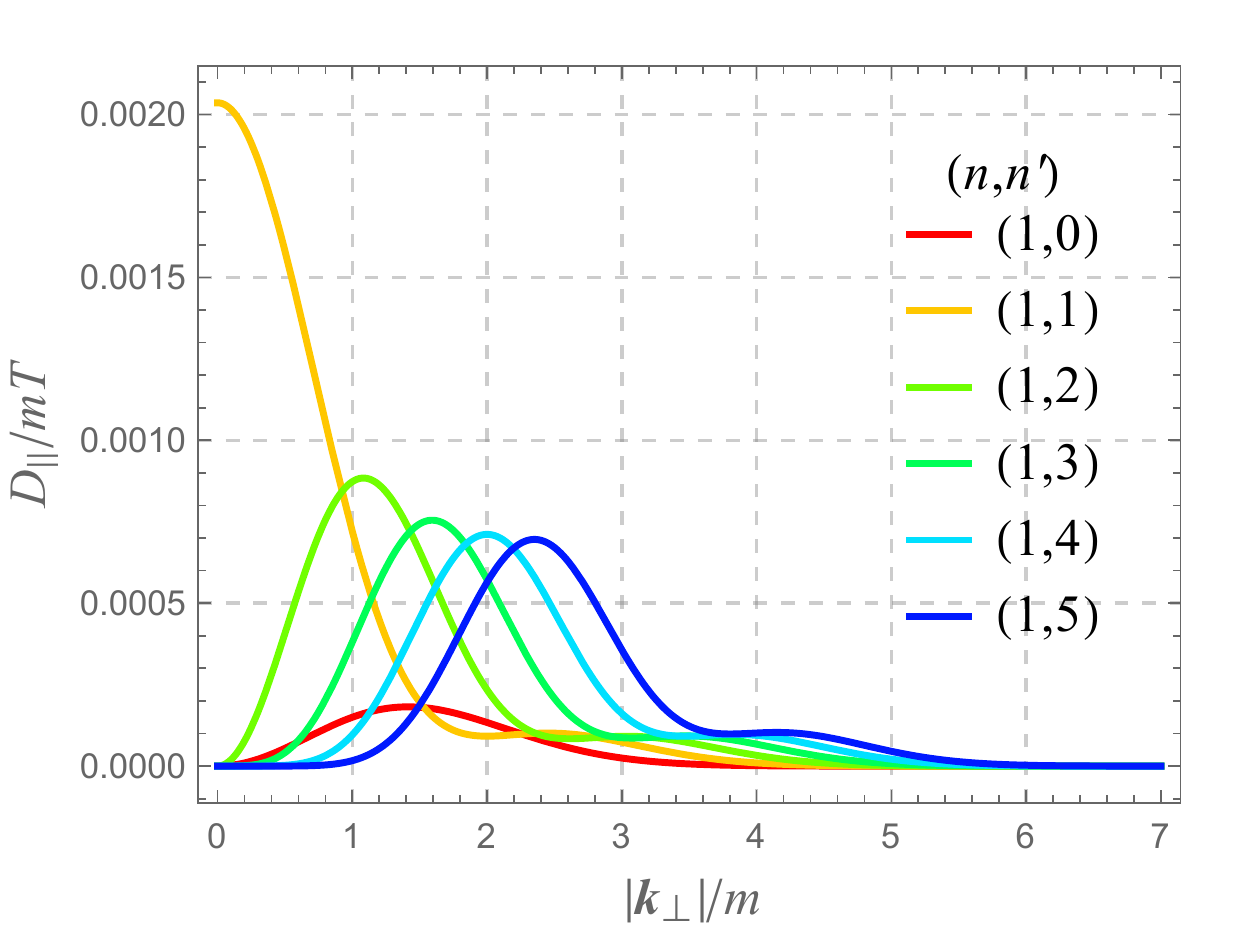}
	\end{center}
\caption{Photon transverse momentum $|{\bm k}_\perp|$-dependences of the conversion rates $\rate_\lambda^{n,n'}$.  The Landau level of the fermion is fixed at $ n=1 $, while that of the anti-fermion runs on $n'=0,1,2,3,4,5$.  Other parameters are fixed at $|qB|/m^2 =1$, $k_0/m=10$, and $k_z/m=0$.  
}
\label{fig:Npm}
\end{figure}

We provide quantitative discussions about the $|{\bm k}_\perp|$-dependence of the conversion rates (\ref{eq:N-lambda}), which is determined by the scalar form factor $|{\Gam}_{n,n'}|^2$ (i.e., the overlap between the produced fermion and anti-fermion transverse wave functions); see Sec.~\ref{sec:3.3} for analytical discussions.  Figure~\ref{fig:Npm} demonstrates that the conversion rate for each pair of Landau levels $ \rate_{\lambda}^{n,n'} $ is suppressed for large values of $ |{\bm k}_\perp| \to \infty$ and exhibits the peaked structure, which is the reminiscent of the transverse momentum conservation modified by the magnetic field.  The peak location is determined by the Landau levels appearing as indices of the Laguerre polynomials in $|{\Gam}_{n,n'}|^2$.  We took $n=1 $ and $n'=0, 1,2,3,4,5 $ just for a demonstration.  Among those cases, one finds that $ \rate_+^{1,0}=0 $ identically due to the reason for spin configurations as we have remarked in Sec.~\ref{sec4.3.1}.  One also finds that the conversion rates vanish in the limit of $|{\bm k}_\perp| \to 0$ except for $D_{+}^{1,2}$, $D_{-}^{1,0}$, and $D_\parallel^{1,1}$.  This is a consequence of the property discussed around Eq.~(\ref{eq:G(k->0)}); in general, only $D_0^{n,n}$, $D_+^{n,n+1}$, $D_-^{n+1,n}$, and $D_\parallel^{n,n}$ can be nonvanishing in the limit of $|{\bm k}_\perp| \to 0$.  Note that the indices of the form factor $  {\Gam}_{n,n'}$ appearing in the conversion rates (\ref{eq:N-lambda}) are not necessarily $(n,n')$ but are shifted by some terms, because the di-lepton spin configurations are different depending on the photon polarizations.  When one further takes the limit of $k_z\to0$, $D_0^{n,n}$, and thus $ D_0^{1,1} $ in Fig.~\ref{fig:Npm}, vanishes because $D_0^{n,n} \propto \epsilon_n \epsilon'_{n} + p_z p'_z - m^2 -2n|qB| = {\mathcal O}(k_z^2)$, while the other three stay finite.

\begin{figure}
\begin{center}
	\mbox{\small For $|qB|/m^2 = 1$}\\
	\includegraphics[width=0.34\hsize]{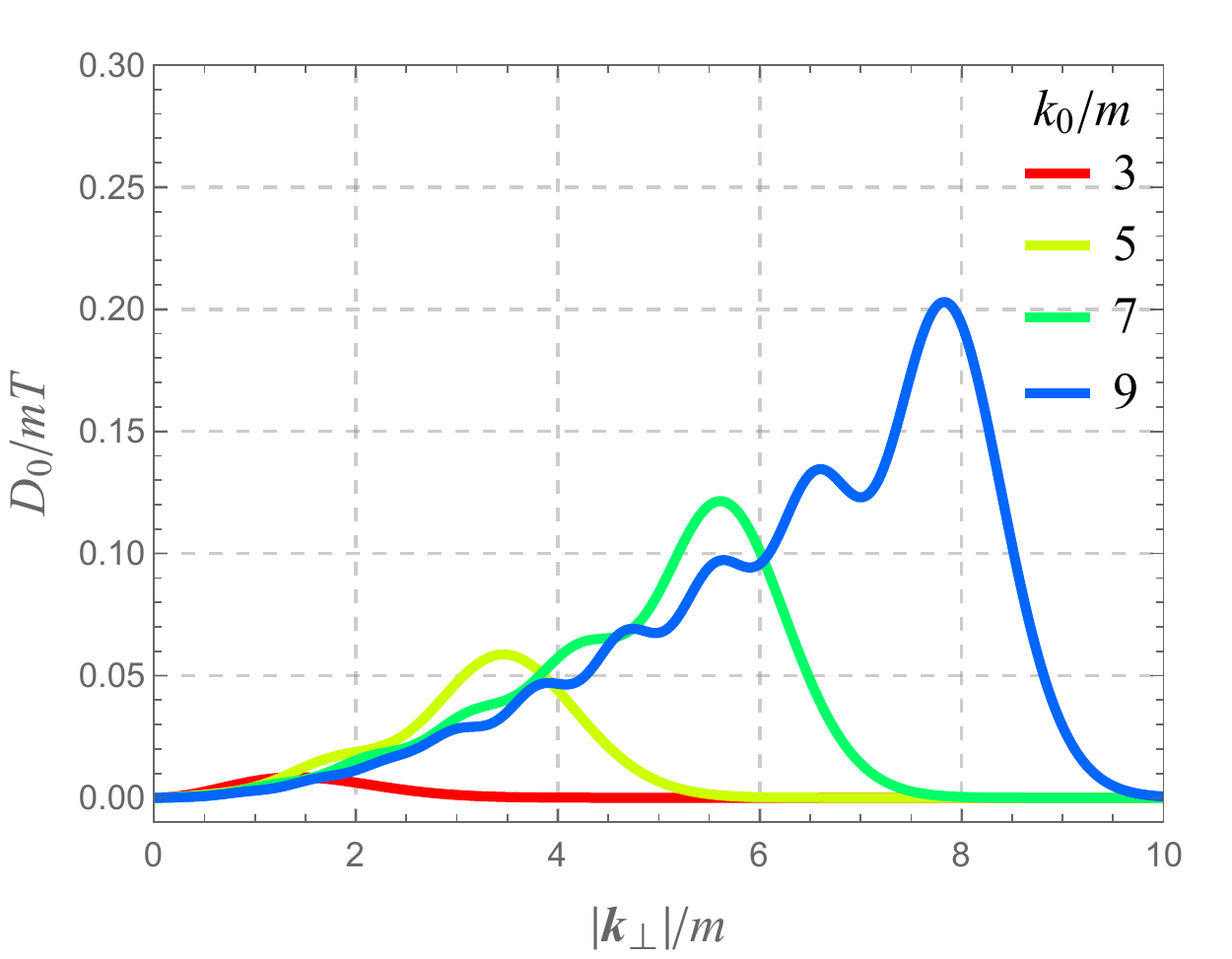} \hspace*{-4mm}
	\includegraphics[width=0.34\hsize]{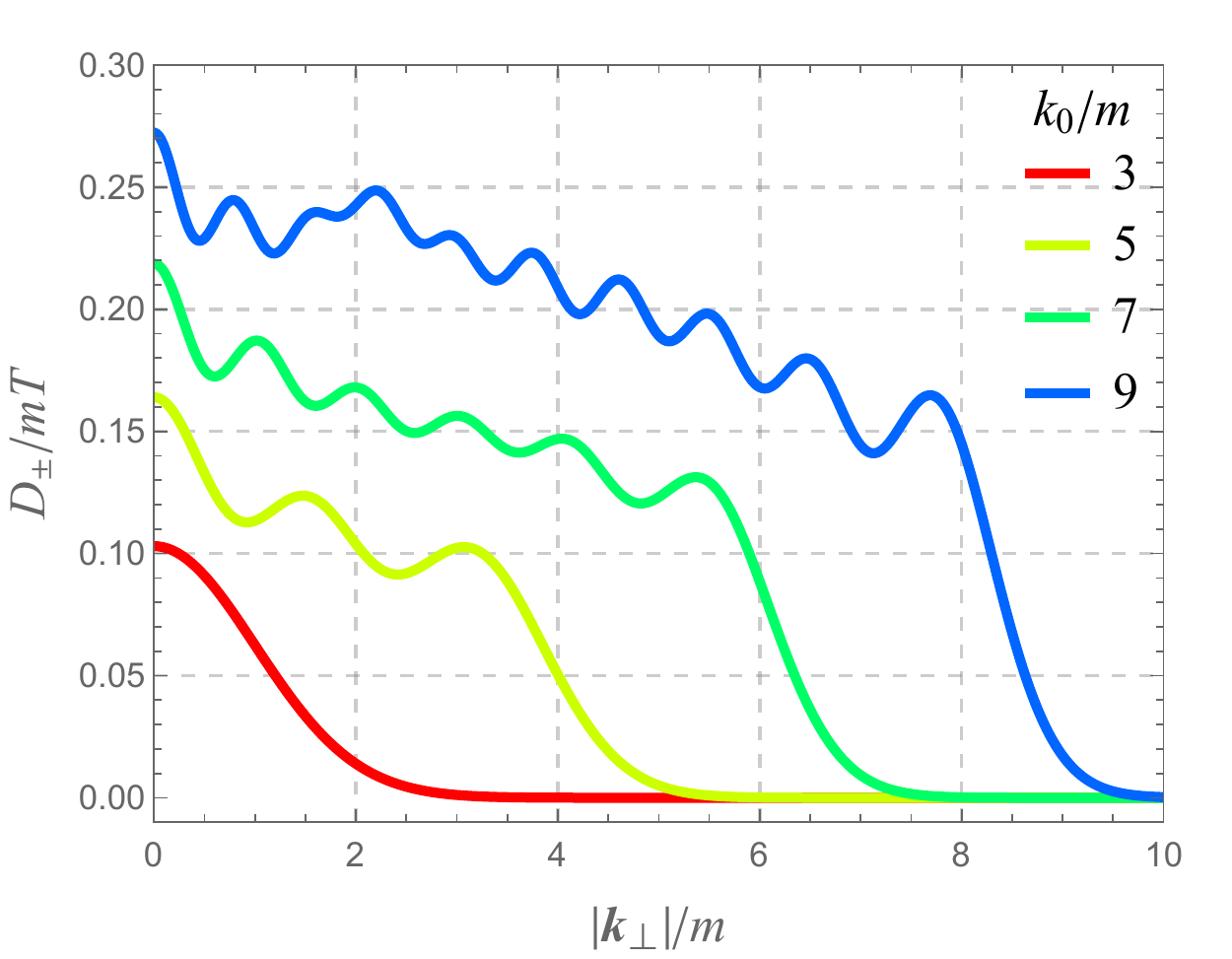} \hspace*{-4mm}
	\includegraphics[width=0.34\hsize]{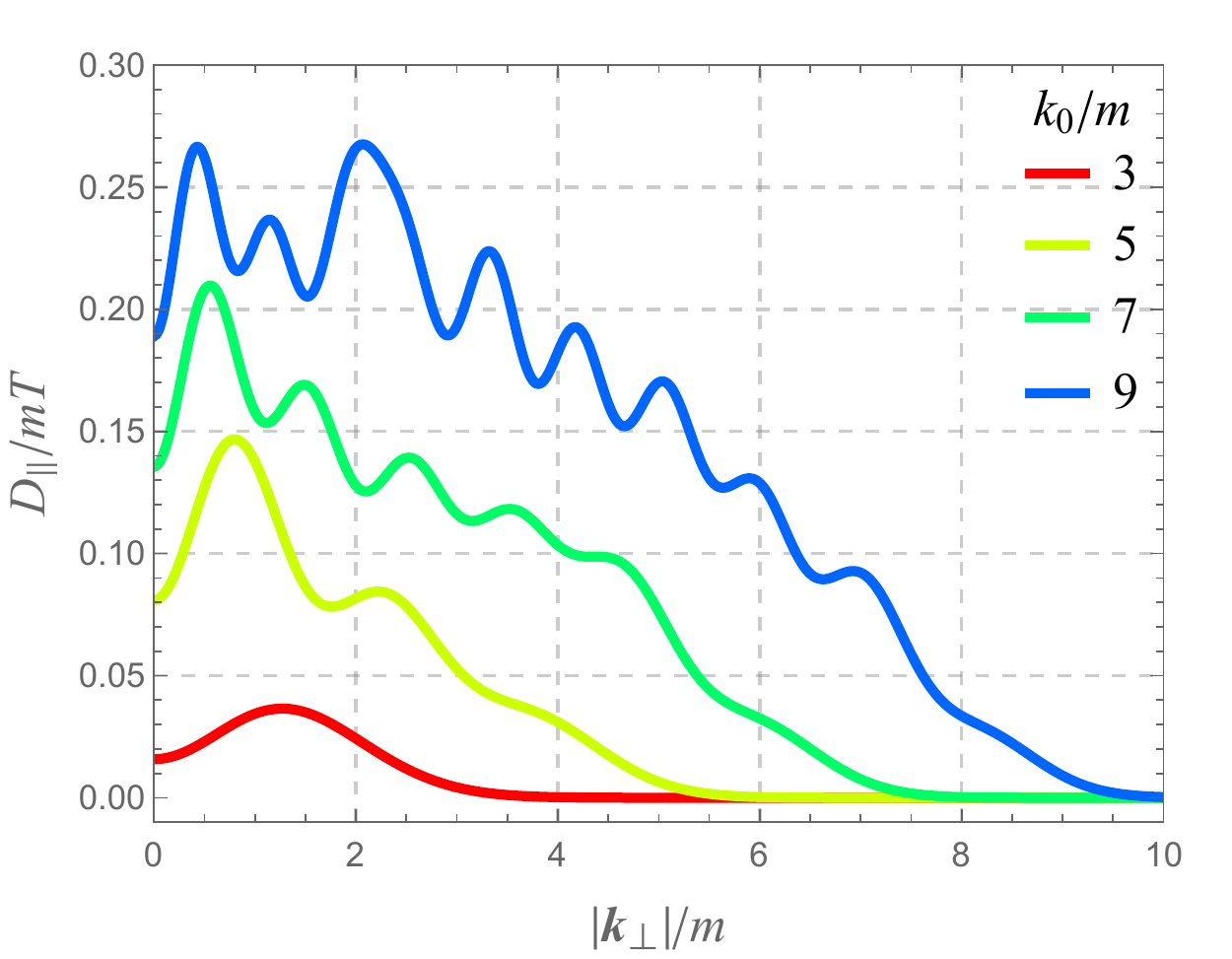} 
\\
	\mbox{\small For $|qB|/m^2 = 3$}\\
	\includegraphics[width=0.34\hsize]{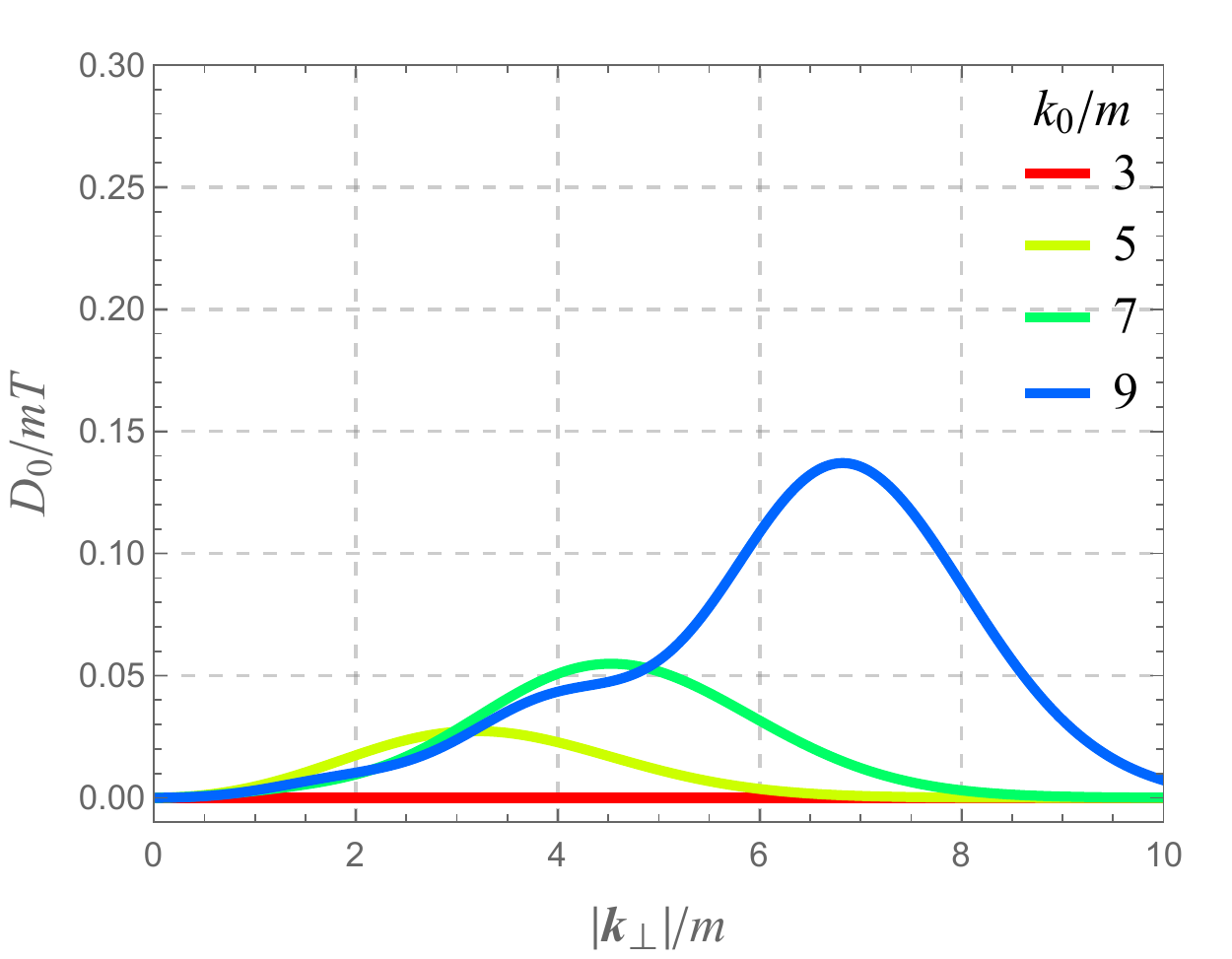} \hspace*{-4mm}
	\includegraphics[width=0.34\hsize]{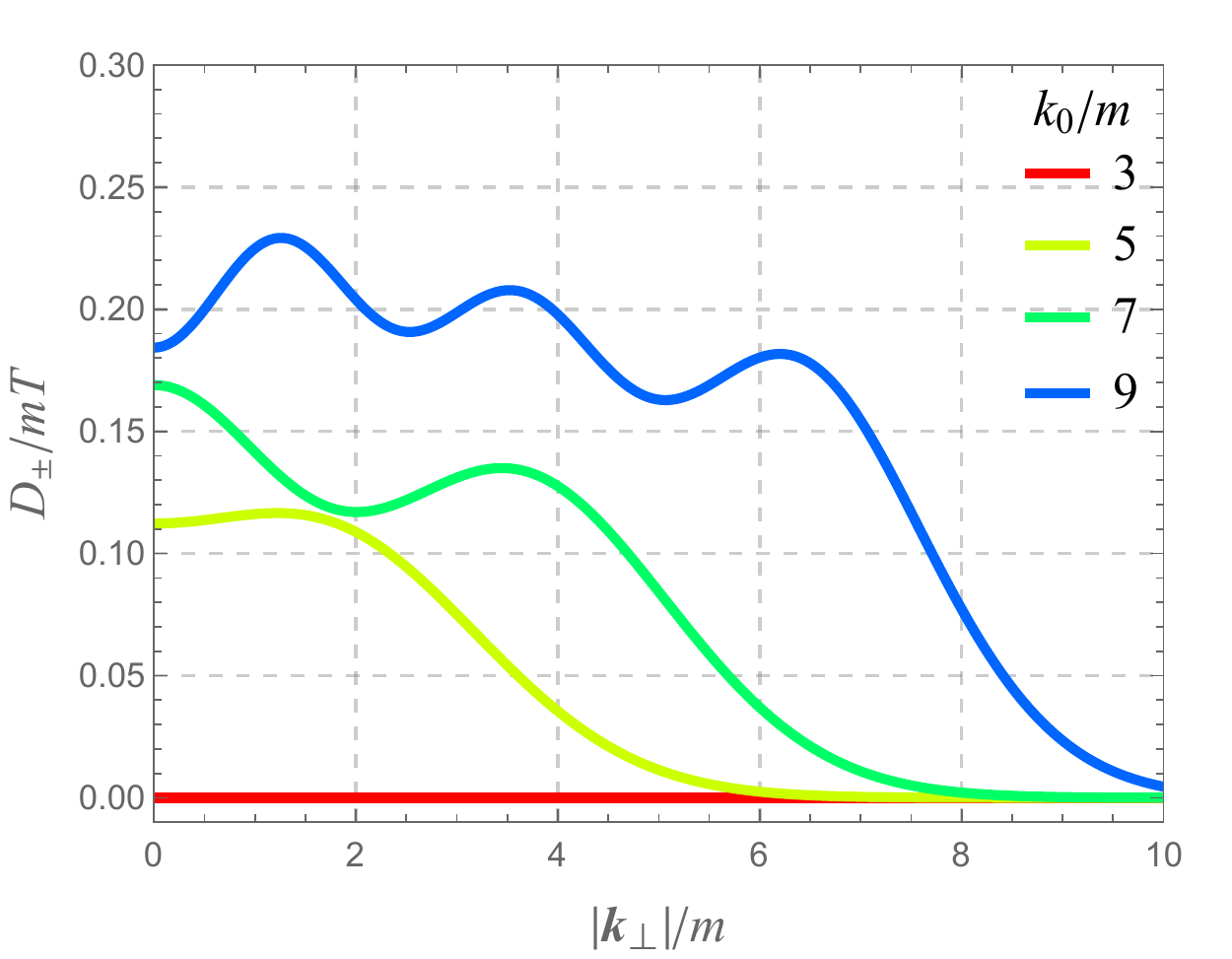} \hspace*{-4mm}
	\includegraphics[width=0.34\hsize]{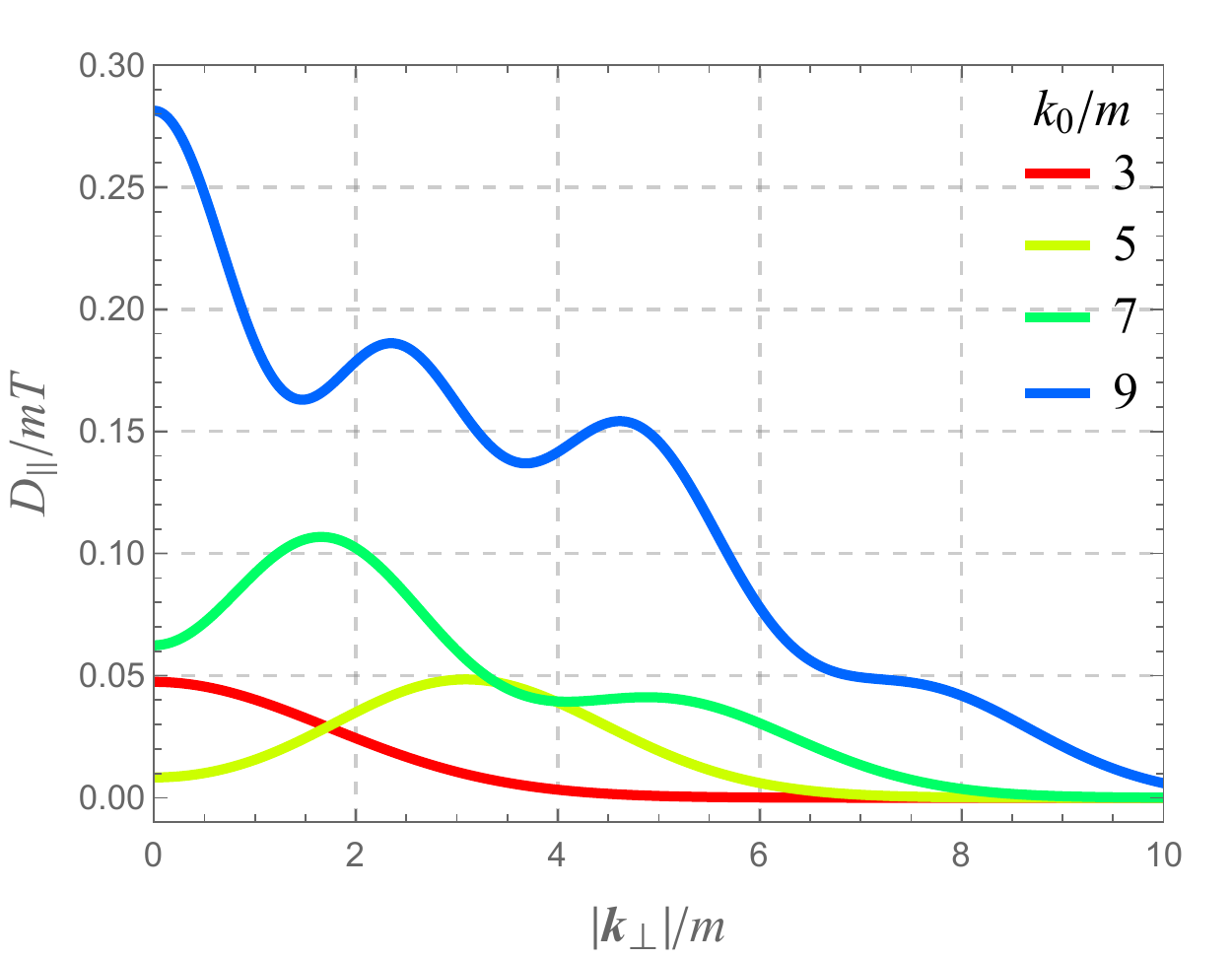}
\end{center}
\caption{The photon transverse momentum $|k_\perp|$-dependences of the conversion rates summed over the Landau levels $ \rate_\lambda$ at $k_z/m=0$.  The lines with different colors distinguish the photon energy $k_0/m=3,5,7$, and $9$.  The magnetic field strengths are taken as $|qB|/m^2 =1  $ (top) and $=3$ (bottom).  
}
\label{fig:Npm-summed}
\end{figure}

The Landau-level summed conversion rates $\rate_{\lambda}$, shown in Fig.~\ref{fig:Npm-summed}, exhibit an oscillating behavior, resulting from superposition of the peaks in each Landau-level contribution $\rate_{\lambda}^{n,n'}$.  The structure of the oscillation changes as the parameters vary, i.e., the oscillation becomes (i) finer for larger photon energies $k_0$ and (ii) more moderate for stronger magnetic fields.  Those changes are attributed to the number of Landau levels that can be excited with a given photon energy.  Namely, the larger photon energy can excite the more higher Landau levels, so that the oscillation acquires a finer structure with contributions of higher modes.  Also, the number of contributing Landau levels decreases as we increase $|qB|/m^2 $, with which the Landau-level spacing increases.  In particular, only the lowest-lying pair of the Landau levels, $(n,n')=(1,0)$ or $(0,1)$, can contribute when the photon energy is small and the magnetic field is strong (e.g., the red line in the bottom right panel for $k_0/m =3$ and $|qB|/m^2=3  $).  Note that the conversion rates $\rate_{\lambda}$ fall off in a large $ |{\bm k}_\perp|/m $, and they fall off slower for larger $k_0$.  This is because larger $k_0$ can excite higher Landau levels having large $|\Delta n|$, which are favorably produced with a large $|{\bm k}_\perp|$ because of the reminiscent of the transverse momentum conservation.

\subsubsection{Magnetic-field dependence}

\label{sec:B-dep}

\begin{figure}
\begin{center}
	\includegraphics[width=0.34\hsize]{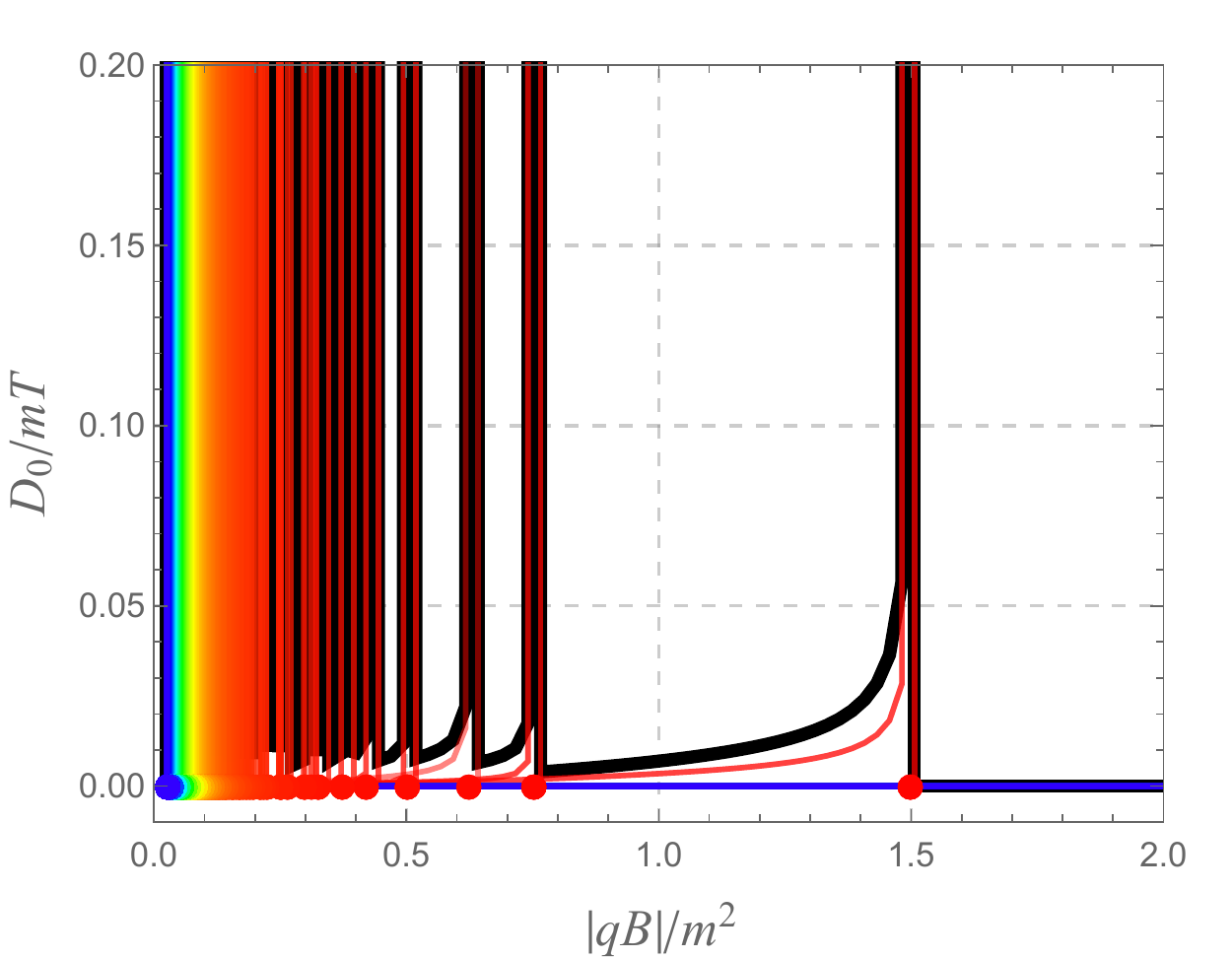} \hspace*{-4mm}
	\includegraphics[width=0.34\hsize]{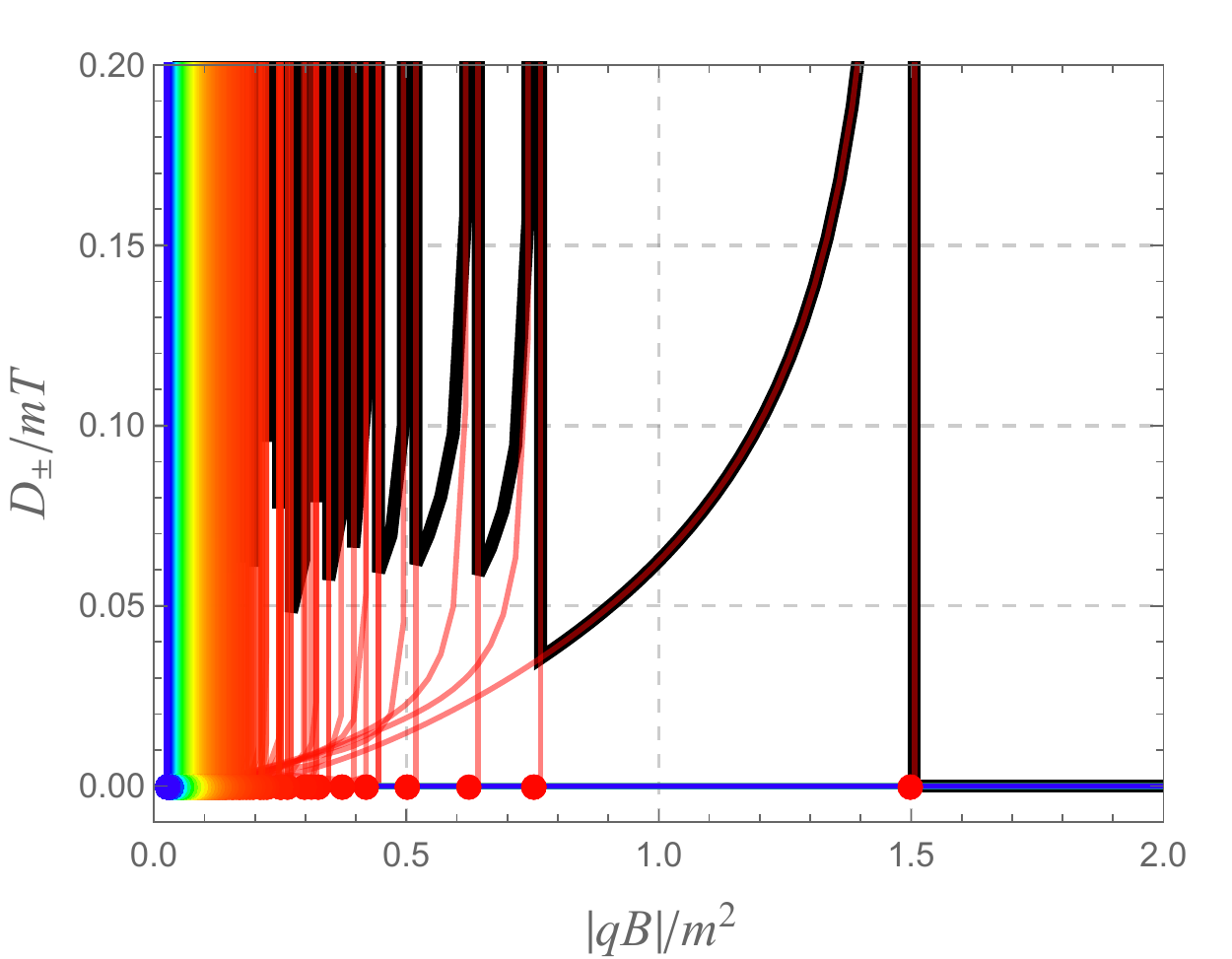} \hspace*{-4mm}
	\includegraphics[width=0.34\hsize]{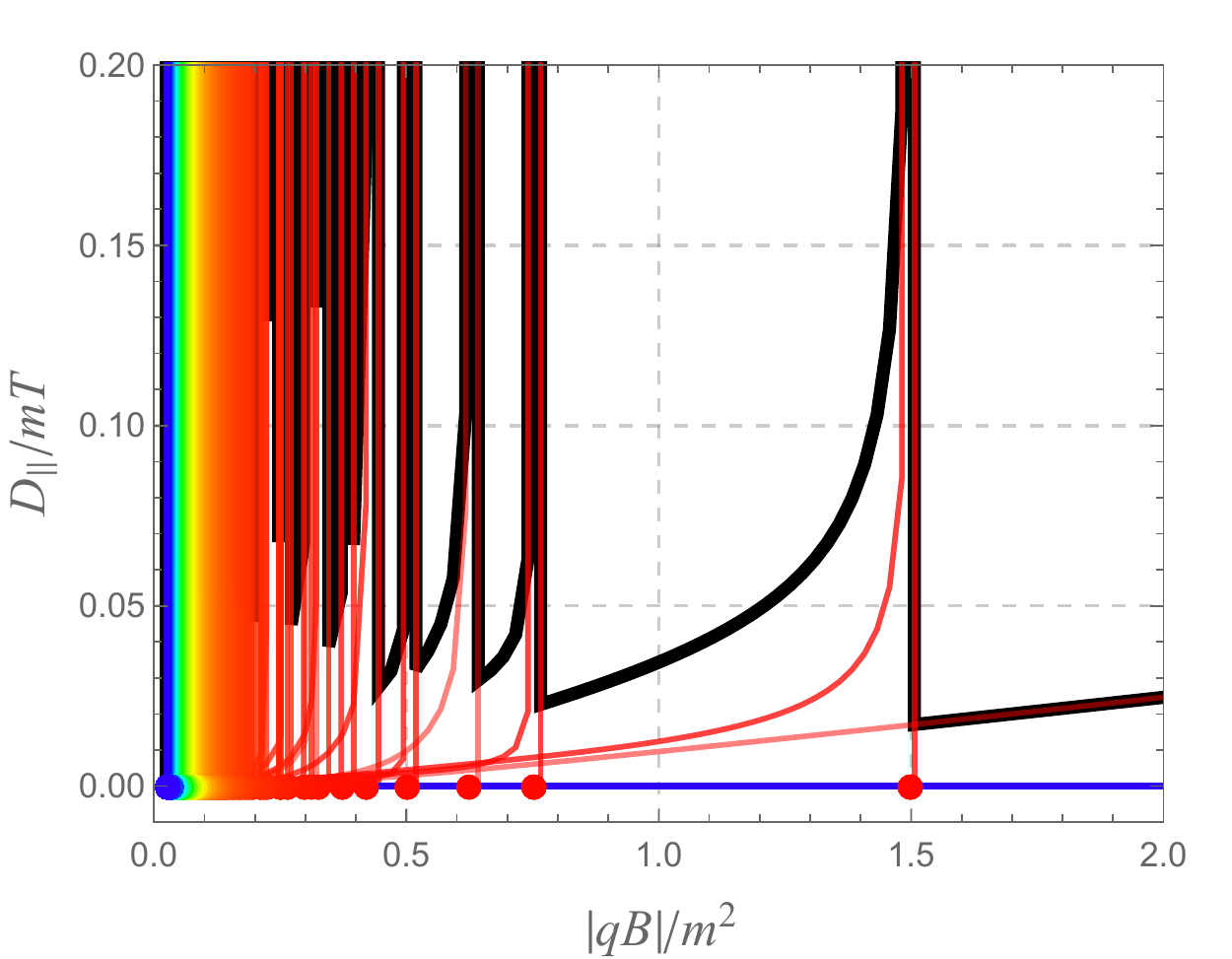}
\end{center}
\caption{The magnetic-field dependence of the conversion rates $ \rate_0$ (left), $ \rate_\pm$ (middle), and $ \rate_\parallel$ (right).  The parameters are fixed as $k_0/m=3$, $|{\bm k}_\perp|/m=1 $ and $k_z/m=0$.  The colored dots on the horizontal axes indicate the upper limits for the magnetic field strength $|qB|_{\rm max}$ (\ref{eq:upper}) in descending order from red (which is always for the lowest Landau level pair $n=n'=0$) to blue.  The colored lines originating from each dot show the contributions from each Landau level pair $ \rate_\lambda^{n,n'} $, and the black lines the total value summed over the Landau levels $ \rate_\lambda$. 
}
\label{fig:8}
\end{figure}

We discuss dependences on the magnetic field strength.  As shown in Fig.~\ref{fig:8}, the conversion rates $D_\lambda$ have spike structures with respect to the magnetic field strength $|qB|$ as well and there exists an upper limit $|qB|_{\rm max}$ for each pair of Landau levels above which the production is prohibited.  Those behaviors are determined by the threshold condition (\ref{eq:thres}), which tells us that fewer Landau levels can contribute to the production as the magnetic field strength $|qB|$, and accordingly the Landau-level spacing, is increased.  Namely, when $|qB|$ is increased from a certain value with a fixed photon energy $k_0$ (or equivalently $k_\parallel^2$), peaks appear when every pair of higher Landau levels stop contributing to the di-lepton production at $|qB|=|qB|_{\rm max}$.  In the end (i.e, in the strong field limit), only the lowest-lying pair with $n=n'=0$ can satisfy the threshold condition (\ref{eq:thres}), which is independent of $|qB|$ for this pair and is satisfied as long as $k_\parallel^2 \geq 4m$.  In other words, there is no upper limit $|qB|_{\rm max}$ for $n=n'=0$.  For the other pairs of Landau levels, one can find the upper limit $|qB|_{\rm max}$ by solving the threshold condition~(\ref{eq:thres}) in terms of $|qB|$ as 
\begin{align}
	|qB|_{\rm max}
		= \left\{ \begin{array}{ll}  
				\displaystyle \frac{ k_\parallel^2-4m^2}{8n} & \ \ {\rm for}\ n=n'\neq 0 \\[10pt]
				\displaystyle \frac{(n+n') k_\parallel^2 - 2\sqrt{(n-n')^2m^2 k_\parallel^2 + nn' k_\parallel^4 }}{2(n-n')^2} &\ \  {\rm for}\ n\neq n' 
			\end{array} \right.\, , \label{eq:upper}
\end{align}
where $ k_\parallel^4  = (k_\parallel^2)^2 $ and $  k_\parallel^2 > 4m^2$ are understood.  $|qB|_{\rm max}$ is an increasing function of $k_0$ or $k_\parallel^2$, meaning that more energetic photons can excite more energetic di-leptons under stronger $|qB|$.  Note that the positivity of the upper limits $ |qB|_{\rm max}$ (\ref{eq:upper}) is guaranteed as long as $  k_\parallel^2 > 4m^2$.  

The summed conversion rates $D_\lambda$ as well as the contributions from each pair of the Landau levels $D_\lambda^{n,n'}$ increase with $qB$, roughly, linearly.  This is because the phase-space volume in a constant magnetic field is proportional to the magnetic field strength as $\int d^2p_\perp \propto \frac{|qB|}{2\pi} \sum_n$, where the factor $|qB|/2\pi$ is the so-called the Landau degeneracy factor and is a manifestation of the Landau quantization.

\subsubsection{Massless limit} \label{sec434}

\begin{figure}
\begin{center}
	\includegraphics[width=0.34\hsize]{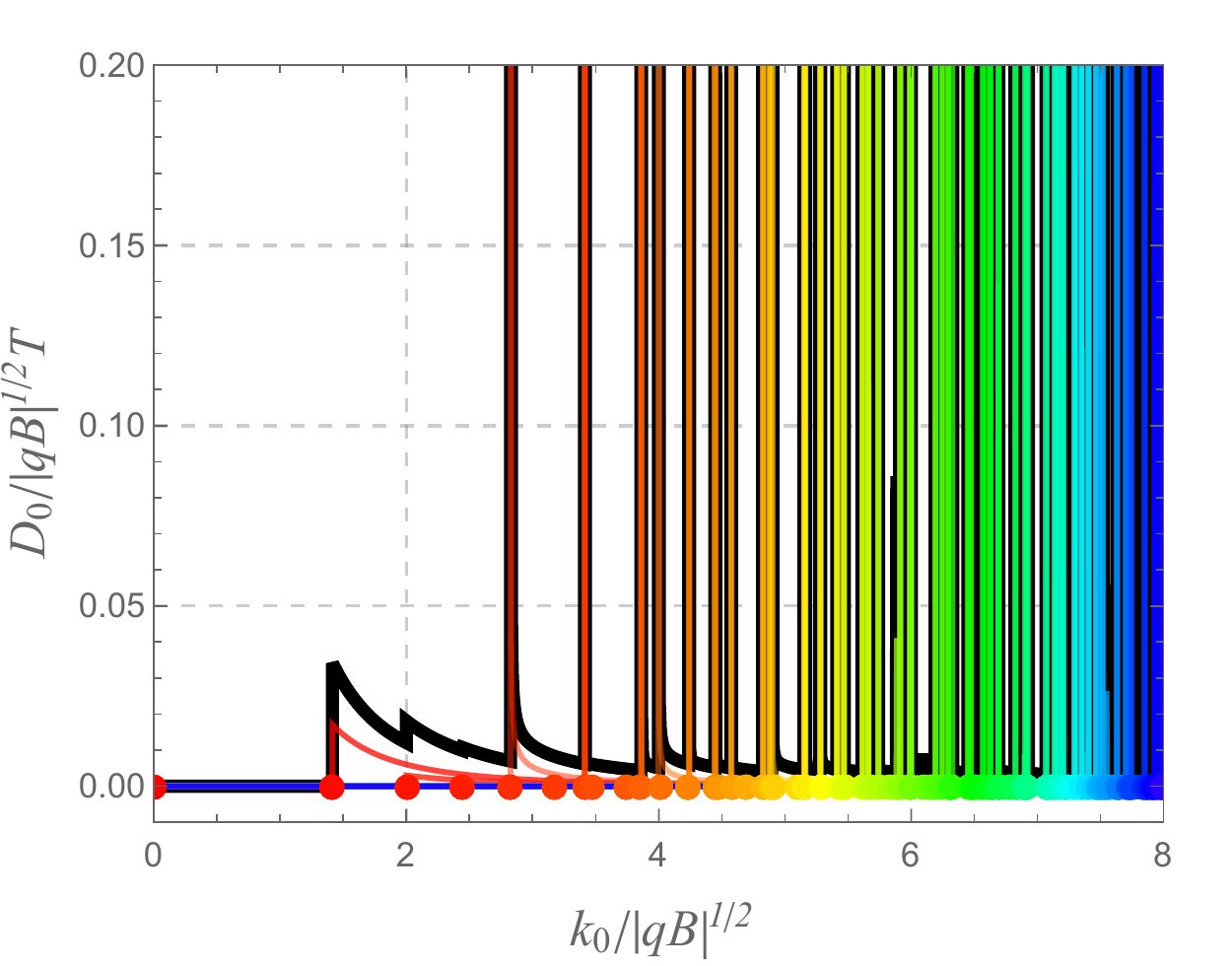} \hspace*{-4mm}
	\includegraphics[width=0.34\hsize]{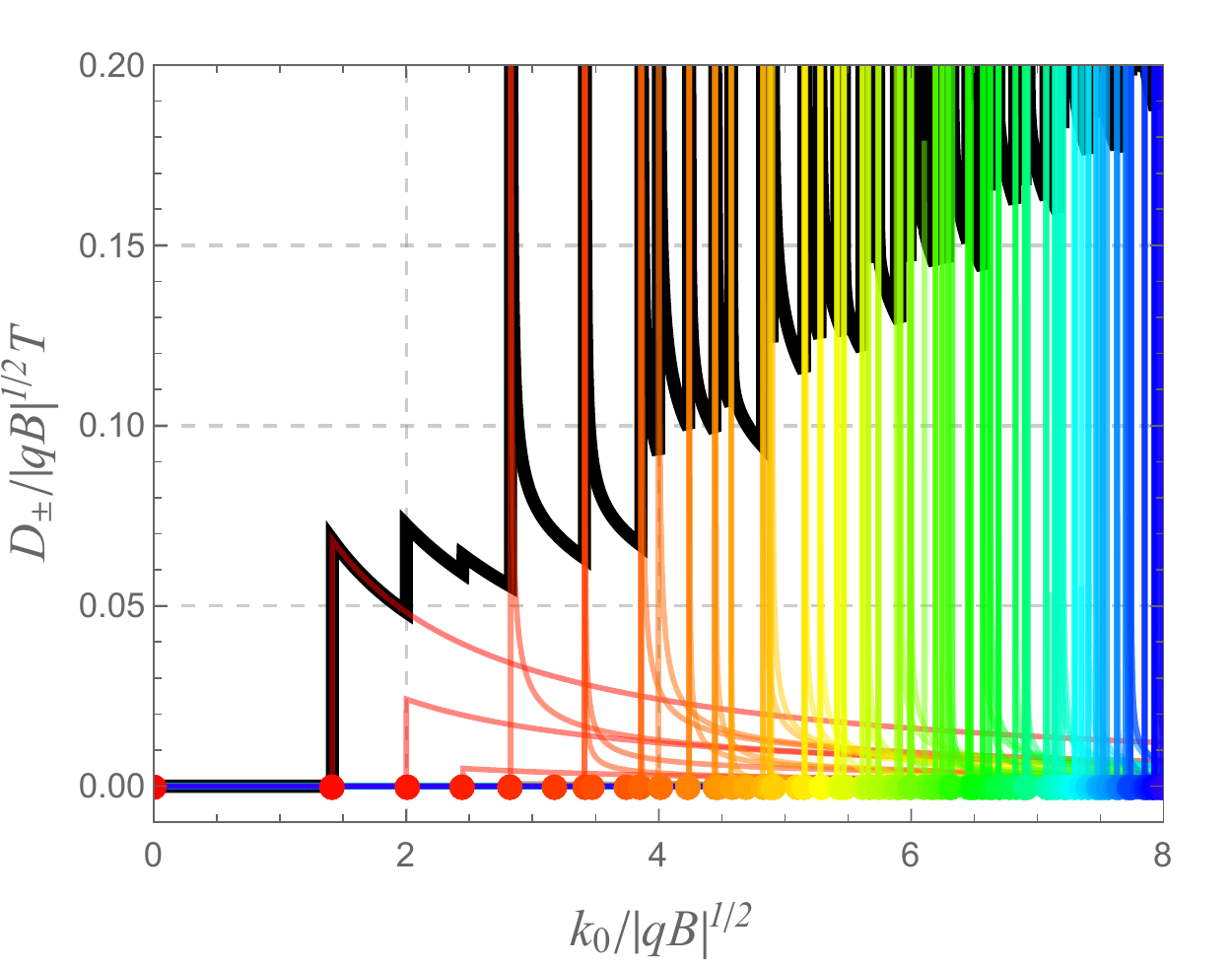} \hspace*{-4mm}
	\includegraphics[width=0.34\hsize]{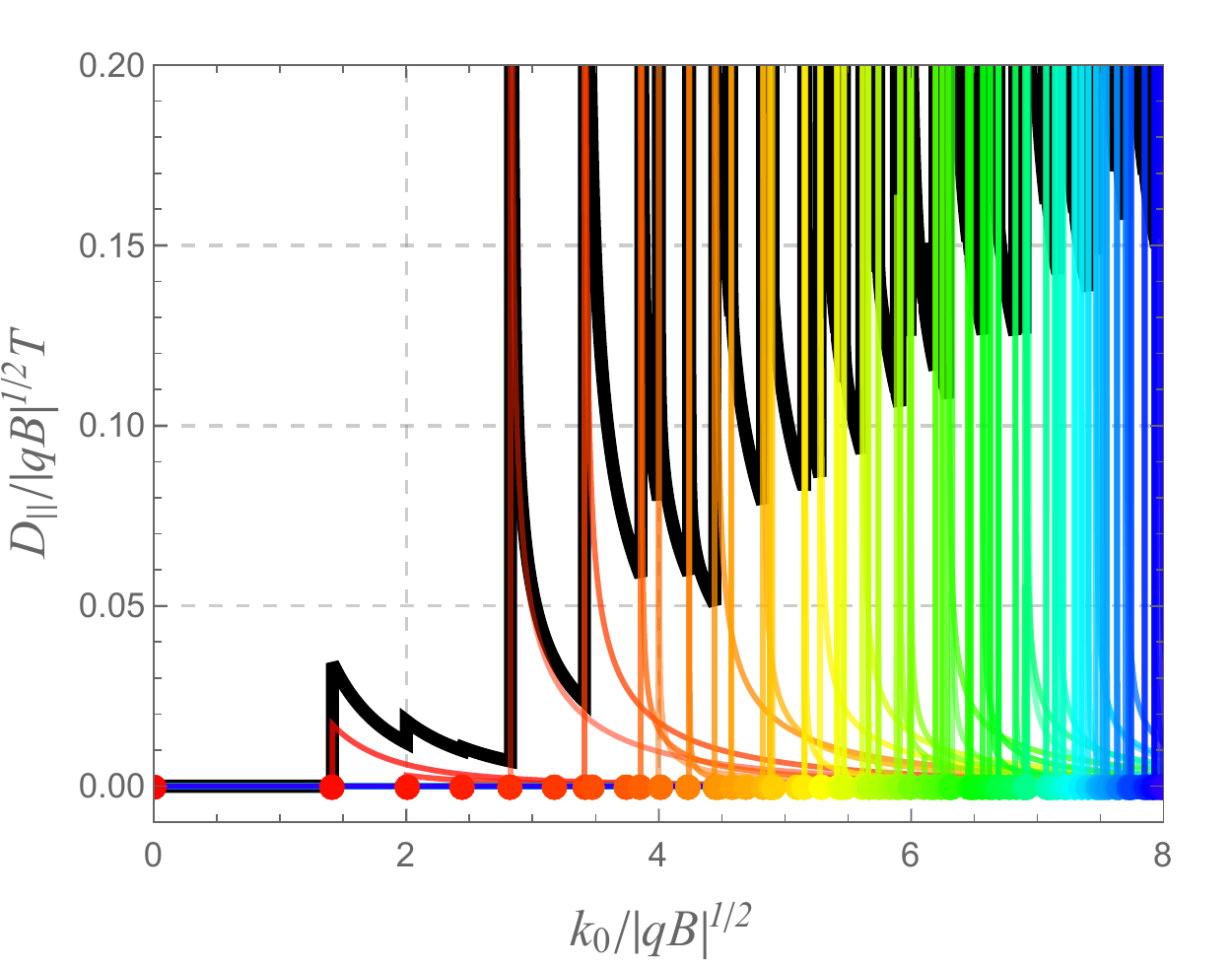} 
\end{center}
\caption{The conversion rates $ \rate_0$ (left), $ \rate_\pm$ (middle), and $ \rate_\parallel$ (right) in the massless limit $m = 0$, plotted against the photon energy $k_0/|qB|^{1/2}$.  The other parameters are fixed at $|{\bm k}_\perp|/|qB|^{1/2}=1 $ and $k_z/|qB|^{1/2}=0$.  As in Fig.~\ref{fig:Npm(k0)}, the colored dots on the horizontal axes indicate the threshold energies (\ref{eq:thres}) in ascending order from red to blue.  The colored lines originating from each dot show the contributions from each Landau level pair $ \rate_\lambda^{n,n'} $, and the black lines the total value summed over the Landau levels $ \rate_\lambda$.  
}
\label{fig:9}
\end{figure}

So far, we have used the fermion mass parameter $m$ just to make the other dimensionful quantities (such as the photon energy $k_0$ and the magnetic field strength $|qB|$) dimensionless.  Such a treatment makes sense only when $m\neq 0$, and we here discuss the massless limit ($m \to 0$) separately.  While the basic features of the di-lepton production are unchanged in the massless limit, we highlight several differences as demonstrated in Fig.~\ref{fig:9}.

In the massless limit, the di-lepton production in the lowest Landau level pair $n=n'=0$ is strictly prohibited, 
regardless of the photon polarization $ \lambda $.  
The physical reason behind this prohibition is the absence of the chirality mixing in a strictly massless theory.  Namely, fermions and anti-fermions in the lowest Landau level belong to different chirality eigenstates in a massless theory, and are not directly coupled with each other.  One can understand this statement from the spin polarization and the kinematics.  Spin in the lowest Landau level is polarized due to the Zeeman effect, and thus the spins of fermions and anti-fermions are polarized in the opposite direction to each other.  In addition, if di-lepton production occurred, the produced fermion and anti-fermion would recede from each other in a back-to-back direction in the center-of-momentum frame of the di-lepton,\footnote{
One can safely take the center-of-momentum frame $k_z=0$, since the incident photon 
must be time-like in the $(1+1)$-dimensional sense $k_\parallel^2>0$ to produce di-leptons because of the threshold condition Eq.~(\ref{eq:thres}). 
} 
indicating that the fermion and anti-fermion would have the same helicity.  Indeed, a fermion and an anti-fermion carrying the {\it same} helicity are created/annihilated by Weyl spinors in {\it different} chirality eigenstates.  Therefore, the di-lepton production in the lowest Landau level cannot occur, unless there is mixing between the right and left chirality eigenstates via a finite mass term.  In case of vector theories, di-fermion production in this massless channel also contradicts with the chiral symmetry or the conservation of axial charge (at the classical level).  A similar prohibition mechanism is known as ``helicity suppression'' in the leptonic decay of charged pions \cite{Donoghue:1992dd, Zyla:2020zbs}. 
At the algebraic level, this statement can be recognized as follows: 
There are three scalar form factors, ${{\Gam}}_{n,n'}$, ${{\Gam}}_{n-1,n'}$, and ${{\Gam}}_{n-1,n'-1}$, 
in the conversion rates (\ref{eq:N-lambda}), among which only ${\Gam}_{n,n'}$ can be nonvanishing for $n=n'=0$.  
However, the coefficients in front of $|{{\Gam}}_{0,0}|^2$ identically vanishes 
in the massless limit, i.e., $  \epsilon_0 \epsilon'_0  + p_z p'_z \pm m^2  = {\mathcal O}(m^2)$. 
Thus, $D^{0,0}_{\lambda} = 0$ holds for any photon polarization $\lambda$. 
This limiting behavior coincides with the above prohibition mechanism in a strictly massless theory 
(in spite of the fact that the naive massless limit 
does not reproduce the correct dispersion relation of the lowest Landau level 
in a massless theory $ \epsilon_0 =\pm p_z $ with signs for the right and left chirality).

Another distinct feature in the massless limit is that the resonances at the thresholds take finite values, unlike $m \neq 0$ case that we have discussed in Sec.~\ref{sec4.3.1}.  Remember the discussion below Eq.~(\ref{eq:denom-expand}), in which we claimed that even if the common denominator factor in the conversion rates (\ref{eq:N-lambda}) is vanishing $|p_z \epsilon'_{n'}-(k_z-p_z)\epsilon_n| = {\mathcal O}(\delta k_\parallel)$, the rates are not necessarily divergent if the other factors in the numerator are also vanishing.  This is actually the case in the massless limit $m\to 0$ when either fermion or anti-fermion is in the lowest Landau level (i.e., $n$ or $n'=0$), as shown in Fig.~\ref{fig:9}.  Indeed, the common numerator factors $\epsilon_n \epsilon'_{n'} + p_z p'_z \pm m^2$ become ${\mathcal O}(\delta k_\parallel)$ because of the linear dispersion relation of the lowest Landau level in the massless limit $\epsilon_0 = |p_{0,n'}^\pm| \to {\mathcal O}(\delta k_\parallel)$ and $\epsilon'_0 = |p_{n,0}^{\prime\pm}| \to {\mathcal O}(\delta k_\parallel)$ at the thresholds.

\section{Summary and outlook} \label{summary}

\subsection{Summary}

We have studied the di-lepton production rate from a single photon under a constant strong magnetic field.  In Sec.~\ref{sec-3}, we have analytically evaluated the squared matrix element for the photon--to--di-lepton conversion vertex by the use of the Ritus-basis formalism (reviewed in Sec.~\ref{sec-2}), in which the mode expansion is organized with the eigenstates of the Dirac operator.  This means that the Dirac operator is diagonalized without any perturbative expansion with respect to the interaction with the strong magnetic field.  Therefore, we have taken into account the interactions between the di-lepton and the strong magnetic field to all orders in the QED coupling constant non-perturbatively (see Fig.~\ref{fig:diagram}).  This treatment is necessary when the magnetic field is so strong that its strength compensates the smallness of the coupling constant.  On the other hand, we have included the coupling of the incident dynamical photon to the di-lepton at the leading order in the coupling constant and regards mutual interactions between the fermion and anti-fermion as higher-order corrections.  These perturbative treatments are justified as long as the QED coupling constant is small, as in the usual perturbation theory.

The squared matrix element (\ref{eq:sq}) is given as a product of the lepton tensor $L^{\mu\nu}_{n,n'}$ and delta functions, accounting for the kinematics of the production process (i.e., the energy and momentum conservation).  We have established the analytical expression of the lepton tensor (\ref{eq:L}) together with the scalar form factor (\ref{eq:FF-Landau}) to all order of the Landau-level summation.  Those analytical expressions enable us to write down an explicit formula for the conversion rates of a single photon into a di-lepton $\rate$ [e.g., Eq.~(\ref{eq:N-lambda}) for the polarization-projected ones].  Notably, we have confirmed in Appendix~\ref{app-b} that the obtained lepton tensor and thus the conversion rates possess the rotational invariance with respect to the direction of the magnetic field and the gauge invariances with respect to both the incident dynamical photon and the external magnetic field, although our calculation has been carried out in the Landau gauge that superficially breaks those symmetries in intermediate steps of the calculation.  Those rigorous consistency checks qualify our results. 

In Sec.~\ref{sec--4}, we have discussed quantitative aspects of the di-lepton production.  First, we have discussed how the kinematics of the production process affects the di-lepton spectrum in the final state.  We have shown that not only the transverse momentum of the produced fermions $|{\bm p}_\perp| \to \sqrt{2n|qB|}$ but also the longitudinal one $p_z \to p^\pm_{n,n'}$ is discretized, because of the energy conservation.  As a result, the di-lepton spectrum has spike structures in the longitudinal-momentum distribution as well as in the (Landau-quantized) transverse-momentum distribution.  We have also discussed the di-lepton spectrum in terms of the zenith angle $\phi$ that is defined as the emission angle of the fermions measured from the magnetic-field direction and thus succeeds the spike structures. 

Finally, we have investigated the inclusive conversion rates $D$ of a single photon, carrying a fixed polarization and momentum, into di-leptons.  The conversion rates exhibit spikes located at the threshold photon energy specified by each pair of the Landau levels.  In case of a massive fermion ($m \neq 0$), the height of the spikes is infinite for any pair of Landau levels $(n,n')$; this is a typical threshold behavior in the $(1+1)$-dimensions.  On the other hand, in the massless case, the height is finite when either of a pair is in the lowest Landau level ($n=0$ or $n'=0$).  In particular, the di-lepton production is strictly prohibited for massless fermions when both of a pair is in the lowest Landau levels, which is an analogue of the so-called helicity suppression.  We have confirmed all these fundamental behaviors with analytic expressions.

\subsection{Outlook}

Having established the fundamental formulas with clear physical interpretations, one can proceed to investigate phenomenological consequences in, e.g., relativistic heavy-ion collisions, neutron stars/magnetars, and high-intensity lasers. We emphasize that our di-lepton production rate predicts not only the photon attenuation rate captured by a complex-valued refractive index but also the entire di-lepton spectrum within a constant magnetic field with the complete resolution of the photon-polarization dependences.  This differential information is necessary, for example, to consider a cascade (or avalanche) process in which the photon--to--di-lepton conversion and its reciprocal reaction occur successively together with other magnetic-field-induced processes such as the cyclotron radiation.  Our results pave the way toward tracking energy and momentum distributions all the way through a cascade process induced by a strong magnetic field.  More specifically, if one uses, for example, a kinetic equation to describe the cascade process, our results provide a collision kernel.  Such a cascade process is not only interesting in its own as characteristic dynamics in a strong magnetic field but also important for understanding actual physics observables as we mention briefly below.

One of the most interesting applications of our results is relativistic heavy-ion collisions.  Relativistic heavy-ion collisions induces the ever strongest magnetic field in the present universe, which is of the order of $|qB| = {\mathcal O}(1\;{\rm GeV}^2)$ just after a collision of the two ions.  The magnetic-field strength depends on the collision geometry and becomes larger for smaller impact parameters $b\searrow$, larger collision energies $\sqrt{s} \nearrow$ and larger atomic numbers of the nuclei $Z \nearrow$ (see, e.g., a review article \cite{Hattori:2016emy} and references therein).  In particular, ultra-peripheral events provide clear cuts of electromagnetic processes without contaminations from QCD processes and/or medium effects (see Refs.~\cite{Zha:2018tlq, Klein:2018fmp, Li:2019yzy, Li:2019sin,  Xiao:2020ddm, Klein:2020jom} for recent theoretical proposals and Refs.~\cite{Adams:2004rz, Aaboud:2017bwk, Sirunyan:2018fhl, Aad:2019ock, Adam:2019mby} for significant progress in the recent measurements with RHIC and the LHC).  Among others, one can study the longitudinal momentum and/or the zenith-angle $\phi$ distribution of di-leptons with respect to those parameters.  Also, the fermion-mass dependence of the di-lepton production rate may be an interesting signature of the magnetic-field effects in analogue with the helicity suppression which explains the dominance of the muon channel over the electron channel in charged-pion decay modes \cite{Donoghue:1992dd, Zyla:2020zbs}.  In relativistic heavy-ion collisions, electrons can be regarded as massless particles as compared to the typical energy scales of the problem, whereas the muon mass is comparable in magnitude as typical QCD scales such as the pion mass.  Therefore, the ``helicity suppression'' of the electron pair production in the lowest-Landau levels could give rise to significant modifications in the low-energy di-lepton spectra.  In analogy with the pion decay modes, muons are more abundantly produced than electrons in between the lowest and the second-lowest energy thresholds specified by the muon mass and the magnetic-field strength, respectively.  We will report quantitative estimates of those effects in a forthcoming paper.

Another interesting application is neutron stars.  Neutron stars, in particular the so-called magnetars, may have stable strong magnetic fields close to or beyond the critical field strength of QED $eB_{\rm cr} \equiv m_e^2$ in their magnetospheres with $m_e  $ being the electron mass \cite{Harding:2006qn,Harding:2013ij, Enoto:2019vcg}.  Our results imply a strong polarization dependence in photons emitted from the stars with strong magnetic fields $eB \gtrsim eB_{\rm cr}$, where only the parallel polarization mode ($\lambda = \parallel$) can produce di-leptons, while the strong polarization dependence may be smeared out in the stars with weaker magnetic fields $eB \lesssim eB_{\rm cr}$.  This could serve as a complementary method to estimate strengths and/or distributions of magnetic fields, which have been commonly estimated via the so-called P-$ \dot {\rm P} $ diagram from observation \cite{Harding:2006qn,Harding:2013ij, Enoto:2019vcg}.  Also, our results indicate that strong magnetic-field effects such as the threshold effects become more prominent near the low-lying Landau-level thresholds than the higher thresholds. 
To get quantitative understanding of the above expectations, however, it is important to take into account the aforementioned cascade process induced by a strong magnetic field, which may amplify the strong-magnetic field effects.  We leave this as a future work.

Finally, we discuss implications for laser physics.  Thanks to the recent developments in lasers (e.g., chirped pulse amplification technique \cite{DiPiazza:2011tq, Zhang:2020lxl, Ejlli:2020yhk}), available electromagnetic field strength is rapidly rising and may reach the critical value $eB_{\rm cr}$ in the future.  One of the possible experimental setups to test our predictions is to combine an intense laser with an electron accelerator, by which energetic photons are supplied via Compton backscatterings (cf. Ref.~\cite{Bragin:2017yau}).  At the present, the available magnetic field strength is limited to $eB \lesssim 10^{-3} \times eB_{\rm cr}$ \cite{Yanovsky:08}.  In this weak-field regime, the vacuum dichroism may be controlled solely by the quantum non-linearity parameter $\chi \equiv \sqrt{|k^\mu F_{\mu\nu}|^2}/m^3$ rather than the field strength $eB$ (or the Lorentz invariants ${\mathcal F}\equiv F_{\mu\nu} F^{\mu\nu}/2m^4 = ({\bm B}^2 - {\bm E}^2)/m^4$ and $F_{\mu\nu} \tilde{F}^{\mu\nu}/4m^4 = {\bm E}\cdot{\bm B}/m^4$) and may be described essentially within the locally constant crossed field approximation, in which $\chi \neq 0, {\mathcal F} = {\mathcal G}= 0$ \cite{Ritus:thesis}.  
As the magnetic field strength increases, 
the pair of the Lorentz invariants approaches a different class such that ${\mathcal F} \gtrsim 1, {\mathcal G} =0$. 
In this class, such an approximate treatment breaks down and our calculation becomes more appropriate.  One then would observe the strong-magnetic field effects such as the discrete spectra and squeezed zenith-angle distributions of di-leptons.  In general, the cascade process would take place, giving rise to some modifications to our lowest order prediction. Such modifications become important for laser fields with a sufficiently large spatial extension, while it would be suppressed for those with a small spatial extension comparable to or smaller than the typical mean-free path for radiative processes under strong magnetic fields.

\section*{Acknowledgments}

The authors thank Xu-Guang~Huang for his contributions in the early stage of this work and useful discussions.  K.~H. thanks Kazunori~Itakura for discussions, which were useful to improve Sections~\ref{sec-2} and \ref{sec-3}, and also Yoshimasa~Hidaka for useful discussions.  S.~Y. is grateful for hospitality at Yukawa Institute for Theoretical Physics, Kyoto University where a part of the initial ideas was discussed.  K.~H. and H.~T. benefited from the international molecule-type workshops at Yukawa Institute for Theoretical Physics (YITP) ``Quantum kinetic theories in magnetic and vortical fields (YITP-T-19-06)'' and ``Potential Toolkit to Attack Nonperturbative Aspects of QFT -- Resurgence and related topics -- (YITP-T-20-03)'', where they had fruitful discussions.  K.~H. is supported in part by JSPS KAKENHI under grant No.~JP20K03948.  S.~Y. is supported by the research startup funding at South China Normal University, National Natural Science Foundation in China (NSFC) under grant No.~11950410495, and Guangdong Natural Science Foundation under No.~2020A1515010794.

\appendix

\section{Dirac fermions in magnetic fields} \label{app-a}

\subsection{Wave function in the Landau gauge}

\label{sec:wf_Landau}

We derive the wave function in the Landau gauge (\ref{eq:Landau-gauge}) on the basis of the coordinate representation of the creation and annihilation operators (\ref{eq:aadagger}).  Since a constant magnetic field does not affect the dynamics in the parallel direction, we focus on the transverse part of the wave function $ \phi$.

First, notice that the gauge field (\ref{eq:Landau-gauge}) is independent of the $ y $ coordinate, 
and the canonical momentum $ p_y $ is a conserved quantity.  Therefore, $ p_y $ is a good quantum number that can be used to label the energy eigenstates.  Consequently, the plane-wave part can be factorized as 
\begin{align}
	\phi_{n,p_y} (x, y) =  \langle x,y \vert n, p_y \rangle 
	=  e^{i  p_y y  } \tilde \phi_{n,p_y}(x)\, .
\end{align}
Nevertheless, the energy spectrum (\ref{eq:fermion-rela}) is independent of this conserved momentum and thus is infinite-fold degenerated for any value of $ p_y $.  This is naturally expected from the gauge invariance of the energy spectrum since the conservation of $  p_y$ is a specific property in the Landau gauge.  [Remember that we obtained the energy spectrum (\ref{eq:fermion-rela}) in a gauge-invariant way.]  Such a degeneracy originates from the fact that the translation of the cyclotron orbits in perpendicular to a constant magnetic field does not cost energy.  As seen below, the conserved canonical momentum divided by the field strength $x_c = p_y/qB$ serves as the center coordinate of the cyclotron orbits aligned in the $  x$ direction.

We shall find the explicitly form $  \tilde \phi_{0,p_y}(x)=\langle x  \vert 0 , p_y \rangle  $.  The wave function of the ground state can be obtained by solving an equation 
\begin{align}
	\langle x \vert \hat a \vert 0 , p_y \rangle = 0\, .
\end{align}
In the Landau gauge, the coordinate representation of the annihilation operator is given by 
\begin{align}
	\hat a 
		= - i \frac{\ell}{\sqrt{2} } \left( \frac{ \partial \ }{\partial x} +  \ell^{-2} (\hat x - x_c) \right)
		= - i \frac{ 1 }{\sqrt{2}  } e^{- \frac{ \xi^2 }{2} }  \frac{\partial \ }{\partial \xi} e^{ \frac{ \xi^2 } {2} }\, ,
\end{align}
where $ \xi = (x - x_c)/\ell$ and $\ell = 1/\sqrt{|qB|}$.  The derivative is assumed to act on what follows on the right as well as on the Gaussian.  The explicit form of the above condition reads 
\begin{align}
	\frac{\partial \ }{\partial \xi} \left( \, e^{ \frac{\xi^2}{2} } \tilde \phi_{0,p_y}  \, \right) = 0\, ,
\end{align}
which one can solve as 
\begin{align}
	\tilde \phi_{0,p_y} (\xi) = C_L  e^{ - \frac{\xi^2}{2} }\, .
\end{align}
The factor of $ C_L=( \ell \pi^{\frac{1}{2}} )^{-1/2}$ comes from the normalization 
\begin{align}
	\int dx \vert \tilde \phi_{0,p_y} (\xi) \vert^2 = 1\, .
\end{align}

As familiar in quantum mechanics, the wave function of higher Landau levels can be obtained by multiplying the creation operators as 
\begin{align}
\tilde \phi_{n,p_y} (\xi)   = \langle x \vert \frac{(a^\dagger)^n}{\sqrt{n!}} \vert 0 , p_y \rangle\, ,
\end{align}
with the coordinate representation of the creation operator 
\begin{align}
	\hat a^\dagger 
		= - i \frac{\ell_f}{\sqrt{2} } \left( 
\frac{ \partial \ }{\partial x} -  \ell_f^{-2} (\hat x - x_c) \right)
		= - i \frac{1}{\sqrt{2} } e^{ \frac{\xi^2}{2} }  \frac{\partial \ }{\partial \xi} e^{ -\frac{ \xi^2 } {2}}\, .
\end{align}
Therefore, the wave function for the general Landau level is obtained as
\begin{align}
	\tilde \phi_{n,p_y} (\xi)
		= \frac{1}{\sqrt{n!}} \left( - \frac{i}{\sqrt{2}  }\right)^n e^{ \frac{\xi^2}{2} } 
\frac{\partial^n \ }{\partial \xi^n } \left( \, e^{ - \frac{\xi^2}{2} } \tilde \phi_{0,p_y} (\xi)  \, \right)
		= C_L \frac{ i^n }{\sqrt{ 2^n n!}  } e^{ - \frac{\xi^2}{2} } H_n(\xi)\, ,
\end{align}
where the Hermite polynomial $H_n$ is defined by $H_n(\xi) = ( -1)^n e^{+ \xi^2 }  \frac{\partial^n \ }{\partial \xi^n } e^{ - \xi^2}$

Summarizing, we have obtained the wave function for the general Landau level $  n$: 
\begin{align}
	\phi_{n,p_y} (\xi,y) 
		= e^{i p_y y }\,  i^n \sqrt{ \frac{1}{ 2^n n! \pi^{\frac{1}{2}} \ell_f} } H_n (\xi)  \, ,
\end{align}
which is Eq.~(\ref{eq:WF_Landau}) in the main text.  Notice that $ x_c $ is the center coordinate of the harmonic oscillation in the $  x$ direction.  The other center coordinate $ y_c $ does not appear in the wave function.  Besides an obvious reason specific to the Landau gauge (\ref{eq:Landau-gauge}), a more essential reason for the absence of $ y_c $ is that the translation of the cyclotron orbits in the $  x$ and $y  $ directions do not commute with each other due to the Aharonov-Bohm phase.  Therefore, one can take only one polynomial of $ x_c $ and $ y_c $ to label the energy eigenstates.  The choice of either $ x_c $ or $ y_c $ corresponds to the Landau gauge, while $ x_c^2 + y_c^2 $ corresponds to another popular gauge, that is, the symmetric gauge.

The density of degenerated states can be counted in a finite box.  Since $0 \leq  x_c \leq L_x $, we have $ 0 \leq p_y \leq |qB| L_x$ when $\sgn(qB)>0 $ and $ -|qB| L_x \leq p_y \leq 0$ when $ \sgn(qB) <0 $.  Therefore, we get the density of states in each Landau level 
\begin{align}
	\frac{1}{ L_x} \int_0^{|qB| L_x} \!\! \frac{dp_y}{2\pi  }  = \frac{|qB|}{2\pi}\, .\label{eq:DoS}
\end{align}
In addition, one should attach the spin degeneracy factor $2$ for higher Landau levels.

\subsection{Canonical anti-commutation relation for the Dirac field}

\label{sec:commutation}

We show that the anti-commutation relations for the fermion creation and annihilation operators (\ref{eq:commutation}) are consistent with the equal-time canonical anti-commutation relation for the Dirac field.  This can be straightforwardly shown as follows.  Plugging the Ritus-basis mode expansion (\ref{eq:Ritus-mode}) into the equal-time anti-commutator, we have
\begin{align}
	\{ \psi (x), \psi^\dagger (x') \}_{x^0={x'}^0}
		&= \sum_{\k=1,2} \sum_{n=0}^{\infty} \int \frac{d p_z}{2\pi} \int  \frac{dp_y}{2\pi} \frac{1}{2\epsilon_n} \Ritus_{n,p_y} (x_\perp)  \nonumber \\
		&\quad \times [   e^{ - i p_\parallel \cdot (x-x') } u_\k (p_n)  \bar u_\k (p_n) +  e^{i p_\parallel \cdot (x-x')}    v_\k (\bar p_n)  \bar v_\k (\bar p_n) ] \Ritus_{n,p_y}^\dagger  (x'_\perp)  \gamma^0\, .
\end{align}
Then, by using the completeness of the wave function (\ref{eq:spin-sum}), one can arrange the Dirac spinors as 
\begin{align}
\label{eq:canonical}
	&\{ \psi (x), \psi^\dagger (x') \}_{x^0={x'}^0}\nonumber \\
	&= \sum_{n=0}^{\infty} \int \frac{d p_z}{2\pi} \int  \frac{dp_y}{2\pi}  \frac{1}{2\epsilon_n} \Ritus_{n,p_y}  (x_\perp)   [   e^{- i p_\parallel \cdot (x-x')}  ( \slashed p_n + m)  +  e^{  i p_\parallel \cdot (x-x') }      ( \bar {\slashed p}_n -m )  ]  \Ritus_{n,p_y}^\dagger  (x'_\perp)  \gamma^0 \nonumber \\
	&= \sum_{n=0}^{\infty} \int \frac{d p_y}{2\pi} \Ritus _{n,p_y}  (x_\perp)   \Ritus_{n,p_y}^\dagger (x'_\perp) \int  \frac{dp_z}{2\pi}  e^{ - i p_\parallel \cdot (x-x') } \nonumber \\
  &= 
  \delta^{(3)} ( {\bm x} - {\bm x}') I_n
  \, ,
\end{align}
where we introduced a ``unit matrix'' 
\begin{align}
	I_n = \left\{
				\begin{array}{ll}
 					\prj_+ & (n=0) \\
					\id_{\rm spinor} & (n \geq1)
				\end{array}
			\right. \, .
\end{align}
To reach the last line, we used an identity 
\begin{align}
	&\sum_{n=0}^{\infty} \int \frac{d p_y}{2\pi}  \Ritus _{n,p_y}  (x_\perp)   \Ritus_{n,p_y}^\dagger (x'_\perp) \nonumber\\
	&= \sum_{n=0}^{\infty} \int \frac{d p_y}{2\pi}  [  \phi_{n,p_y} (x_\perp)  \phi_{n,p_y}^\ast (x'_\perp) \prj_+ +  \phi_{n-1,p_y} (x_\perp)  \phi_{n-1,p_y}^\ast (x'_\perp) \prj_- ] \nonumber\\
	&= \delta^{(2)}(x_\perp - x'_\perp) I_n\, ,
\end{align}
where we explicitly wrote the label $p_y$ in the fermion transverse fermion function $\phi_{n,p_y}$ for clarity [which was suppressed in the main text for simplicity, as stated below Eq.~(\ref{eq:wave-functions})].  This is a manifestation of the completeness in the Landau quantization with the Landau gauge (\ref{eq:Landau-gauge}), 
\begin{align}
	\sum_{n=0}^\infty\int \frac{d p_y}{2\pi} \phi_{n,p_y} (x_\perp)  \phi_{n,p_y}^\ast (x'_\perp) 
	= \sum_{n=0}^\infty\int \frac{d p_y}{2\pi}  \langle x_\perp | n , p_y \rangle  \langle n, p_y | x'_\perp \rangle 
	=  \delta^{(2)}(x_\perp - x'_\perp) \, ,
\end{align}
and of the spin projection operators $ \prj_+ + \prj_- = 1 $ in all the energy levels but the ground state. 
Therefore, Eq.~(\ref{eq:canonical}) indicates that the Dirac field $\psi$ satisfies the canonical anti-commutation relation for a Dirac fermion in the higher Landau levels ($ n\geq1 $) and that for a Weyl fermion in the ground state ($ n=0 $).  In the latter, either upper or lower component of the Dirac spinor is projected out by $ \prj_+ $ on the both sides of the equation, depending on $\sgn(qB)$.

Incidentally, one can also show an identity 
\begin{align}
	&\int d^2 x_\perp  \Ritus _{n,p_y}  (x_\perp)   \Ritus_{n',p_y'}^\dagger (x_\perp) \nonumber\\
	&= \int d^2 x_\perp [  \phi_{n,p_y} (x_\perp)  \phi_{n',p_y'}^\ast (x_\perp) \prj_+ 
	+  \phi_{n-1,p_y} (x_\perp)  \phi_{n'-1,p_y'}^\ast (x_\perp) \prj_- ]\nonumber\\
	&= 2\pi \delta( p_y - p'_y) \delta_{n,n'} I_n \, ,
\end{align}
according to the orthogonality among the Landau levels: 
\begin{align}
	\int d^2 x_\perp \phi_{n,p_y} (x_\perp)  \phi_{n',p_y'}^\ast (x_\perp) 
	= \int d^2 x_\perp  \langle x_\perp | n , p_y \rangle  \langle n', p'_y | x_\perp \rangle
	= 2\pi\delta(p_y-p'_y) \delta_{n,n'} \, .
\end{align}

\section{Consistency checks for the conversion amplitude}\label{app-b}

\subsection{The Ward identity at each pair of Landau levels} \label{app-b1}

We discuss the Ward identity for the amplitude ${\mathcal M}^\mu$.  In general, according to the Dirac equation in an external field,
\begin{align}
	( i \slashed D_{\rm ext} - m) \psi = 0\, , \quad
	\bar \psi ( i \overleftarrow{  \slashed D}_{\rm ext}^\ast + m)  = 0, 
\end{align}
we find the current conservation law,  
\begin{align}
	i \partial_\mu ( \bar \psi \gamma^\mu \psi) 
		= \bar \psi ( - q A_{\rm ext}^\ast - m) \psi + \bar \psi (  q A _{\rm ext}+ m) \psi 
		= 0 \, , \label{eq:currentconservation}
\end{align}
where we used $ A_{\rm ext}^\ast  =A_{\rm ext} $.  Comparing the current conservation law (\ref{eq:currentconservation}) with the definition of the amplitude ${\mathcal M}^\mu$ (\ref{eq:matrixelement}), one understands
\begin{align}
	0 = k_\mu {\mathcal M}^\mu . \label{eq:wardamp}
\end{align}
This is the Ward identity for the amplitude ${\mathcal M}^\mu$.

We explicitly confirm the Ward identity for the amplitude (\ref{eq:wardamp}) after the Landau-level decomposition is performed.  Contracting the amplitude ${\mathcal M}^\mu$ with the photon momentum $k_\mu$, we get
\begin{align}
	k_\mu {\mathcal M}^\mu
		&=  (2\pi)^3 \delta^{(2)}( k_\parallel - p_\parallel - p'_\parallel) \delta(k_y - p_y + p'_y) \nonumber\\
			&\quad \times \Big[ \, \bar u_\k (p_n) \slashed p_\parallel \H_0 v_{\k'} (\bar p_{n'}^\prime) + \bar u_\k(p_n) \H_0 \slashed p_\parallel^{\prime} v_{\k'}(\bar p_{n'}^\prime) +  \bar u_\k(p_n) \slashed k_\perp  \H_1 v_{\k'}(\bar p_{n'}^\prime) \, \Big] \\
		&=  (2\pi)^3 \delta^{(2)}( k_\parallel - p_\parallel - p'_\parallel) \delta(k_y - p_y + p'_y) \times  \sqrt{2 |qB|} \bar u_\k(p_n) \gamma^1  ( \Xi^+_{n,n'} \prj_+ + \Xi^-_{n,n'} \prj_-  ) v_{\k'}(\bar p_{n'}^\prime) \,, \nonumber
\end{align}
where we used the Dirac equations (\ref{eq:free_u}) and (\ref{eq:free_v}), the eigenvalue relation of the spin operator $ i\,\sgn(qB) \gamma^1\gamma^2 \prj_\pm = \pm \prj_\pm $, and $ [ \gamma_\parallel^\mu, \H_{0,1} ] =0 $.  In the second line, the coefficients of $  \prj_\pm$ can be arranged as 
\begin{subequations}
\label{eq:id}
\begin{align}
\label{eq:id1}
	\Xi^+_{n,n'} 
		&\equiv \sqrt{n}  {\Gam}_{n, n'} - \sqrt{n'} {\Gam}_{n-1,n'-1} - |\bar {\bm k}_\perp| e^{ i \,\sgn(qB) \theta_\perp } {\Gam}_{n-1,n'} \, , \\
\label{eq:id2}
	\Xi^-_{n,n'} 
		&\equiv \sqrt{n} {\Gam}_{n-1,n'-1} - \sqrt{n'} {\Gam}_{n,n'} - |\bar {\bm k}_\perp| e^{ -i \,\sgn(qB) \theta_\perp }    {\Gam}_{n,n'-1} \, .
\end{align}
\end{subequations}
Below, we prove two identities: 
\begin{align}
	\Xi^+_{n,n'} = 0 \, , \quad \Xi^-_{n,n'} =0\, , \label{eq:identities}
\end{align}
which will indicate that the amplitude $ {\mathcal M}^\mu $ satisfies the Ward identity (\ref{eq:wardamp}) for each pair of $ n $ and $ n' $.  To do this, we use the explicit form of the scalar form factor ${{\Gam}}_{n,n'}$ (\ref{eq:FF-Landau}), with which one can arrange the first two terms in Eq.~(\ref{eq:id1}) as
\begin{align}
\label{eq:proof1}
	&\sqrt{n}  {\Gam}_{n, n'} - \sqrt{n'}   {\Gam}_{n-1,n'-1}  \nonumber \\
	&= e^{  i \frac{k_x(p_y + p^\prime _y)}{2 qB} } \, e^{-  \frac{1}{2} |\bar {\bm k}_\perp|^2} (-1)^{\Delta n} e^{ -i\, \sgn(qB) \Delta n \, \theta_\perp} | \bar {\bm k}_\perp|^{|\Delta n|} \sqrt{\frac{n!}{ n^\prime!}}  \frac{1 }{\sqrt{n}} \left[ n  L_{ n }^{\Delta n}  -  n' L_{ n-1 }^{\Delta n}  \right] \nonumber \\
	&= e^{  i \frac{k_x(p_y + p^\prime _y)}{2 qB} } \, e^{-  \frac{1}{2} |\bar {\bm k}_\perp|^2} (-1)^{\Delta n}  e^{ - i \, \sgn(qB)\Delta n \, \theta_\perp} | \bar {\bm k}_\perp|^{|\Delta n|} \sqrt{\frac{n!}{ n^\prime!}}  \frac{1}{\sqrt{n}} \left[   -  |\bar {\bm k}_\perp|^2  L_{ n-1 }^{\Delta n + 1}  \right] \nonumber \\
	&= | \bar {\bm k}_\perp|  e^{  i\, \sgn(qB)  \theta_\perp} {\Gam}_{n-1,n'}  \, ,
\end{align}
where, to reach the second line, we used \cite{AssociatedLaguerrePolynomial}
\begin{align}
\label{eq:formula1}
	z L_{n-1}^{k+1} (z) = - \left[ n L_n^k (z)  - (n+k) L_{n-1}^k (z)  \right] 
.
\end{align}
This proves the identity $  \Xi^+_{n,n'} = 0 $ in Eq.~(\ref{eq:identities}).  Similarly, by arranging the first two terms in Eq.~(\ref{eq:id2}), one obtains
\begin{align}
	\sqrt{n} {\Gam}_{n-1, n'-1} - \sqrt{n'}  {\Gam}_{n,n'} 
		&= g (-1)^{\Delta n}  e^{ - i\,\sgn(qB) \Delta n \, \theta_\perp} | \bar {\bm k}_\perp|^{|\Delta n|} \sqrt{\frac{n!}{ n'! }} (- \sqrt{n'}) \left[  L_{ n }^{\Delta n}  -  L_{ n-1 }^{\Delta n}  \right] \nonumber \\
		&= g (-1)^{\Delta n-1}  e^{ - i \,\sgn(qB) \Delta n \, \theta_\perp}  | \bar {\bm k}_\perp|^{|\Delta n|} \sqrt{\frac{n!}{(n'-1)! }}   L_{ n }^{ \Delta n -1 } \nonumber \\
		&=  | \bar {\bm k}_\perp|  e^{  - i \,\sgn(qB) \theta_\perp} {\Gam}_{ n, n'-1 } \, ,
\end{align}
where we used \cite{AssociatedLaguerrePolynomial}
\begin{align}
	L^k_n (z) = L^{k+1}_n (z) - L^{k+1}_{n-1} (z) \, . \label{eq:formula2}
\end{align}
Therefore, we have proven the remaining identity $ \Xi^-_{n,n'} = 0 $ in Eq.~(\ref{eq:identities}).  This completes the proof of the Ward identity (\ref{eq:wardamp}).

\subsection{Gauge and rotational invariances of the squared amplitude} \label{app-b2}

We confirm the following three points as consistency checks for the main formula (\ref{eq:L}): (i) the amplitude square $ |\epsilon_\mu {\mathcal M}^\mu |^2 \propto \varepsilon_\mu \varepsilon_\nu^\ast L^{\mu\nu} $ is a real-valued quantity; (ii) the lepton tensor $ L^{\mu\nu}  $ satisfies the Ward identity, i.e., $k_\mu  L^{\mu\nu} = k_\nu  L^{\mu\nu}=0  $ [which follows from the current conservation law (\ref{eq:currentconservation})], and (iii) $ \varepsilon_\mu \varepsilon_\nu^\ast L^{\mu\nu}  $ has the rotational symmetry in the transverse plane for an arbitrary photon polarization $ \varepsilon_\mu $.

(i) The condition that $  \varepsilon_\mu \varepsilon^{\ast}_\nu L^{\mu\nu}$ be a real-valued quantity for an arbitrary photon polarization (without a polarization sum) is satisfied if $  \varepsilon_\mu \varepsilon^{\ast}_\nu L^{\mu\nu} =  [ \varepsilon_\mu \varepsilon^{\ast}_\nu L^{\mu\nu}]^\ast$.  This requires that the lepton tensor $L^{\mu\nu}$ to be an Hermitian matrix: 
\begin{align}
	L^{ \mu\nu} = L^{\ast \nu\mu}\, . \label{eq:Hermitian}
\end{align}
Since we have $ L^{\ast\nu\mu}_\parallel = L_\parallel^{\mu\nu} $, $ \varepsilon_\pm^{\mu\ast} = \varepsilon_\mp^\mu $, and $ \Q_\pm^{\ast \nu\mu} = \Q_\pm^{\mu\nu} $, our lepton tensor $L^{\mu\nu}$ (\ref{eq:L}) transforms as a Hermitian matrix as required.

(ii) Next, we examine the Ward identity for the lepton tensor $k_\mu  L^{\mu\nu} = k_\nu  L^{\mu\nu}=0  $.  The contraction between $ L^{\mu\nu} $ and the photon momentum $k_\mu$ results in the following combinations 
\begin{subequations}
\begin{align}
	k_\mu L_\parallel^{\mu\nu} 
		&= 4 |qB| ( n  p_\parallel^{\prime\nu} +  n' p_\parallel^{\nu}  ) \, , \\
	k \cdot p_\parallel 
		&= ( p_\parallel \cdot p_\parallel^{\prime} + m^2 ) + 2n|qB| \, , \\
	k \cdot p'_\parallel 
		&=( p_\parallel \cdot p_\parallel^{\prime} + m^2 ) + 2n' |qB| \, , \\
	k \cdot \varepsilon_\pm 
		&= - \sqrt{ |qB| \, }  |\bar {\bm k}_\perp| e^{ \pm i \, \sgn(qB) \theta_\perp } \, , \\
	k _\mu \Q_\pm^{\mu\nu} 
		&= \sqrt{ |qB| \, }  |\bar {\bm k}_\perp| e^{  \pm i \, \sgn(qB) \theta_\perp } \varepsilon_\pm^{ \nu \ast}\, ,
\end{align}
\end{subequations}
where we used the on-shell conditions $ p_\parallel^2 = m^2 + 2n|qB| $ and $ p_\parallel^{\prime2} = m^2 + 2n'|qB| $.  By using those expressions, we obtain 
\begin{align}
	k_\mu L^{\mu\nu} 
		&= - 2 \sqrt{2n' |qB|} \ ( \Xi^+_{n,n'} {\Gam}_{n-1,n'-1}^\ast + \Xi^-_{n,n'}  \Gam_{n,n'}^\ast  ) p_\parallel^{\nu} \nonumber \\
		&\quad + 2 \sqrt{2n |qB|}  \ (\Xi^+_{n,n'}  {\Gam}_{n,n'}^\ast + \Xi^-_{n,n'} {\Gam}_{n-1,n'-1}^\ast  )p_\parallel^{\prime\nu} \nonumber \\
		&\quad + 4\sqrt{|qB|} \  {\Gam}_{n,n'-1} ^\ast \big[ (p_\parallel \cdot p'_\parallel + m^2)   \Xi^-_{n,n'}  - 2|qB|\sqrt{nn'} \Xi^+_{n,n'}   \big]  \varepsilon_+^{ \nu}  \nonumber\\
		&\quad +  4 \sqrt{|qB|} \ {\Gam}_{n-1,n'} ^\ast  \big[  (p_\parallel \cdot p'_\parallel + m^2)  \Xi^+_{n,n'} - 2|qB|\sqrt{nn'} \Xi^-_{n,n'}   \big]   \varepsilon_-^{\nu}  \, .
\end{align}
Since all those terms are proportional to $ \Xi^\pm_{n n'} $ that satisfy the identities (\ref{eq:identities}), the contraction between the lepton tensor and the photon momentum vanishes, i.e., 
\begin{align}
	k_\mu L^{\mu\nu} = 0\, .
\end{align}
This also implies that $ k_\nu  L^{\mu\nu} = ( k_\nu  L^{\nu\mu} )^\ast =0 $ for a real-valued photon momentum $ k^\mu $, according to the Hermitian property in Eq.~(\ref{eq:Hermitian}).

(iii) Finally, notice that all the $ \theta_\perp $ dependences in the lepton tensor $L^{\mu\nu}$ (\ref{eq:L}) come in the following combination [see also the explicit form of the scalar form factor ${\Gam}_{n,n'}$ (\ref{eq:FF-Landau}) and use $ \varepsilon_\pm^{\mu\ast} =   \varepsilon_\mp^\mu$]
\begin{align}
	|{\bm k}_\perp|  e^{ -  i \, \sgn(qB) \theta_\perp }  \varepsilon_+^\nu A_\nu 
	= \frac{1}{\sqrt{2}} k_\perp^\ast A_\perp \, ,
\end{align}
where $ A_\mu $ is a photon field to be contracted with the lepton tensor.  On the right-hand side of the above equation, we defined $  k_\perp^\ast = k_x - i \, \sgn(qB) k_y$, $A_\perp = A_x + i \, \sgn(qB) A_y  $.  From this expression, it is clear that the lepton tensor $L^{\mu\nu}$ has the rotational invariance in the transverse plane when contracted with an arbitrary photon polarization.  While the wave function in the Landau gauge explicitly breaks the rotational invariance, it has been restored in the physical quantity as expected, serving as a consistency check of the gauge invariance.

\bibliographystyle{apsrev4-1}
\bibliography{bib}

\begin{thebibliography}{59}%
\makeatletter
\providecommand \@ifxundefined [1]{%
 \@ifx{#1\undefined}
}%
\providecommand \@ifnum [1]{%
 \ifnum #1\expandafter \@firstoftwo
 \else \expandafter \@secondoftwo
 \fi
}%
\providecommand \@ifx [1]{%
 \ifx #1\expandafter \@firstoftwo
 \else \expandafter \@secondoftwo
 \fi
}%
\providecommand \natexlab [1]{#1}%
\providecommand \enquote  [1]{``#1''}%
\providecommand \bibnamefont  [1]{#1}%
\providecommand \bibfnamefont [1]{#1}%
\providecommand \citenamefont [1]{#1}%
\providecommand \href@noop [0]{\@secondoftwo}%
\providecommand \href [0]{\begingroup \@sanitize@url \@href}%
\providecommand \@href[1]{\@@startlink{#1}\@@href}%
\providecommand \@@href[1]{\endgroup#1\@@endlink}%
\providecommand \@sanitize@url [0]{\catcode `\\12\catcode `\$12\catcode
  `\&12\catcode `\#12\catcode `\^12\catcode `\_12\catcode `\%12\relax}%
\providecommand \@@startlink[1]{}%
\providecommand \@@endlink[0]{}%
\providecommand \url  [0]{\begingroup\@sanitize@url \@url }%
\providecommand \@url [1]{\endgroup\@href {#1}{\urlprefix }}%
\providecommand \urlprefix  [0]{URL }%
\providecommand \Eprint [0]{\href }%
\providecommand \doibase [0]{http://dx.doi.org/}%
\providecommand \selectlanguage [0]{\@gobble}%
\providecommand \bibinfo  [0]{\@secondoftwo}%
\providecommand \bibfield  [0]{\@secondoftwo}%
\providecommand \translation [1]{[#1]}%
\providecommand \BibitemOpen [0]{}%
\providecommand \bibitemStop [0]{}%
\providecommand \bibitemNoStop [0]{.\EOS\space}%
\providecommand \EOS [0]{\spacefactor3000\relax}%
\providecommand \BibitemShut  [1]{\csname bibitem#1\endcsname}%
\let\auto@bib@innerbib\@empty
%</preamble>
\bibitem [{\citenamefont {Heisenberg}\ and\ \citenamefont
  {Euler}(1936)}]{Heisenberg:1935qt}%
  \BibitemOpen
  \bibfield  {author} {\bibinfo {author} {\bibfnamefont {W.}~\bibnamefont
  {Heisenberg}}\ and\ \bibinfo {author} {\bibfnamefont {H.}~\bibnamefont
  {Euler}},\ }\href {\doibase 10.1007/BF01343663} {\bibfield  {journal}
  {\bibinfo  {journal} {Z. Phys.}\ }\textbf {\bibinfo {volume} {98}},\ \bibinfo
  {pages} {714} (\bibinfo {year} {1936})},\ \Eprint
  {http://arxiv.org/abs/physics/0605038} {arXiv:physics/0605038 [physics]}
  \BibitemShut {NoStop}%
%%CITATION = PHYSICS/0605038;%%
\bibitem [{\citenamefont {Sauter}(1931)}]{sauter1931behavior}%
  \BibitemOpen
  \bibfield  {author} {\bibinfo {author} {\bibfnamefont {F.}~\bibnamefont
  {Sauter}},\ }\href@noop {} {\bibfield  {journal} {\bibinfo  {journal} {Zeit.
  f. Phys}\ }\textbf {\bibinfo {volume} {69}},\ \bibinfo {pages} {742}
  (\bibinfo {year} {1931})}\BibitemShut {NoStop}%
\bibitem [{\citenamefont {Euler}\ and\ \citenamefont
  {Kockel}(1935)}]{Euler:1935zz}%
  \BibitemOpen
  \bibfield  {author} {\bibinfo {author} {\bibfnamefont {H.}~\bibnamefont
  {Euler}}\ and\ \bibinfo {author} {\bibfnamefont {B.}~\bibnamefont {Kockel}},\
  }\href {\doibase 10.1007/BF01493898} {\bibfield  {journal} {\bibinfo
  {journal} {Naturwiss.}\ }\textbf {\bibinfo {volume} {23}},\ \bibinfo {pages}
  {246} (\bibinfo {year} {1935})}\BibitemShut {NoStop}%
%%CITATION = NATWA,23,246;%%
\bibitem [{\citenamefont {Weisskopf}(1936)}]{Weisskopf:1996bu}%
  \BibitemOpen
  \bibfield  {author} {\bibinfo {author} {\bibfnamefont {V.}~\bibnamefont
  {Weisskopf}},\ }\href@noop {} {\bibfield  {journal} {\bibinfo  {journal}
  {Kong. Dan. Vid. Sel. Mat. Fys. Med.}\ }\textbf {\bibinfo {volume} {14N6}},\
  \bibinfo {pages} {1} (\bibinfo {year} {1936})}\BibitemShut {NoStop}%
%%CITATION = KDVSA,14N6,1;%%
\bibitem [{\citenamefont {Schwinger}(1951)}]{Schwinger:1951nm}%
  \BibitemOpen
  \bibfield  {author} {\bibinfo {author} {\bibfnamefont {J.~S.}\ \bibnamefont
  {Schwinger}},\ }\href {\doibase 10.1103/PhysRev.82.664} {\bibfield  {journal}
  {\bibinfo  {journal} {Phys. Rev.}\ }\textbf {\bibinfo {volume} {82}},\
  \bibinfo {pages} {664} (\bibinfo {year} {1951})}\BibitemShut {NoStop}%
%%CITATION = PHRVA,82,664;%%
\bibitem [{\citenamefont {Toll}(1952)}]{Toll:1952rq}%
  \BibitemOpen
  \bibfield  {author} {\bibinfo {author} {\bibfnamefont {J.~S.}\ \bibnamefont
  {Toll}},\ }\emph {\bibinfo {title} {{The Dispersion relation for light and
  its application to problems involving electron pairs}}},\ \href@noop {}
  {Ph.D. thesis},\ \bibinfo  {school} {Princeton University} (\bibinfo {year}
  {1952})\BibitemShut {NoStop}%
%%CITATION = RX-1535;%%
\bibitem [{\citenamefont {Adler}(1971)}]{Adler:1971wn}%
  \BibitemOpen
  \bibfield  {author} {\bibinfo {author} {\bibfnamefont {S.~L.}\ \bibnamefont
  {Adler}},\ }\href {\doibase 10.1016/0003-4916(71)90154-0} {\bibfield
  {journal} {\bibinfo  {journal} {Annals Phys.}\ }\textbf {\bibinfo {volume}
  {67}},\ \bibinfo {pages} {599} (\bibinfo {year} {1971})}\BibitemShut
  {NoStop}%
%%CITATION = APNYA,67,599;%%
\bibitem [{\citenamefont {Tsai}\ and\ \citenamefont
  {Erber}(1974)}]{Tsai:1974fa}%
  \BibitemOpen
  \bibfield  {author} {\bibinfo {author} {\bibfnamefont {W.-y.}\ \bibnamefont
  {Tsai}}\ and\ \bibinfo {author} {\bibfnamefont {T.}~\bibnamefont {Erber}},\
  }\href {\doibase 10.1103/PhysRevD.10.492} {\bibfield  {journal} {\bibinfo
  {journal} {Phys. Rev. D}\ }\textbf {\bibinfo {volume} {10}},\ \bibinfo
  {pages} {492} (\bibinfo {year} {1974})}\BibitemShut {NoStop}%
\bibitem [{\citenamefont {Tsai}\ and\ \citenamefont
  {Erber}(1975)}]{Tsai:1975iz}%
  \BibitemOpen
  \bibfield  {author} {\bibinfo {author} {\bibfnamefont {W.-y.}\ \bibnamefont
  {Tsai}}\ and\ \bibinfo {author} {\bibfnamefont {T.}~\bibnamefont {Erber}},\
  }\href {\doibase 10.1103/PhysRevD.12.1132} {\bibfield  {journal} {\bibinfo
  {journal} {Phys. Rev.}\ }\textbf {\bibinfo {volume} {D12}},\ \bibinfo {pages}
  {1132} (\bibinfo {year} {1975})}\BibitemShut {NoStop}%
%%CITATION = PHRVA,D12,1132;%%
\bibitem [{\citenamefont {Shabad}(1975)}]{Shabad:1975ik}%
  \BibitemOpen
  \bibfield  {author} {\bibinfo {author} {\bibfnamefont {A.}~\bibnamefont
  {Shabad}},\ }\href {\doibase 10.1016/0003-4916(75)90144-X} {\bibfield
  {journal} {\bibinfo  {journal} {Annals Phys.}\ }\textbf {\bibinfo {volume}
  {90}},\ \bibinfo {pages} {166} (\bibinfo {year} {1975})}\BibitemShut
  {NoStop}%
\bibitem [{\citenamefont {Melrose}\ and\ \citenamefont
  {Stoneham}(1976)}]{Melrose:1976dr}%
  \BibitemOpen
  \bibfield  {author} {\bibinfo {author} {\bibfnamefont {D.}~\bibnamefont
  {Melrose}}\ and\ \bibinfo {author} {\bibfnamefont {R.}~\bibnamefont
  {Stoneham}},\ }\href {\doibase 10.1007/BF02730208} {\bibfield  {journal}
  {\bibinfo  {journal} {Nuovo Cim. A}\ }\textbf {\bibinfo {volume} {32}},\
  \bibinfo {pages} {435} (\bibinfo {year} {1976})}\BibitemShut {NoStop}%
\bibitem [{\citenamefont {Urrutia}(1978)}]{Urrutia:1977xb}%
  \BibitemOpen
  \bibfield  {author} {\bibinfo {author} {\bibfnamefont {L.}~\bibnamefont
  {Urrutia}},\ }\href {\doibase 10.1103/PhysRevD.17.1977} {\bibfield  {journal}
  {\bibinfo  {journal} {Phys. Rev. D}\ }\textbf {\bibinfo {volume} {17}},\
  \bibinfo {pages} {1977} (\bibinfo {year} {1978})}\BibitemShut {NoStop}%
\bibitem [{\citenamefont {Heyl}\ and\ \citenamefont
  {Hernquist}(1997)}]{Heyl:1997hr}%
  \BibitemOpen
  \bibfield  {author} {\bibinfo {author} {\bibfnamefont {J.~S.}\ \bibnamefont
  {Heyl}}\ and\ \bibinfo {author} {\bibfnamefont {L.}~\bibnamefont
  {Hernquist}},\ }\href {\doibase 10.1088/0305-4470/30/18/022} {\bibfield
  {journal} {\bibinfo  {journal} {J. Phys. A}\ }\textbf {\bibinfo {volume}
  {30}},\ \bibinfo {pages} {6485} (\bibinfo {year} {1997})},\ \Eprint
  {http://arxiv.org/abs/hep-ph/9705367} {arXiv:hep-ph/9705367} \BibitemShut
  {NoStop}%
\bibitem [{\citenamefont {Baier}\ and\ \citenamefont
  {Katkov}(2007)}]{Baier:2007dw}%
  \BibitemOpen
  \bibfield  {author} {\bibinfo {author} {\bibfnamefont {V.}~\bibnamefont
  {Baier}}\ and\ \bibinfo {author} {\bibfnamefont {V.}~\bibnamefont {Katkov}},\
  }\href {\doibase 10.1103/PhysRevD.75.073009} {\bibfield  {journal} {\bibinfo
  {journal} {Phys. Rev. D}\ }\textbf {\bibinfo {volume} {75}},\ \bibinfo
  {pages} {073009} (\bibinfo {year} {2007})},\ \Eprint
  {http://arxiv.org/abs/hep-ph/0701119} {arXiv:hep-ph/0701119} \BibitemShut
  {NoStop}%
\bibitem [{\citenamefont {Shabad}\ and\ \citenamefont
  {Usov}(2010)}]{Shabad:2010hx}%
  \BibitemOpen
  \bibfield  {author} {\bibinfo {author} {\bibfnamefont {A.~E.}\ \bibnamefont
  {Shabad}}\ and\ \bibinfo {author} {\bibfnamefont {V.~V.}\ \bibnamefont
  {Usov}},\ }\href {\doibase 10.1103/PhysRevD.81.125008} {\bibfield  {journal}
  {\bibinfo  {journal} {Phys. Rev. D}\ }\textbf {\bibinfo {volume} {81}},\
  \bibinfo {pages} {125008} (\bibinfo {year} {2010})},\ \Eprint
  {http://arxiv.org/abs/1002.1813} {arXiv:1002.1813 [hep-th]} \BibitemShut
  {NoStop}%
\bibitem [{\citenamefont {Hattori}\ and\ \citenamefont
  {Itakura}(2013{\natexlab{a}})}]{Hattori:2012je}%
  \BibitemOpen
  \bibfield  {author} {\bibinfo {author} {\bibfnamefont {K.}~\bibnamefont
  {Hattori}}\ and\ \bibinfo {author} {\bibfnamefont {K.}~\bibnamefont
  {Itakura}},\ }\href {\doibase 10.1016/j.aop.2012.11.010} {\bibfield
  {journal} {\bibinfo  {journal} {Annals Phys.}\ }\textbf {\bibinfo {volume}
  {330}},\ \bibinfo {pages} {23} (\bibinfo {year} {2013}{\natexlab{a}})},\
  \Eprint {http://arxiv.org/abs/1209.2663} {arXiv:1209.2663 [hep-ph]}
  \BibitemShut {NoStop}%
%%CITATION = ARXIV:1209.2663;%%
\bibitem [{\citenamefont {Hattori}\ and\ \citenamefont
  {Itakura}(2013{\natexlab{b}})}]{Hattori:2012ny}%
  \BibitemOpen
  \bibfield  {author} {\bibinfo {author} {\bibfnamefont {K.}~\bibnamefont
  {Hattori}}\ and\ \bibinfo {author} {\bibfnamefont {K.}~\bibnamefont
  {Itakura}},\ }\href {\doibase 10.1016/j.aop.2013.03.016} {\bibfield
  {journal} {\bibinfo  {journal} {Annals Phys.}\ }\textbf {\bibinfo {volume}
  {334}},\ \bibinfo {pages} {58} (\bibinfo {year} {2013}{\natexlab{b}})},\
  \Eprint {http://arxiv.org/abs/1212.1897} {arXiv:1212.1897 [hep-ph]}
  \BibitemShut {NoStop}%
%%CITATION = ARXIV:1212.1897;%%
\bibitem [{\citenamefont {Di~Piazza}\ \emph {et~al.}(2012)\citenamefont
  {Di~Piazza}, \citenamefont {Muller}, \citenamefont {Hatsagortsyan},\ and\
  \citenamefont {Keitel}}]{DiPiazza:2011tq}%
  \BibitemOpen
  \bibfield  {author} {\bibinfo {author} {\bibfnamefont {A.}~\bibnamefont
  {Di~Piazza}}, \bibinfo {author} {\bibfnamefont {C.}~\bibnamefont {Muller}},
  \bibinfo {author} {\bibfnamefont {K.}~\bibnamefont {Hatsagortsyan}}, \ and\
  \bibinfo {author} {\bibfnamefont {C.}~\bibnamefont {Keitel}},\ }\href
  {\doibase 10.1103/RevModPhys.84.1177} {\bibfield  {journal} {\bibinfo
  {journal} {Rev. Mod. Phys.}\ }\textbf {\bibinfo {volume} {84}},\ \bibinfo
  {pages} {1177} (\bibinfo {year} {2012})},\ \Eprint
  {http://arxiv.org/abs/1111.3886} {arXiv:1111.3886 [hep-ph]} \BibitemShut
  {NoStop}%
\bibitem [{\citenamefont {Zhang}\ \emph {et~al.}(2020)\citenamefont {Zhang},
  \citenamefont {Bulanov}, \citenamefont {Seipt}, \citenamefont {Arefiev},\
  and\ \citenamefont {Thomas}}]{Zhang:2020lxl}%
  \BibitemOpen
  \bibfield  {author} {\bibinfo {author} {\bibfnamefont {P.}~\bibnamefont
  {Zhang}}, \bibinfo {author} {\bibfnamefont {S.}~\bibnamefont {Bulanov}},
  \bibinfo {author} {\bibfnamefont {D.}~\bibnamefont {Seipt}}, \bibinfo
  {author} {\bibfnamefont {A.}~\bibnamefont {Arefiev}}, \ and\ \bibinfo
  {author} {\bibfnamefont {A.}~\bibnamefont {Thomas}},\ }\href {\doibase
  10.1063/1.5144449} {\bibfield  {journal} {\bibinfo  {journal} {Phys.
  Plasmas}\ }\textbf {\bibinfo {volume} {27}},\ \bibinfo {pages} {050601}
  (\bibinfo {year} {2020})},\ \Eprint {http://arxiv.org/abs/2001.00957}
  {arXiv:2001.00957 [physics.plasm-ph]} \BibitemShut {NoStop}%
\bibitem [{\citenamefont {Ejlli}\ \emph {et~al.}(2020)\citenamefont {Ejlli},
  \citenamefont {Della~Valle}, \citenamefont {Gastaldi}, \citenamefont
  {Messineo}, \citenamefont {Pengo}, \citenamefont {Ruoso},\ and\ \citenamefont
  {Zavattini}}]{Ejlli:2020yhk}%
  \BibitemOpen
  \bibfield  {author} {\bibinfo {author} {\bibfnamefont {A.}~\bibnamefont
  {Ejlli}}, \bibinfo {author} {\bibfnamefont {F.}~\bibnamefont {Della~Valle}},
  \bibinfo {author} {\bibfnamefont {U.}~\bibnamefont {Gastaldi}}, \bibinfo
  {author} {\bibfnamefont {G.}~\bibnamefont {Messineo}}, \bibinfo {author}
  {\bibfnamefont {R.}~\bibnamefont {Pengo}}, \bibinfo {author} {\bibfnamefont
  {G.}~\bibnamefont {Ruoso}}, \ and\ \bibinfo {author} {\bibfnamefont
  {G.}~\bibnamefont {Zavattini}},\ }\href {\doibase
  10.1016/j.physrep.2020.06.001} {\bibfield  {journal} {\bibinfo  {journal}
  {Phys. Rept.}\ }\textbf {\bibinfo {volume} {871}},\ \bibinfo {pages} {1}
  (\bibinfo {year} {2020})},\ \Eprint {http://arxiv.org/abs/2005.12913}
  {arXiv:2005.12913 [physics.optics]} \BibitemShut {NoStop}%
\bibitem [{\citenamefont {Fermi}(1924)}]{Fermi}%
  \BibitemOpen
  \bibfield  {author} {\bibinfo {author} {\bibfnamefont {E.}~\bibnamefont
  {Fermi}},\ }\href@noop {} {\bibfield  {journal} {\bibinfo  {journal} {Z.
  Phys. 29, 315-327}\ } (\bibinfo {year} {1924})}\BibitemShut {NoStop}%
\bibitem [{\citenamefont {von Weizsacker}(1934)}]{Weizsacker}%
  \BibitemOpen
  \bibfield  {author} {\bibinfo {author} {\bibfnamefont {C.}~\bibnamefont {von
  Weizsacker}},\ }\href@noop {} {\bibfield  {journal} {\bibinfo  {journal} {Z.
  Phys. 88, 612-625}\ } (\bibinfo {year} {1934})}\BibitemShut {NoStop}%
\bibitem [{\citenamefont {Williams}(1935)}]{Williams}%
  \BibitemOpen
  \bibfield  {author} {\bibinfo {author} {\bibfnamefont {E.}~\bibnamefont
  {Williams}},\ }\href@noop {} {\bibfield  {journal} {\bibinfo  {journal} {Kgl.
  Danske Videnskab. Selskab Mat.-fys. Medd. 13, No. 4}\ } (\bibinfo {year}
  {1935})}\BibitemShut {NoStop}%
\bibitem [{\citenamefont {Breit}\ and\ \citenamefont
  {Wheeler}(1934)}]{Breit:1934zz}%
  \BibitemOpen
  \bibfield  {author} {\bibinfo {author} {\bibfnamefont {G.}~\bibnamefont
  {Breit}}\ and\ \bibinfo {author} {\bibfnamefont {J.~A.}\ \bibnamefont
  {Wheeler}},\ }\href {\doibase 10.1103/PhysRev.46.1087} {\bibfield  {journal}
  {\bibinfo  {journal} {Phys. Rev.}\ }\textbf {\bibinfo {volume} {46}},\
  \bibinfo {pages} {1087} (\bibinfo {year} {1934})}\BibitemShut {NoStop}%
\bibitem [{\citenamefont {Adams}\ \emph {et~al.}(2004)\citenamefont {Adams}
  \emph {et~al.}}]{Adams:2004rz}%
  \BibitemOpen
  \bibfield  {author} {\bibinfo {author} {\bibfnamefont {J.}~\bibnamefont
  {Adams}} \emph {et~al.} (\bibinfo {collaboration} {STAR}),\ }\href {\doibase
  10.1103/PhysRevC.70.031902} {\bibfield  {journal} {\bibinfo  {journal} {Phys.
  Rev. C}\ }\textbf {\bibinfo {volume} {70}},\ \bibinfo {pages} {031902}
  (\bibinfo {year} {2004})},\ \Eprint {http://arxiv.org/abs/nucl-ex/0404012}
  {arXiv:nucl-ex/0404012} \BibitemShut {NoStop}%
\bibitem [{\citenamefont {Aaboud}\ \emph {et~al.}(2017)\citenamefont {Aaboud}
  \emph {et~al.}}]{Aaboud:2017bwk}%
  \BibitemOpen
  \bibfield  {author} {\bibinfo {author} {\bibfnamefont {M.}~\bibnamefont
  {Aaboud}} \emph {et~al.} (\bibinfo {collaboration} {ATLAS}),\ }\href
  {\doibase 10.1038/nphys4208} {\bibfield  {journal} {\bibinfo  {journal}
  {Nature Phys.}\ }\textbf {\bibinfo {volume} {13}},\ \bibinfo {pages} {852}
  (\bibinfo {year} {2017})},\ \Eprint {http://arxiv.org/abs/1702.01625}
  {arXiv:1702.01625 [hep-ex]} \BibitemShut {NoStop}%
\bibitem [{\citenamefont {Sirunyan}\ \emph {et~al.}(2019)\citenamefont
  {Sirunyan} \emph {et~al.}}]{Sirunyan:2018fhl}%
  \BibitemOpen
  \bibfield  {author} {\bibinfo {author} {\bibfnamefont {A.~M.}\ \bibnamefont
  {Sirunyan}} \emph {et~al.} (\bibinfo {collaboration} {CMS}),\ }\href
  {\doibase 10.1016/j.physletb.2019.134826} {\bibfield  {journal} {\bibinfo
  {journal} {Phys. Lett. B}\ }\textbf {\bibinfo {volume} {797}},\ \bibinfo
  {pages} {134826} (\bibinfo {year} {2019})},\ \Eprint
  {http://arxiv.org/abs/1810.04602} {arXiv:1810.04602 [hep-ex]} \BibitemShut
  {NoStop}%
\bibitem [{\citenamefont {Aad}\ \emph {et~al.}(2019)\citenamefont {Aad} \emph
  {et~al.}}]{Aad:2019ock}%
  \BibitemOpen
  \bibfield  {author} {\bibinfo {author} {\bibfnamefont {G.}~\bibnamefont
  {Aad}} \emph {et~al.} (\bibinfo {collaboration} {ATLAS}),\ }\href {\doibase
  10.1103/PhysRevLett.123.052001} {\bibfield  {journal} {\bibinfo  {journal}
  {Phys. Rev. Lett.}\ }\textbf {\bibinfo {volume} {123}},\ \bibinfo {pages}
  {052001} (\bibinfo {year} {2019})},\ \Eprint
  {http://arxiv.org/abs/1904.03536} {arXiv:1904.03536 [hep-ex]} \BibitemShut
  {NoStop}%
\bibitem [{\citenamefont {Adam}\ \emph {et~al.}(2019)\citenamefont {Adam} \emph
  {et~al.}}]{Adam:2019mby}%
  \BibitemOpen
  \bibfield  {author} {\bibinfo {author} {\bibfnamefont {J.}~\bibnamefont
  {Adam}} \emph {et~al.} (\bibinfo {collaboration} {STAR}),\ }\href@noop {} {\
  (\bibinfo {year} {2019})},\ \Eprint {http://arxiv.org/abs/1910.12400}
  {arXiv:1910.12400 [nucl-ex]} \BibitemShut {NoStop}%
\bibitem [{\citenamefont {Kouveliotou}\ \emph {et~al.}(2003)\citenamefont
  {Kouveliotou}, \citenamefont {Duncan},\ and\ \citenamefont
  {Thompson}}]{Kouveliotou:2003tb}%
  \BibitemOpen
  \bibfield  {author} {\bibinfo {author} {\bibfnamefont {C.}~\bibnamefont
  {Kouveliotou}}, \bibinfo {author} {\bibfnamefont {R.}~\bibnamefont {Duncan}},
  \ and\ \bibinfo {author} {\bibfnamefont {C.}~\bibnamefont {Thompson}},\
  }\href@noop {} {\bibfield  {journal} {\bibinfo  {journal} {Sci. Am.}\
  }\textbf {\bibinfo {volume} {288N2}},\ \bibinfo {pages} {24} (\bibinfo {year}
  {2003})}\BibitemShut {NoStop}%
\bibitem [{\citenamefont {Harding}\ and\ \citenamefont
  {Lai}(2006)}]{Harding:2006qn}%
  \BibitemOpen
  \bibfield  {author} {\bibinfo {author} {\bibfnamefont {A.~K.}\ \bibnamefont
  {Harding}}\ and\ \bibinfo {author} {\bibfnamefont {D.}~\bibnamefont {Lai}},\
  }\href {\doibase 10.1088/0034-4885/69/9/R03} {\bibfield  {journal} {\bibinfo
  {journal} {Rept. Prog. Phys.}\ }\textbf {\bibinfo {volume} {69}},\ \bibinfo
  {pages} {2631} (\bibinfo {year} {2006})},\ \Eprint
  {http://arxiv.org/abs/astro-ph/0606674} {arXiv:astro-ph/0606674} \BibitemShut
  {NoStop}%
\bibitem [{\citenamefont {Staubert}\ \emph {et~al.}(2019)\citenamefont
  {Staubert} \emph {et~al.}}]{Staubert:2018xgw}%
  \BibitemOpen
  \bibfield  {author} {\bibinfo {author} {\bibfnamefont {R.}~\bibnamefont
  {Staubert}} \emph {et~al.},\ }\href {\doibase 10.1051/0004-6361/201834479}
  {\bibfield  {journal} {\bibinfo  {journal} {Astron. Astrophys.}\ }\textbf
  {\bibinfo {volume} {622}},\ \bibinfo {pages} {A61} (\bibinfo {year}
  {2019})},\ \Eprint {http://arxiv.org/abs/1812.03461} {arXiv:1812.03461
  [astro-ph.HE]} \BibitemShut {NoStop}%
\bibitem [{\citenamefont {Bignami}\ \emph {et~al.}(2003)\citenamefont
  {Bignami}, \citenamefont {Caraveo}, \citenamefont {De~Luca},\ and\
  \citenamefont {Mereghetti}}]{bignami2003magnetic}%
  \BibitemOpen
  \bibfield  {author} {\bibinfo {author} {\bibfnamefont {G.}~\bibnamefont
  {Bignami}}, \bibinfo {author} {\bibfnamefont {P.}~\bibnamefont {Caraveo}},
  \bibinfo {author} {\bibfnamefont {A.}~\bibnamefont {De~Luca}}, \ and\
  \bibinfo {author} {\bibfnamefont {S.}~\bibnamefont {Mereghetti}},\
  }\href@noop {} {\bibfield  {journal} {\bibinfo  {journal} {Nature}\ }\textbf
  {\bibinfo {volume} {423}},\ \bibinfo {pages} {725} (\bibinfo {year}
  {2003})}\BibitemShut {NoStop}%
\bibitem [{\citenamefont {Grasso}\ and\ \citenamefont
  {Rubinstein}(2001)}]{Grasso:2000wj}%
  \BibitemOpen
  \bibfield  {author} {\bibinfo {author} {\bibfnamefont {D.}~\bibnamefont
  {Grasso}}\ and\ \bibinfo {author} {\bibfnamefont {H.~R.}\ \bibnamefont
  {Rubinstein}},\ }\href {\doibase 10.1016/S0370-1573(00)00110-1} {\bibfield
  {journal} {\bibinfo  {journal} {Phys. Rept.}\ }\textbf {\bibinfo {volume}
  {348}},\ \bibinfo {pages} {163} (\bibinfo {year} {2001})},\ \Eprint
  {http://arxiv.org/abs/astro-ph/0009061} {arXiv:astro-ph/0009061 [astro-ph]}
  \BibitemShut {NoStop}%
%%CITATION = ASTRO-PH/0009061;%%
\bibitem [{\citenamefont {Giovannini}(2004)}]{Giovannini:2003yn}%
  \BibitemOpen
  \bibfield  {author} {\bibinfo {author} {\bibfnamefont {M.}~\bibnamefont
  {Giovannini}},\ }\href {\doibase 10.1142/S0218271804004530} {\bibfield
  {journal} {\bibinfo  {journal} {Int. J. Mod. Phys.}\ }\textbf {\bibinfo
  {volume} {D13}},\ \bibinfo {pages} {391} (\bibinfo {year} {2004})},\ \Eprint
  {http://arxiv.org/abs/astro-ph/0312614} {arXiv:astro-ph/0312614 [astro-ph]}
  \BibitemShut {NoStop}%
%%CITATION = ASTRO-PH/0312614;%%
\bibitem [{\citenamefont {Kandus}\ \emph {et~al.}(2011)\citenamefont {Kandus},
  \citenamefont {Kunze},\ and\ \citenamefont {Tsagas}}]{Kandus:2010nw}%
  \BibitemOpen
  \bibfield  {author} {\bibinfo {author} {\bibfnamefont {A.}~\bibnamefont
  {Kandus}}, \bibinfo {author} {\bibfnamefont {K.~E.}\ \bibnamefont {Kunze}}, \
  and\ \bibinfo {author} {\bibfnamefont {C.~G.}\ \bibnamefont {Tsagas}},\
  }\href {\doibase 10.1016/j.physrep.2011.03.001} {\bibfield  {journal}
  {\bibinfo  {journal} {Phys. Rept.}\ }\textbf {\bibinfo {volume} {505}},\
  \bibinfo {pages} {1} (\bibinfo {year} {2011})},\ \Eprint
  {http://arxiv.org/abs/1007.3891} {arXiv:1007.3891 [astro-ph.CO]} \BibitemShut
  {NoStop}%
\bibitem [{\citenamefont {Sadooghi}\ and\ \citenamefont
  {Taghinavaz}(2017)}]{Sadooghi:2016jyf}%
  \BibitemOpen
  \bibfield  {author} {\bibinfo {author} {\bibfnamefont {N.}~\bibnamefont
  {Sadooghi}}\ and\ \bibinfo {author} {\bibfnamefont {F.}~\bibnamefont
  {Taghinavaz}},\ }\href {\doibase 10.1016/j.aop.2016.11.008} {\bibfield
  {journal} {\bibinfo  {journal} {Annals Phys.}\ }\textbf {\bibinfo {volume}
  {376}},\ \bibinfo {pages} {218} (\bibinfo {year} {2017})},\ \Eprint
  {http://arxiv.org/abs/1601.04887} {arXiv:1601.04887 [hep-ph]} \BibitemShut
  {NoStop}%
%%CITATION = ARXIV:1601.04887;%%
\bibitem [{\citenamefont {Ghosh}\ and\ \citenamefont
  {Chandra}(2018)}]{Ghosh:2018xhh}%
  \BibitemOpen
  \bibfield  {author} {\bibinfo {author} {\bibfnamefont {S.}~\bibnamefont
  {Ghosh}}\ and\ \bibinfo {author} {\bibfnamefont {V.}~\bibnamefont
  {Chandra}},\ }\href {\doibase 10.1103/PhysRevD.98.076006} {\bibfield
  {journal} {\bibinfo  {journal} {Phys. Rev. D}\ }\textbf {\bibinfo {volume}
  {98}},\ \bibinfo {pages} {076006} (\bibinfo {year} {2018})},\ \Eprint
  {http://arxiv.org/abs/1808.05176} {arXiv:1808.05176 [hep-ph]} \BibitemShut
  {NoStop}%
\bibitem [{\citenamefont {Wang}\ \emph {et~al.}(2020)\citenamefont {Wang},
  \citenamefont {Shovkovy}, \citenamefont {Yu},\ and\ \citenamefont
  {Huang}}]{Wang:2020dsr}%
  \BibitemOpen
  \bibfield  {author} {\bibinfo {author} {\bibfnamefont {X.}~\bibnamefont
  {Wang}}, \bibinfo {author} {\bibfnamefont {I.~A.}\ \bibnamefont {Shovkovy}},
  \bibinfo {author} {\bibfnamefont {L.}~\bibnamefont {Yu}}, \ and\ \bibinfo
  {author} {\bibfnamefont {M.}~\bibnamefont {Huang}},\ }\href {\doibase
  10.1103/PhysRevD.102.076010} {\bibfield  {journal} {\bibinfo  {journal}
  {Phys. Rev. D}\ }\textbf {\bibinfo {volume} {102}},\ \bibinfo {pages}
  {076010} (\bibinfo {year} {2020})},\ \Eprint
  {http://arxiv.org/abs/2006.16254} {arXiv:2006.16254 [hep-ph]} \BibitemShut
  {NoStop}%
\bibitem [{\citenamefont {Ghosh}\ \emph {et~al.}(2020)\citenamefont {Ghosh},
  \citenamefont {Chaudhuri}, \citenamefont {Sarkar},\ and\ \citenamefont
  {Roy}}]{Ghosh:2020xwp}%
  \BibitemOpen
  \bibfield  {author} {\bibinfo {author} {\bibfnamefont {S.}~\bibnamefont
  {Ghosh}}, \bibinfo {author} {\bibfnamefont {N.}~\bibnamefont {Chaudhuri}},
  \bibinfo {author} {\bibfnamefont {S.}~\bibnamefont {Sarkar}}, \ and\ \bibinfo
  {author} {\bibfnamefont {P.}~\bibnamefont {Roy}},\ }\href {\doibase
  10.1103/PhysRevD.101.096002} {\bibfield  {journal} {\bibinfo  {journal}
  {Phys. Rev. D}\ }\textbf {\bibinfo {volume} {101}},\ \bibinfo {pages}
  {096002} (\bibinfo {year} {2020})},\ \Eprint
  {http://arxiv.org/abs/2004.09203} {arXiv:2004.09203 [nucl-th]} \BibitemShut
  {NoStop}%
\bibitem [{\citenamefont {Ritus}(1972)}]{Ritus:1972ky}%
  \BibitemOpen
  \bibfield  {author} {\bibinfo {author} {\bibfnamefont {V.~I.}\ \bibnamefont
  {Ritus}},\ }\href {\doibase 10.1016/0003-4916(72)90191-1} {\bibfield
  {journal} {\bibinfo  {journal} {Annals Phys.}\ }\textbf {\bibinfo {volume}
  {69}},\ \bibinfo {pages} {555} (\bibinfo {year} {1972})}\BibitemShut
  {NoStop}%
%%CITATION = APNYA,69,555;%%
\bibitem [{\citenamefont {Ritus}(1978)}]{Ritus:1978cj}%
  \BibitemOpen
  \bibfield  {author} {\bibinfo {author} {\bibfnamefont {V.~I.}\ \bibnamefont
  {Ritus}},\ }\href@noop {} {\bibfield  {journal} {\bibinfo  {journal} {Sov.
  Phys. JETP}\ }\textbf {\bibinfo {volume} {48}},\ \bibinfo {pages} {788}
  (\bibinfo {year} {1978})},\ \bibinfo {note} {[Zh. Eksp. Teor.
  Fiz.75,1560(1978)]}\BibitemShut {NoStop}%
%%CITATION = SPHJA,48,788;%%
\bibitem [{\citenamefont {Hattori}\ \emph {et~al.}()\citenamefont {Hattori},
  \citenamefont {Itakura},\ and\ \citenamefont {Ozaki}}]{HIO}%
  \BibitemOpen
  \bibfield  {author} {\bibinfo {author} {\bibfnamefont {K.}~\bibnamefont
  {Hattori}}, \bibinfo {author} {\bibfnamefont {K.}~\bibnamefont {Itakura}}, \
  and\ \bibinfo {author} {\bibfnamefont {S.}~\bibnamefont {Ozaki}},\
  }\href@noop {} {\bibinfo  {journal} {To be published}\ }\BibitemShut
  {NoStop}%
\bibitem [{\citenamefont {Peskin}\ and\ \citenamefont
  {Schroeder}(1995)}]{Peskin:1995ev}%
  \BibitemOpen
\bibfield  {journal} {  }\bibfield  {author} {\bibinfo {author} {\bibfnamefont
  {M.~E.}\ \bibnamefont {Peskin}}\ and\ \bibinfo {author} {\bibfnamefont
  {D.~V.}\ \bibnamefont {Schroeder}},\ }\href
  {http://www.slac.stanford.edu/spires/find/books/www?cl=QC174.45%3AP4} {\emph
  {\bibinfo {title} {{An Introduction to quantum field theory}}}}\ (\bibinfo
  {year} {1995})\BibitemShut {NoStop}%
%%CITATION = INSPIRE-407703;%%
\bibitem [{\citenamefont {Donoghue}\ \emph {et~al.}(1992)\citenamefont
  {Donoghue}, \citenamefont {Golowich},\ and\ \citenamefont
  {Holstein}}]{Donoghue:1992dd}%
  \BibitemOpen
  \bibfield  {author} {\bibinfo {author} {\bibfnamefont {J.~F.}\ \bibnamefont
  {Donoghue}}, \bibinfo {author} {\bibfnamefont {E.}~\bibnamefont {Golowich}},
  \ and\ \bibinfo {author} {\bibfnamefont {B.~R.}\ \bibnamefont {Holstein}},\
  }\href {\doibase 10.1017/CBO9780511524370} {\bibfield  {journal} {\bibinfo
  {journal} {Camb. Monogr. Part. Phys. Nucl. Phys. Cosmol.}\ }\textbf {\bibinfo
  {volume} {2}},\ \bibinfo {pages} {1} (\bibinfo {year} {1992})},\ \bibinfo
  {note} {[Camb. Monogr. Part. Phys. Nucl. Phys. Cosmol.35(2014)]}\BibitemShut
  {NoStop}%
%%CITATION = CMPCE,2,1;%%
\bibitem [{\citenamefont {Zyla}\ \emph {et~al.}(2020)\citenamefont {Zyla} \emph
  {et~al.}}]{Zyla:2020zbs}%
  \BibitemOpen
  \bibfield  {author} {\bibinfo {author} {\bibfnamefont {P.}~\bibnamefont
  {Zyla}} \emph {et~al.} (\bibinfo {collaboration} {Particle Data Group}),\
  }\href {\doibase 10.1093/ptep/ptaa104} {\bibfield  {journal} {\bibinfo
  {journal} {PTEP}\ }\textbf {\bibinfo {volume} {2020}},\ \bibinfo {pages}
  {083C01} (\bibinfo {year} {2020})}\BibitemShut {NoStop}%
\bibitem [{\citenamefont {Hattori}\ and\ \citenamefont
  {Huang}(2017)}]{Hattori:2016emy}%
  \BibitemOpen
  \bibfield  {author} {\bibinfo {author} {\bibfnamefont {K.}~\bibnamefont
  {Hattori}}\ and\ \bibinfo {author} {\bibfnamefont {X.-G.}\ \bibnamefont
  {Huang}},\ }\href {\doibase 10.1007/s41365-016-0178-3} {\bibfield  {journal}
  {\bibinfo  {journal} {Nucl. Sci. Tech.}\ }\textbf {\bibinfo {volume} {28}},\
  \bibinfo {pages} {26} (\bibinfo {year} {2017})},\ \Eprint
  {http://arxiv.org/abs/1609.00747} {arXiv:1609.00747 [nucl-th]} \BibitemShut
  {NoStop}%
%%CITATION = ARXIV:1609.00747;%%
\bibitem [{\citenamefont {Zha}\ \emph {et~al.}(2020)\citenamefont {Zha},
  \citenamefont {Brandenburg}, \citenamefont {Tang},\ and\ \citenamefont
  {Xu}}]{Zha:2018tlq}%
  \BibitemOpen
  \bibfield  {author} {\bibinfo {author} {\bibfnamefont {W.}~\bibnamefont
  {Zha}}, \bibinfo {author} {\bibfnamefont {J.~D.}\ \bibnamefont
  {Brandenburg}}, \bibinfo {author} {\bibfnamefont {Z.}~\bibnamefont {Tang}}, \
  and\ \bibinfo {author} {\bibfnamefont {Z.}~\bibnamefont {Xu}},\ }\href
  {\doibase 10.1016/j.physletb.2019.135089} {\bibfield  {journal} {\bibinfo
  {journal} {Phys. Lett. B}\ }\textbf {\bibinfo {volume} {800}},\ \bibinfo
  {pages} {135089} (\bibinfo {year} {2020})},\ \Eprint
  {http://arxiv.org/abs/1812.02820} {arXiv:1812.02820 [nucl-th]} \BibitemShut
  {NoStop}%
\bibitem [{\citenamefont {Klein}\ \emph {et~al.}(2019)\citenamefont {Klein},
  \citenamefont {Mueller}, \citenamefont {Xiao},\ and\ \citenamefont
  {Yuan}}]{Klein:2018fmp}%
  \BibitemOpen
  \bibfield  {author} {\bibinfo {author} {\bibfnamefont {S.}~\bibnamefont
  {Klein}}, \bibinfo {author} {\bibfnamefont {A.}~\bibnamefont {Mueller}},
  \bibinfo {author} {\bibfnamefont {B.-W.}\ \bibnamefont {Xiao}}, \ and\
  \bibinfo {author} {\bibfnamefont {F.}~\bibnamefont {Yuan}},\ }\href {\doibase
  10.1103/PhysRevLett.122.132301} {\bibfield  {journal} {\bibinfo  {journal}
  {Phys. Rev. Lett.}\ }\textbf {\bibinfo {volume} {122}},\ \bibinfo {pages}
  {132301} (\bibinfo {year} {2019})},\ \Eprint
  {http://arxiv.org/abs/1811.05519} {arXiv:1811.05519 [hep-ph]} \BibitemShut
  {NoStop}%
\bibitem [{\citenamefont {Li}\ \emph {et~al.}(2019)\citenamefont {Li},
  \citenamefont {Zhou},\ and\ \citenamefont {Zhou}}]{Li:2019yzy}%
  \BibitemOpen
  \bibfield  {author} {\bibinfo {author} {\bibfnamefont {C.}~\bibnamefont
  {Li}}, \bibinfo {author} {\bibfnamefont {J.}~\bibnamefont {Zhou}}, \ and\
  \bibinfo {author} {\bibfnamefont {Y.-J.}\ \bibnamefont {Zhou}},\ }\href
  {\doibase 10.1016/j.physletb.2019.07.005} {\bibfield  {journal} {\bibinfo
  {journal} {Phys. Lett. B}\ }\textbf {\bibinfo {volume} {795}},\ \bibinfo
  {pages} {576} (\bibinfo {year} {2019})},\ \Eprint
  {http://arxiv.org/abs/1903.10084} {arXiv:1903.10084 [hep-ph]} \BibitemShut
  {NoStop}%
\bibitem [{\citenamefont {Li}\ \emph {et~al.}(2020)\citenamefont {Li},
  \citenamefont {Zhou},\ and\ \citenamefont {Zhou}}]{Li:2019sin}%
  \BibitemOpen
  \bibfield  {author} {\bibinfo {author} {\bibfnamefont {C.}~\bibnamefont
  {Li}}, \bibinfo {author} {\bibfnamefont {J.}~\bibnamefont {Zhou}}, \ and\
  \bibinfo {author} {\bibfnamefont {Y.-J.}\ \bibnamefont {Zhou}},\ }\href
  {\doibase 10.1103/PhysRevD.101.034015} {\bibfield  {journal} {\bibinfo
  {journal} {Phys. Rev. D}\ }\textbf {\bibinfo {volume} {101}},\ \bibinfo
  {pages} {034015} (\bibinfo {year} {2020})},\ \Eprint
  {http://arxiv.org/abs/1911.00237} {arXiv:1911.00237 [hep-ph]} \BibitemShut
  {NoStop}%
\bibitem [{\citenamefont {Xiao}\ \emph {et~al.}(2020)\citenamefont {Xiao},
  \citenamefont {Yuan},\ and\ \citenamefont {Zhou}}]{Xiao:2020ddm}%
  \BibitemOpen
  \bibfield  {author} {\bibinfo {author} {\bibfnamefont {B.-W.}\ \bibnamefont
  {Xiao}}, \bibinfo {author} {\bibfnamefont {F.}~\bibnamefont {Yuan}}, \ and\
  \bibinfo {author} {\bibfnamefont {J.}~\bibnamefont {Zhou}},\ }\href@noop {}
  {\  (\bibinfo {year} {2020})},\ \Eprint {http://arxiv.org/abs/2003.06352}
  {arXiv:2003.06352 [hep-ph]} \BibitemShut {NoStop}%
\bibitem [{\citenamefont {Klein}\ \emph {et~al.}(2020)\citenamefont {Klein},
  \citenamefont {Mueller}, \citenamefont {Xiao},\ and\ \citenamefont
  {Yuan}}]{Klein:2020jom}%
  \BibitemOpen
  \bibfield  {author} {\bibinfo {author} {\bibfnamefont {S.}~\bibnamefont
  {Klein}}, \bibinfo {author} {\bibfnamefont {A.}~\bibnamefont {Mueller}},
  \bibinfo {author} {\bibfnamefont {B.-W.}\ \bibnamefont {Xiao}}, \ and\
  \bibinfo {author} {\bibfnamefont {F.}~\bibnamefont {Yuan}},\ }\href@noop {}
  {\  (\bibinfo {year} {2020})},\ \Eprint {http://arxiv.org/abs/2003.02947}
  {arXiv:2003.02947 [hep-ph]} \BibitemShut {NoStop}%
\bibitem [{\citenamefont {Harding}(2013)}]{Harding:2013ij}%
  \BibitemOpen
  \bibfield  {author} {\bibinfo {author} {\bibfnamefont {A.~K.}\ \bibnamefont
  {Harding}},\ }\href {\doibase 10.1007/s11467-013-0285-0} {\bibfield
  {journal} {\bibinfo  {journal} {Front. Phys. (Beijing)}\ }\textbf {\bibinfo
  {volume} {8}},\ \bibinfo {pages} {679} (\bibinfo {year} {2013})},\ \Eprint
  {http://arxiv.org/abs/1302.0869} {arXiv:1302.0869 [astro-ph.HE]} \BibitemShut
  {NoStop}%
\bibitem [{\citenamefont {Enoto}\ \emph {et~al.}(2019)\citenamefont {Enoto},
  \citenamefont {Kisaka},\ and\ \citenamefont {Shibata}}]{Enoto:2019vcg}%
  \BibitemOpen
  \bibfield  {author} {\bibinfo {author} {\bibfnamefont {T.}~\bibnamefont
  {Enoto}}, \bibinfo {author} {\bibfnamefont {S.}~\bibnamefont {Kisaka}}, \
  and\ \bibinfo {author} {\bibfnamefont {S.}~\bibnamefont {Shibata}},\ }\href
  {\doibase 10.1088/1361-6633/ab3def} {\bibfield  {journal} {\bibinfo
  {journal} {Rept. Prog. Phys.}\ }\textbf {\bibinfo {volume} {82}},\ \bibinfo
  {pages} {106901} (\bibinfo {year} {2019})}\BibitemShut {NoStop}%
\bibitem [{\citenamefont {Bragin}\ \emph {et~al.}(2017)\citenamefont {Bragin},
  \citenamefont {Meuren}, \citenamefont {Keitel},\ and\ \citenamefont
  {Di~Piazza}}]{Bragin:2017yau}%
  \BibitemOpen
  \bibfield  {author} {\bibinfo {author} {\bibfnamefont {S.}~\bibnamefont
  {Bragin}}, \bibinfo {author} {\bibfnamefont {S.}~\bibnamefont {Meuren}},
  \bibinfo {author} {\bibfnamefont {C.~H.}\ \bibnamefont {Keitel}}, \ and\
  \bibinfo {author} {\bibfnamefont {A.}~\bibnamefont {Di~Piazza}},\ }\href
  {\doibase 10.1103/PhysRevLett.119.250403} {\bibfield  {journal} {\bibinfo
  {journal} {Phys. Rev. Lett.}\ }\textbf {\bibinfo {volume} {119}},\ \bibinfo
  {pages} {250403} (\bibinfo {year} {2017})},\ \Eprint
  {http://arxiv.org/abs/1704.05234} {arXiv:1704.05234 [hep-ph]} \BibitemShut
  {NoStop}%
\bibitem [{\citenamefont {Yanovsky}\ \emph {et~al.}(2008)\citenamefont
  {Yanovsky}, \citenamefont {Chvykov}, \citenamefont {Kalinchenko},
  \citenamefont {Rousseau}, \citenamefont {Planchon}, \citenamefont {Matsuoka},
  \citenamefont {Maksimchuk}, \citenamefont {Nees}, \citenamefont {Cheriaux},
  \citenamefont {Mourou},\ and\ \citenamefont {Krushelnick}}]{Yanovsky:08}%
  \BibitemOpen
  \bibfield  {author} {\bibinfo {author} {\bibfnamefont {V.}~\bibnamefont
  {Yanovsky}}, \bibinfo {author} {\bibfnamefont {V.}~\bibnamefont {Chvykov}},
  \bibinfo {author} {\bibfnamefont {G.}~\bibnamefont {Kalinchenko}}, \bibinfo
  {author} {\bibfnamefont {P.}~\bibnamefont {Rousseau}}, \bibinfo {author}
  {\bibfnamefont {T.}~\bibnamefont {Planchon}}, \bibinfo {author}
  {\bibfnamefont {T.}~\bibnamefont {Matsuoka}}, \bibinfo {author}
  {\bibfnamefont {A.}~\bibnamefont {Maksimchuk}}, \bibinfo {author}
  {\bibfnamefont {J.}~\bibnamefont {Nees}}, \bibinfo {author} {\bibfnamefont
  {G.}~\bibnamefont {Cheriaux}}, \bibinfo {author} {\bibfnamefont
  {G.}~\bibnamefont {Mourou}}, \ and\ \bibinfo {author} {\bibfnamefont
  {K.}~\bibnamefont {Krushelnick}},\ }\href {\doibase 10.1364/OE.16.002109}
  {\bibfield  {journal} {\bibinfo  {journal} {Opt. Express}\ }\textbf {\bibinfo
  {volume} {16}},\ \bibinfo {pages} {2109} (\bibinfo {year}
  {2008})}\BibitemShut {NoStop}%
\bibitem [{\citenamefont {Ritus}(1985)}]{Ritus:thesis}%
  \BibitemOpen
  \bibfield  {author} {\bibinfo {author} {\bibfnamefont {V.~I.}\ \bibnamefont
  {Ritus}},\ }\href@noop {} {\bibfield  {journal} {\bibinfo  {journal}
  {J.~Sov.~Laser~Res.}\ }\textbf {\bibinfo {volume} {6}},\ \bibinfo {pages}
  {497} (\bibinfo {year} {1985})}\BibitemShut {NoStop}%
\bibitem [{\citenamefont {Weisstein}()}]{AssociatedLaguerrePolynomial}%
  \BibitemOpen
  \bibfield  {author} {\bibinfo {author} {\bibfnamefont {E.~W.}\ \bibnamefont
  {Weisstein}},\ }\href
  {http://mathworld.wolfram.com/AssociatedLaguerrePolynomial.html} {\bibinfo
  {journal} {From MathWorld--A Wolfram Web Resource.
  http://mathworld.wolfram.com/AssociatedLaguerrePolynomial.html}\
  }\BibitemShut {NoStop}%
\end{thebibliography}%

\end{document}